%% file: mn2e_revised2astroph.tex
\documentclass[useAMS,usenatbib]{mn2e}

\usepackage{graphicx}
\usepackage{lscape}
\usepackage{aalongtable}
\usepackage{deluxetable}
\usepackage{ulem}

\title[The s-Process in Low Metallicity Stars. II.]
{The s-Process in Low Metallicity Stars. \\
II. Interpretation of High-Resolution 
   Spectroscopic Observations with AGB models.}
\author[S. Bisterzo, R. Gallino, O. Straniero, S. Cristallo, F. K\"appeler]
{S. Bisterzo$^{1}$\thanks{E-mail: bisterzo@ph.unito.it (AVR); sarabisterzo@gmail.com (ANO)}, 
R. Gallino$^{1,2}$,
O. Straniero$^{2}$,
S. Cristallo$^{3}$
and 
F. K\"appeler$^{4}$\\
$^{1}$Dipartimento di Fisica Generale, Universit\`{a} 
   di Torino, Via P. Giuria 1, 10125 Torino, Italy\\
$^{2}$INAF Osservatorio Astronomico di Collurania, via M. 
   Maggini, 64100 Teramo, Italy\\
$^{3}$Departamento de Fisica Teorica y del Cosmos, Universidad
de Granada, Campus de Fuentenueva, 18071 Granada, Spain\\
$^{4}$Karlsruhe Institute of Technology, Campus Nord, 
Institut f$\ddot{\rm u}$r Kernphysik, D-76021 Karlsruhe, Germany\\}

\begin{document}

\date{Accepted 1988 December 15. Received 1988 December 14; in original form 1988 October 11}

\pagerange{\pageref{firstpage}--\pageref{lastpage}} \pubyear{2002}

\maketitle

\label{firstpage}

\begin{abstract}

High-resolution spectroscopic observations of a hundred
metal-poor Carbon and $s$-rich stars (CEMP-$s$) collected 
from the literature are compared with the theoretical 
nucleosynthesis models of asymptotic giant branch (AGB) presented 
in Paper I ($M^{\rm AGB}_{\rm ini}$ = 1.3, 1.4, 1.5, 2 $M_{\odot}$, $-$3.6
$\la$ [Fe/H] $\la$ $-$1.5).
The $s$-process enhancement detected in these objects is associated 
to binary systems: the more massive companion evolved faster through
the thermally pulsing AGB phase (TP-AGB), synthesising in the inner 
He-intershell the $s$-elements, which are partly dredged-up to 
the surface during the third dredge-up (TDU) episode.
The secondary observed low mass companion became CEMP-$s$ by mass 
transfer of C and $s$-rich material from the primary AGB.
\\
We analyse the light elements as C, N, O, Na and Mg, as well as the two 
$s$-process indicators, [hs/ls] (where ls = $<$Y, Zr$>$ is the
the light-$s$ peak at N = 50 and hs = $<$La, Nd, Sm$>$ the heavy-$s$ 
peak at N = 82), and [Pb/hs]. 
We distinguish between CEMP-$s$ with high $s$-process enhancement, 
[hs/Fe] $\ga$ 1.5 (CEMP-$s$II), and mild $s$-process enhanced stars,
[hs/Fe] $<$ 1.5 (CEMP-$s$I).
To interpret the observations, a 
range of $s$-process efficiencies at any given metallicity is necessary.
This is confirmed by the high spread observed in [Pb/hs] ($\sim$ 2 dex). 
A degeneration of solutions is found with some exceptions:
most main-sequence CEMP-$s$II stars with low [Na/Fe] 
can only be interpreted with $M^{\rm AGB}_{\rm ini}$ = 1.3 -- 1.4 
$M_{\odot}$. 
Giants having suffered the first dredge-up (FDU) need a dilution 
$\ga$ 1 dex ($dil$ is defined as the mass of the 
convective envelope of the observed star, $M^{\rm obs}_\star$, over the 
material transferred from the AGB to the companion, $M^{\rm trans}_{\rm AGB}$).
Then, AGB models with higher AGB initial masses ($M^{\rm AGB}_{\rm ini}$ = 1.5 
-- 2 $M_{\odot}$) are adopted to interpret CEMP-$s$II giants.
In general, solutions with AGB models in the mass range 
$M^{\rm AGB}_{\rm ini}$ = 1.3 -- 2 $M_{\odot}$ and different dilution 
factors are found for CEMP-$s$I stars.
\\
About half of the CEMP-$s$ stars with europium measurements show
a high $r$-process enhancement (CEMP-$s/r$). 
The scenario for the origin of CEMP-$s/r$ stars is a debated issue. 
We propose that the molecular cloud, from which the binary system 
formed, was previously enriched in $r$-process elements, most likely by
local SN~II pollution.
This initial $r$-enrichment does not affect the $s$-process nucleosynthesis. 
However, for a high $r$-process enrichment ([r/Fe]$^{\rm ini}$ = 2), 
the $r$-process contributions to solar La, Nd and Sm (30\%, 40\%, 70\%)
have to be considered. 
This increases the maximum [hs/ls] up to $\sim$ 0.3 dex. 
CEMP-$s/r$ stars reflect this behaviour, showing 
higher [hs/ls] than observed in CEMP-$s$ on average.
\\
Detailed analyses for individual stars will be provided in Paper III.

\end{abstract}



\section{Introduction}\label{intro}

The chemical compositions observed in metal-poor stars are a test
of stellar models of nucleosynthesis. 
Recent high-resolution spectroscopic surveys\footnote{The ESO Large 
Program First Stars \citep{cayrel04}; the HK-survey \citep{beers92,cohen05,beers07}; 
the Hamburg/ESO Survey \citep{christlieb03}; the SEGUE survey, Sloan 
Extension for Galactic Exploration and Understanding, the SEGUE Stellar 
Parameter Pipeline, SSPP \citep{lee08a,lee08b}; the Sloan Digital Sky 
Survey, SDSS \citep{york00}; 
the Chemical Abundances of Stars in the Halo (CASH) Projects with the 
Hobby-Eberly Telescope, HET \citep{ramsey98}.} have identified a 
sizeable number of metal-poor stars with carbon enhancement (CEMP).
Following the classification by \citet{beers05}, we may distinguish
CEMP stars in: CEMP-$s$, CEMP-$s/r$, CEMP-$r$, CEMP-no.
Stars with enhancement in $s$-process elements (CEMP-$s$) are the 
majority among the CEMP stars ($\sim$ 80\%, \citealt{aoki07}). 
About half of CEMP-$s$ stars, for which Eu (a typical $r$-process element)
has been detected, are enriched in both $s$- and $r$-elements, the 
CEMP-$s/r$ stars.  
Starting with the first discoveries of CEMP-$s/r$ stars by \citet{barbuy97}, 
\citet{hill00} and \citet{cohen03}, one of the most debated issues has begun.
Indeed, CEMP-$s/r$ stars show in their photosphere a competition between two 
neutron capture processes of completely different astrophysical origin, 
the $s$-process and the $r$-process. 
Finally, CEMP-$r$ stars exhibit a strong enhancement in $r$-process 
elements only, and CEMP-no do not show appreciable abundances 
of heavy elements.
\\
\citet{aoki02alpha,aoki04} identified in CS 29498--043 a further subclass, 
the CEMP-$\alpha$ stars, which show a large excess of C, N, O, and $\alpha$-elements. 
Similar characteristics have been detected in the giant CS 22949--037,
previously detected by \citet{norris01,depagne02}.
A few objects, called CEMP-no/$s$ \citep{sivarani06}, show subsolar Sr and 
a moderately enhanced Ba (about 1 dex).
Other candidates are SDSS J1036+1212 \citep{behara10} and BD $-$1$^{\circ}$2582
\citep{simmerer04,roederer10r}. 

This paper is focused on the theoretical interpretation of CEMP-$s$ and 
CEMP-$s/r$ stars.
These are old stars of low initial mass 
($M$ $<$ 0.9 $M_\odot$) 
on the main-sequence or the giant phase.
The most plausible explanation of their observed $s$-enhancement is mass 
transfer by stellar winds in binary systems from a primary companion while 
on its asymptotic giant branch (AGB) phase (now a white dwarf).

The $s$-process nucleosynthesis is mainly ascribed to AGB stars 
during their thermally pulsing (TP) phase.
After a thermal instability, 
the bottom of the convective envelope penetrates into the top layers of the 
region between the H- and He-shells (He-intershell), enriching the 
surface with freshly synthesised $^{12}$C and $s$-process elements.
This recurrent phenomenon is called third dredge-up (TDU).
The major neutron source is the $^{13}$C($\alpha$, n)$^{16}$O reaction.
$^{13}$C burns radiatively in a thin layer of the He-intershell, the
so-called $^{13}$C-pocket, at a temperature of about 0.9 $\times$ 10$^{8}$ K
\citep{straniero95}. During the TDU, a small amount of protons at the base 
of the convective envelope are assumed to penetrate in the top layers 
of the He-intershell, and are captured by the primary $^{12}$C 
directly produced by helium burning during previous TPs 
via the $^{12}$C(p, $\gamma$)$^{13}$N($\beta^{+}$ $\nu$)$^{13}$C reaction 
\citep{iben83}.
A second neutron source, partially activated in low mass AGBs (1.3 $\la$ 
$M/M_\odot$ $<$ 3) during TPs, is the $^{22}$Ne($\alpha$, n)$^{25}$Mg reaction, which
burns more efficiently in intermediate mass AGBs (3 $\la$ $M/M_\odot$ $\la$ 8), 
where a higher temperature is reached at the bottom of the TP.
For a complete discussion on AGB nucleosynthesis the reader may refer
to \citet{busso99,straniero06,SCG08,kaeppeler10rmp}.
\\
Spectroscopic observations of Galactic disc MS, S, C(N) and Ba stars  
show a spread in the distribution of the $s$-process elements for a given
metallicity (\citealt{busso95,busso01,abia01,abia02,gallino05,husti09pasa} and
references therein), which increases in the halo of
 CEMP-$s$ stars \citep{SCG08}. 
This spread involves the three $s$-peaks at the magic neutron numbers,
accumulation points of the $s$-process owing to the low neutron-capture 
cross sections: light-s (ls; Sr, Y, Zr) (N = 50), 
heavy-s (hs; Ba, La, Ce, Nd, Sm) (N = 82) and Pb (N = 126). 
In the halo, a large amount of lead ($^{208}$Pb) is produced, because
the number of neutrons available per iron seed increases
as $^{56}$Fe decreases with the metallicity, while
the $^{13}$C is a primary neutron source directly produced in the
star independently of the metallicity \citep{gallino98,goriely00}.
Then, the neutron flux overcomes the first two peaks 
feeding Pb, which covers a range from about thirty times solar to 
values higher than four thousand times solar (e.g., see HD 189711 by
\citealt{vaneck03} and CS 29497--030 by \citealt{ivans05}).
A range of $s$-process efficiencies is required in order to interpret 
the observations in the halo. 
Starting from the case ST adopted by \citet{gallino98} 
and \citet{arlandini99}, which was shown to reproduce the solar main component
as the average between AGB models of initial masses 1.5 and 3.0 $M_\odot$ at
half solar metallicity, we have multiplied or divided the $^{13}$C 
(and $^{14}$N) abundance in the pocket by different factors.
\\
Our theoretical results are obtained with a post-process nucleosynthesis 
method \citep{gallino98}, based on full evolutionary FRANEC 
(Frascati Raphson-Newton Evolutionary Code, 
\citealt{chieffi89}) models, following the prescriptions by
\citet{straniero03}.
The AGB models with different initial masses (1.3 
$\leq$ $M/M_\odot$ $\leq$ 2), metallicities ($-$3.6 $\leq$ [Fe/H] $\leq$ 
$-$1.5) and $s$-process efficiencies (ST/150 $\leq$ $^{13}$C-pocket $\leq$ 
ST$\times$2) have been presented by \citet{bisterzo10}, hereafter 
Paper I. 
Below the minimum choice of the $^{13}$C-pocket the $s$-process production
is negligible. The case ST$\times$2 corresponds to an upper limit, because 
further proton ingestion leads to the formation of $^{14}$N at expenses
of $^{13}$C. 
We treat the $^{13}$C-pocket as a free parameter, assumed to be constant 
pulse by pulse.  
As discussed in Paper I, the formation of the $^{13}$C-pocket represents a 
significative source of uncertainty affecting AGB models because the properties
of the physical mechanisms involved are not completely understood.
The approximation is adopted to test AGB models through a comparison 
with spectroscopic observations of different stellar populations.
Neutron-capture rates and charged particle reactions are updated to 
2010 (KADoNiS\footnote{Karlsruhe Astrophysical Database of Nucleosynthesis 
in Stars, web address `http://www.kadonis.org' as well 
as further references given in Appendix~A of Paper I.}, \citealt{dillmann06}; 
NACRE\footnote{Web 
address http://pntpm3.ulb.ac.be/Nacre/barre$\_$database.htm} compilation, 
\citealt{angulo99}).
In the halo, AGB models of initial 
masses $M$ = 1.3, 1.4, 1.5 and 2 $M_{\odot}$ suffer 5, 10, 20 and 26 TDUs, 
respectively.
The mass involved in each TDU increases with the AGB initial
mass and with decreasing metallicity (Fig.~1 and~2 of Paper I).
\\
An AGB star produces a huge amount of primary $^{12}$C by partial
He burning in the He-intershell. The stellar envelope becomes progressively
enriched in [C/Fe] by the effect of recurrent TDUs.
An increasing amount of primary $^{22}$Ne 
is also synthesised in the advanced thermal pulses by conversion of 
primary $^{12}$C to primary $^{14}$N during H-burning via 
$^{14}$N($\alpha$, $\gamma$)$^{18}$F($\beta^{+}$ $\nu$)$^{18}$O 
and $^{18}$O($\alpha$, $\gamma$)$^{22}$Ne reactions
(\citealt{mowlavi99}; \citealt{gallino06}; \citealt{husti07}).
This $^{22}$Ne contributes significantly to the primary
production of light isotopes, as $^{23}$Na (via $^{22}$Ne(n, $\gamma$)$^{23}$Na) 
and $^{24,25,26}$Mg (via $^{23}$Na(n, $\gamma$)$^{24}$Mg, $^{22}$Ne($\alpha$, 
n)$^{25}$Mg and $^{22}$Ne($\alpha$, $\gamma$)$^{26}$Mg). 
$^{22}$Ne, together with $^{12}$C, $^{16}$O and $^{23}$Na, is among the major 
neutron poisons in the $^{13}$C-pocket. For higher metallicities this 
effect decreases and becomes negligible at [Fe/H] $\geq$ $-$1 (see also 
\citealt{gallino06}). Moreover, the neutron capture chain starting from $^{22}$Ne(n, 
$\gamma$) extends up to $^{56}$Fe, producing seeds for the $s$-process.
At halo metallicities and at increasing initial AGB mass (resulting in a 
corresponding increase of the temperature at the bottom of the thermal pulse), 
Sr, Y and Zr receive an 
increasing contribution by the $^{22}$Ne($\alpha$, n)$^{25}$Mg neutron source
because of the higher amount of primary $^{22}$Ne.
\\
The behaviour of the three $s$-process peaks\footnote{We defined ls = $<$Y, Zr$>$ 
and hs = $<$La, Nd, Sm$>$ (see Paper I). Sr is excluded from the ls elements and 
Ba from the hs elements because their lines are in general affected by higher 
uncertainties \citep{busso01}.
Examples in CEMP-$s$ stars will be given in Bisterzo et al., submitted,
hereafter Paper III.} ls, hs and Pb with metallicity is not linear, being extremely 
dependent both on the efficiency of the $^{13}$C-pocket and on metallicity.
The two $s$-process indexes [hs/ls] and [Pb/hs] characterise the 
$s$-process distribution independently of the mixing between the 
material transferred from the AGB to the observed companion. 
Once the comparison of the two $s$-process indexes with spectroscopic 
observations of a given star determines the efficiency of the 
$^{13}$C-pocket, one obtains the dilution factor $dil$, defined 
as the mass of the convective envelope of the observed star ($M^{\rm obs}_\star$) 
over the material transferred from the AGB to the companion 
($M^{\rm trans}_{\rm AGB}$):
\begin{equation}
\centering
\label{eq3}
dil = \log\left(\frac{M^{\rm obs}_{\star}}{\Delta M^{\rm trans}_{\rm AGB}}\right).
\end{equation}

The aim of this paper is to interpret the observations in CEMP-$s$ stars 
in order to test the AGB nucleosynthesis models presented in Paper I.
In particular, the large sample of CEMP-$s$ stars collected from the literature
allows us to obtain statistical constraints on theoretical models and to verify
the reliability of the models themselves. 
A general description of the sample is given in Section~\ref{observations}.
In the analysis, we consider the spectroscopic observations of $s$-process
elements belonging to the three $s$-peaks, light elements as C, N, O, Na 
and Mg, as well as Eu to investigate possible $r$-process contributions.
With the improvement of the high quality spectra, the determination of 
the [La/Eu] ratio in these stars provides the most precise tool to distinguish
the respective $r$- and $s$-contributions and to investigate their origin.
Indeed, lanthanum is mainly synthesised by the $s$-process (70\% of solar La, 
\citealt{winckler06}), while europium is an element with a dominant $r$-process 
contribution (less than 6\% of solar Eu comes from $s$-process). 
In Paper I, we predicted a pure $s$-process ratio [La/Eu]$_{\rm s}$ 
= 0.8 -- 1.1. 
In general, accounting of error bars, observed values in the range 0.0 
$\la$ [La/Eu] $\la$ 0.5 indicate CEMP-$s$ stars having 
experienced an important $r$-process contribution \citep{beers05}.
Different hypotheses have been advanced to interpret CEMP-$s/r$ stars. 
After a brief introduction about recent spectroscopic observations
of $r$-process elements in some peculiar low metallicity stars (Section~\ref{r}),
a possible CEMP-$s/r$ scenario is discussed in Section~\ref{CEMPs+r}.
\\
Subsequently, we perform a general analysis by comparing theoretical 
AGB models with spectroscopic observations for [La/Eu] versus metallicity,
[La/Fe] versus [Eu/Fe], [hs/ls] and [Pb/hs] versus [Fe/H] 
(Section~\ref{generalcomparison}).
Then, we discuss three stars with different characteristics,
the CEMP-$s$ giant HD 196944, the main-sequence CEMP-$s/r$ HE 0338--3945,
and a CEMP-$s$ HE 1135+0139 for which no lead is measured.
These stars are used as examples to illustrate the method we adopt 
for the theoretical interpretation with AGB models (Section~\ref{method}).
A detailed analysis of the individual stars is provided in Paper III.
One of the main goals is to highlight possible differences between 
models and observations, providing starting points of debate in which 
spectroscopic and theoretical studies may intervene. 
A summary of the main results is given in Sections~\ref{summary}
and~\ref{conclusions}.


\section{Presentation of the sample} \label{observations}

\begin{figure}
\includegraphics[angle=-90,width=8.5cm]{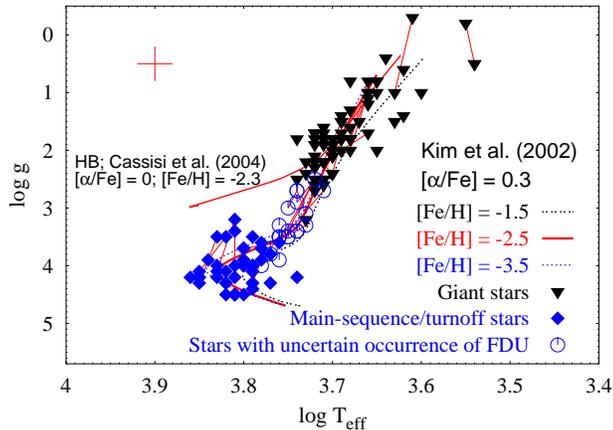}
\caption{Spectroscopic gravities versus effective temperature 
of CEMP-$s$ and CEMP-$s/r$ listed in Table~\ref{tuttestelleparam}. 
Filled circles indicate main-sequence or turnoff stars labelled \textit{`no'} 
in column~7; triangles are giants labelled \textit{`yes'} in column~7. 
Stars with uncertain occurrence of the FDU are represented with empty circles
(see text).
Evolutionary tracks by \citet{kim02} for stars with $M$ = 0.8 $M_\odot$ at 11 Gyr 
and various metallicities are plotted ([Fe/H] = $-$1.5 black line, [Fe/H] 
= $-$2.5 magenta line, [Fe/H] = $-$3.5 blue line). The Horizontal Branch
(HB) track at [Fe/H] = $-$2.3 by \citet{cassisi04} is shown with a solid red line. 
Small changes in the assumed mass have a little effect on the location of the HB.
Different observations of the same star are connected by thin solid lines.
Typical error bars are shown.
(\textit{See the electronic edition of the Journal for a colour version of 
this and the following figures.})}
\label{z1m3z1m4z1m5a2101112GyHB}
\end{figure}

About a hundred CEMP-$s$ and CEMP-$s/r$ stars have been observed in 
the last decade with very high resolution spectroscopy ($R$ $\ga$ 50\,000)
(\citealt{aoki02a,aoki02c,aoki02d,aoki06,aoki07,aoki08}; 
\citealt{barbuy05};
\citealt{cohen03,cohen06};
\citealt{goswami06};
\citealt{GA10};
\citealt{ivans05}
\citealt{ish10};
\citealt{jonsell06}
\citealt{JB02,JB04};
\citealt{lucatello03};
\citealt{lucatello04PhD};
\citealt{masseron06,masseron10};
\citealt{pereira09};
\citealt{PS01};
\citealt{schuler08};
\citealt{thompson08};
\citealt{tsangarides05};
\citealt{vaneck03};
\citealt{zhang09};
\citealt{roederer08}; 
\citealt{roederer10})
including spectroscopic data by 
\citet{barklem05} ($R$ $\sim$ 20\,000; 
$S/N$ = 30 -- 80)\footnote{The signal-to-noise they obtain is from 30 
to 50, with the only exception of HE 0202--2204, for which $S/N$ = 80.};
\citet{behara10} and \citet{sneden03b} ($R$ $\sim$ 30\,000).  
\citet{allen10IAU} provided high resolution spectra for
a new CEMP-$s$ candidate and four new CEMP-$s/r$ stars, but at present
no data are available for further discussions.
Among CEMP-$s$ stars we include objects with [C/Fe] $>$ 0.5.
Note that some authors prefer a distinction at [C/Fe] $>$ 1 \citep{beers05}.
\\
Additional stars with a limited number of spectroscopic observations
are CEMP-$s$ candidates: HE 0322--1504, HE 0507--1430, HE 1045--1434, 
CS 22947--187, CS 22949--008 and HD 187216 
\citep{beers07,MW95,rossi05,johnson07,preston09pasa,kipper94}. 
An intense search for very metal-poor candidates is the Galaxy is underway,
from which a large number CEMP-$s$ and CEMP-$s/r$ star are to be expected
\citep{beers07bra}.
\\
A radial velocity study by \citet{lucatello05}
suggests that all CEMP-$s$ stars belong to double (or multiple) systems
(see also \citealt{tsangarides05,preston09pasa}).
Mass transfer occurs mostly through efficient stellar winds, because
the distance between the primary AGB star and the secondary are 
in general large, with periods $P$ $>$ 200 d.
Only two CEMP-$s$ with known radial velocities are short 
amplitude binaries, for which the accretion occurred via Roche-Lobe
outflow: HE 0024$-$2523 \citep{lucatello03} ($P$ = 3.41 d),
and SDSS 1707+58 \citep{aoki08}. 
When available, radial velocities variations of CEMP-$s$ stars, their 
period and the relative references are listed in Appendix~A, 
Table~A1, online material.
\\
In Table~\ref{tuttestelleparam}, we collect all CEMP-$s$ and CEMP-$s/r$
stars discussed here, with their metallicity, atmospheric parameters, 
and evolutionary status. 
In bold are marked the references considered in our analysis (column~2).
All stars have metallicities in the range $-$3.5 $\leq$ [Fe/H] $\leq$ 
$-$1.7, with five exceptions: CS 29503--010 ([Fe/H] = $-$1.06; \citealt{aoki07}), 
HD 26 ([Fe/H] = $-$1.25,$-$1.02; \citealt{vaneck03}, \citealt{masseron10}), 
HD 206983 ([Fe/H] = $-$0.99, $-$1.43; \citealt{JP01}, \citealt{masseron10}),
HE 0507--1653 ([Fe/H] = $-$1.38, 
$-$1.42; \citealt{aoki07,schuler08}), HE 1152--0355 ([Fe/H] = $-$1.27; 
\citealt{goswami06}). 
These stars may be considered as a link between Ba stars and CEMP-$s$ stars,
and will be discussed in a separate Section in Paper III.
The most metal$-$poor stars among CEMP-$s$ and CEMP-$s/r$
are CS 22960--053 \citep{aoki07}, CS 30322--023 
\citep{aoki07,masseron06}, HE 1005--1439 \citep{aoki07,schuler08}, HE 1410--0004
\citep{cohen06}, SDSS 0126+06 \citep{aoki08}, and SDSS J1349--0229 \citep{behara10}, 
with [Fe/H] $\la$ $-$3.
\\
In Fig.~\ref{z1m3z1m4z1m5a2101112GyHB}, we plot log $g$
versus log $T_{\rm eff}$ for the stars listed in Table~\ref{tuttestelleparam}. 
By comparison, we overlap evolutionary tracks of models with initial mass 
0.8 $M_\odot$ at 11 Gyr and three metallicities ([Fe/H] = $-$1.5, $-$2.5 and $-$3.5) by
\citet{kim02} and the Horizontal Branch (HB) track at [Fe/H] = $-$2.3 by 
\citet{cassisi04}. 
Thirty stars are located on the main-sequence or close to 
the turnoff.
Twenty-one lie on the subgiant phase. 
All the remaining stars are giants.
This distinction is important for the following discussion,
in relation to the occurrence of the first dredge-up (FDU) episode.
This large mixing between the convective envelope and the inner
layers of the star modifies the chemical composition of 
the surface (see also Section~\ref{carbon}).
In case of binary systems with mass transfer, 
this mixing also dilutes the C and $s$-rich material previously 
transferred from the AGB companion. \\
Unfortunately, the estimate of the atmospheric parameters at which 
the FDU occurs, in particular the effective temperature, is
uncertain.
For [Fe/H] $\la$ $-$2.0, a main-sequence star with initial mass of $\sim$ 
0.8 $M_\odot$ has a very thin convective envelope ($\sim$ 10$^{-3}$ $M_\odot$), 
thus a negligible dilution by convection is expected.
After mass accretion, owing to the larger radiative opacity
of the C-rich material deposited on the stellar surface, the depth of the
convective envelope may eventually increase. However, due to 
gravitational settling, when the star attains the turnoff, the
external layers will be again deprived of heavy elements, so that the
convective envelope is even smaller than that at the zero age main-sequence.
Indeed, gravitational settling, 
which acts on time-scales of billions of years (for stars on the main-sequence), 
depletes the heavy elements from the thin convective envelope of the observed 
CEMP-$s$ star, which becomes H-rich.
For primary stars with a typical initial mass of 1.5 $M_\odot$, mass transfer occurred
about ten billion years ago and the heavy elements accumulated just below 
the convective envelope (deeper gravitational settling would require longer time-scales). 
The larger is the mass accreted, the longer is the time required to reach the inner 
layers of the envelope with a primordial chemical composition. 
Then, a CEMP-$s$ star at the turnoff would not show $s$-enhancement if the 
gravitational settling was efficient during the main-sequence.\\
Once at the turnoff, the convective envelope starts immediately to advance 
in depth and recovers what was slowly lost during the previous phase.
Consequently, the material transferred from the primary star is progressively 
diluted. The efficiency of this mixing strongly depends on the amount of mass 
accreted from the primary. 
Indeed, the effective temperature on the subgiant branch at which the 
convective envelope reaches the layers with a primordial chemical composition
decreases by increasing the amount of mass transferred.
For an extreme case in which the mass accreted is $\sim$ 0.1 $M_\odot$, no 
dilution will be observed at the beginning of the subgiant phase. \\
Besides gravitational settling, additional processes, as 

\input{table1_revised2.tex}

\input{table2.tex}
\input{table3.tex}
\input{table4.tex}

\noindent radiative acceleration
and thermohaline mixing, should be included in the analysis.
Radiative acceleration may impede gravitational settling 
 (see e.g., \citealt{richard02}).
Instead, mixing induced by thermohaline instabilities may reach deep layers in 
a shorter time scale ($\sim$ millions of years; see e.g., \citealt{stancliffe07}) 
if not prevented by the other two processes. \\
The resulting efficiency of all these mixing is very difficult to estimate 
\citep{vauclair04,eggleton06,charbonnel07,charbonnel10,stancliffe08,stancliffe10,denissenkov08,denissenkov09,denissenkov10,cantiello10,thompson08,angelou11}.
It is even more problematic if rotation or magnetic fields are included
in the analysis.
In conclusion, the depth of the mixing after the turnoff phase can not be clearly 
established: consequently, we can not theoretically establish the dilution
for CEMP-$s$ stars with effective temperature in the range $T_{\rm eff}$ $\sim$ 
(5700 $\pm$ 150) K, because we do not know the level of depth of the convective 
envelope. In these cases, CEMP-$s$ stars are labelled 
as `no?' in Table~\ref{tuttestelleparam}. 
The FDU is technically defined as the point of the maximum 
sinking of the convective envelope during the subgiant phase.
The FDU involves about 80\% of the mass of the star, and erases
all processes occurred during the previous phases.
Therefore, CEMP-$s$ having suffered the FDU ($T_{\rm eff}$ $\la$ 5500 K) need 
a dilution of the order of 1 dex or more.
These giants are labelled `yes' in column~7 of 
Table~\ref{tuttestelleparam} (triangles in Fig.~\ref{z1m3z1m4z1m5a2101112GyHB}). 
Main-sequence or turnoff stars having not suffered the FDU are labeled `no' 
in column~7 of Table~\ref{tuttestelleparam} (filled circles in 
Fig.~\ref{z1m3z1m4z1m5a2101112GyHB}). 
The possible need of dilution in order to interpret observations of
main-sequence/turnoff stars suggests that mixing (as thermohaline)
were at play.
Note that the relative ratio of two elements [El$_{\rm 1}$/El$_{\rm 2}$]
is not affected by the dilution and will be adopted as useful constraint
for AGB models (Section~\ref{generalcomparison}).

The stars are divided into two groups. CEMP-$s$ and CEMP-$s/r$ stars with 
several observed $s$-element are listed in Table~\ref{table5_sindicator}. 
If a limited number of elements is measured, in particular only Sr among ls
or Ba among hs, the star is listed in Table~\ref{tablestellemancanti}. 
This distinction is made because, in general, Sr and Ba are 
affected by higher uncertainties with respect to the other $s$-elements 
\citep{mashonkina08,andr09,short06}. 
When available, in Tables~\ref{table5_sindicator} and~\ref{tablestellemancanti} 
we report [Na/Fe], [Mg/Fe], [ls/Fe], [hs/Fe] and [Pb/Fe], 
the $s$-process indicators [hs/ls] and [Pb/hs], 
as well as [La/Fe], [Eu/Fe], [La/Eu].
References and labels in columns~2 and~3 are the same as in
 Table~\ref{tuttestelleparam}. 
In both Tables, we further distinguish between different classes of stars,
following their abundance pattern:
\begin{itemize}
	\item CEMP-$s$II are stars with a high $s$-process enhancement, 
	[hs/Fe] $\ga$ 1.5 (labeled as `sII' in column~15); 
	\item CEMP-$s$II also showing an $r$-enhancement are called CEMP-$s$II$/r$
              (in general with [La/Eu]$_{\rm obs}$ $\sim$ 0.0 $\div$ 0.5). 
	     Similarly to the classification based on the $s$-process enhancement, 
            we may distinguish between:\\
             -- CEMP-$s$II/$r$II with [r/Fe]$^{\rm ini}$ $\sim$ 1.5 $\div$ 2.0 
                     (labelled as `sII/rII' in column~15) and\\
             -- CEMP-$s$II/$r$I with [r/Fe]$^{\rm ini}$ $\sim$ 1.0 
                   (labelled as `sII/rI' in column~15),\\
	     following our definition of the $r$-process enhancement based on 
             AGB model predictions, as we will discuss in Section~\ref{CEMPs+r};            
 	\item CEMP-$s$I are stars with a mild $s$ enrichment, [hs/Fe] $<$ 1.5
 (labeled as `sI' in column~15\footnote{\citet{barklem05} 
first called `s-II' stars three CEMP-$s$ stars with high $s$-enhancement: 
HE 0131--3953, HE 0338--3945, afterwards studied by \citet{jonsell06}, and HE 1105+0027.
Another distinction is provided by \citet{masseron10}, who used
``CEMP-low-$s$'' to denote stars with a low Ba enhancement, but with [Ba/Eu] 
showing evidence of contamination by $s$-process material. 
They found four CEMP-low-$s$ stars, all discussed here as
CEMP-$s$I stars: CS 30322--023 by \citet{masseron06}, HK II 17435--00532
by \citet{roederer08}, HE 1001--0243 and HE 1419-1324.}).
\end{itemize}
An additional class of CEMP-$s$I stars with mild $r$-process contribution
would be expected, the CEMP-$s$I/$r$I stars.
None of the stars of our sample belong to this category,  
likely because of our definition of CEMP-s/r stars, which considers $r$-rich 
those stars having [r/Fe]$^{\rm ini}$ $\geq$ 1 (see Section~\ref{CEMPs+r}).
Moreover, we classify stars without Eu measurements as CEMP-$s$I/$-$ 
or CEMP-$s$II/$-$ (labeled `sI/$-$' or `sII/$-$' in column~15).
The degree of the $s$-enhancement may depend on different factors:
firstly it is affected by the $s$-process efficiency and by the initial
mass of the primary AGB; 
afterwards, the orbital parameters of the binary system
(e.g., the distance between the two stars), the efficiency of the stellar 
winds and the degree of mixing with the envelope of the observed companion
influence the final $s$-distribution.
At the end of Tables~\ref{table5_sindicator} and~\ref{tablestellemancanti} 
we report the range covered by the observations.
The number of stars belonging to different classes is given in 
Table~\ref{class}, where stars from Tables~\ref{table5_sindicator}
and~\ref{tablestellemancanti} are considered in columns~2 and~3, respectively.
As in Table~\ref{tuttestelleparam}, we distinguish between stars before 
or after the FDU (`no' or `yes', respectively).
\\
Sodium is measured in 53 CEMP-$s$ and CEMP-$s/r$ stars (column~5), 23 of them
have been studied by \citet{aoki07}.
[Na/Fe] is very high in some CEMP-$s$ stars (CS 29528--028 
by \citealt{aoki07} has [Na/Fe] = 2.3; SDSS 1707+58 by \citealt{aoki08} 
has [Na/Fe] = 2.7).
As recalled in Section~\ref{intro}, Na is synthesised via neutron capture starting 
from the large amount of primary $^{22}$Ne produced at low metallicities
($^{22}$Ne(n, $\gamma$)$^{23}$Ne($\beta^{-}$ $\nu$)$^{23}$Na). 
Therefore, [Na/Fe] increases with the number of TPs, providing an important 
constraint of the AGB initial mass in CEMP-$s$ stars.
Unfortunately, in very metal-poor stars Na may be affected by strong 
uncertainties due to non-local thermodynamic equilibrium (NLTE)
corrections or three-dimensional (3D) hydrodynamical model atmospheres. 
Recent studies by \citet{andr07} show that the NLTE effects
may decrease [Na/Fe] by 0.7 dex 
(see also \citealt{barbuy05} and \citealt{aoki07}).
Most of the [Na/Fe] measurements in Tables~\ref{table5_sindicator} 
and~\ref{tablestellemancanti} account of NLTE corrections or
they are provided by using two subordinate lines (5682 and 5688 
${\rm \AA}$), which are weak and exhibit small NLTE corrections \citep{takeda03}.
\\
Magnesium is detected in several stars, covering a range from solar up
to [Mg/Fe] $\sim$ 1.7 dex.
NLTE corrections may increase the final [Mg/Fe] by about 0.3 dex \citep{andr10}.
In most stars, [Mg/Fe] agrees with observations in
field stars \citep{andr10,mashonkina03,gehren06}, 0.2 $\la$ [Mg/Fe] $\la$ 0.6. 
The only exceptions are CS 29497--34, CS 29528--028,
LP 625--44, SDSS 1707+58 with [Mg/Fe] $>$ 1.0. 
In these stars, [Na/Fe] is enhanced as well, supporting the hypothesis that Mg
is produced starting from the primary $^{22}$Ne through $^{22}$Ne(n, 
$\gamma$)$^{23}$Ne($\beta^{-}$$\nu$)$^{23}$Na(n, 
$\gamma$)$^{24}$Na($\beta^{-}$$\nu$)$^{24}$Mg.
In addition, $^{25,26}$Mg are synthesised via the $^{22}$Ne($\alpha$, n)$^{25}$Mg and
$^{22}$Ne($\alpha$, $\gamma$)$^{26}$Mg reactions.
\\ 
The ratios [ls/Fe] and [hs/Fe] are reported in columns~7 and~8. 
For stars listed in Table~\ref{table5_sindicator}, if a given element 
among Y and Zr or among La, Nd and Sm is not observed we adopt the AGB prediction.
It means that we consider in the ls and hs average the values of the missing 
elements estimated with the best theoretical AGB interpretation. 
The best AGB model that interprets the observations is calibrated on the basis
of an accurate analysis, as we will describe in Section~\ref{method}.
Note that, in general, with this method the [ls/Fe] and [hs/Fe] ratios agree
within 0.15 dex with the values provided without including our AGB predictions.
Major details on the adopted models (listed in Tables~\ref{summary1} and ~\ref{summary2})
will be discussed in Paper III for individual stars.
When available, [ls/Fe] and [hs/Fe] account for the number of lines 
detected for each element.
In Table~\ref{tablestellemancanti}, stars with a limited number of
$s$-process observations are listed:
in several cases only Sr among ls or Ba among hs are detected.
If not differently specified (see caption), we adopt for these stars the 
different notation [ls$\dag$/Fe] = [Sr/Fe] and [hs$\dag$/Fe] = [Ba/Fe].
The $s$-process indicators [hs/ls] and [Pb/hs] are reported in 
columns~9 and~11, respectively.
Most stars cover a range between 0.4 $\la$ [hs/ls] $\la$ 1.0.
Negative values are observed in two CEMP-$s$I, CS 22942--019
(Table~\ref{table5_sindicator}) and CS 22956--28 (Table~\ref{tablestellemancanti}).
Excluding the upper limits, the range of [Pb/hs]  covers 
about 2 dex. Five stars have negative [Pb/hs] (HD 189711, HE 1305+0007 
and V Ari from Table~\ref{table5_sindicator}; CS 22891--171 and CS 
29495--42 from Table~\ref{tablestellemancanti}).
\\
The observed [La/Fe], [Eu/Fe] and their ratio [La/Eu] (columns~11 to~13) 
provide, in general,
a good indicator of the $s$- and $r$-process contribution in stars. 
A detailed discussion about CEMP-$s/r$ stars and their theoretical
interpretation will be given in the following Sections.
Eu is detected in 38 stars of Table~\ref{table5_sindicator};
20 of them are CEMP-$s/r$ with an observed [La/Eu] $\sim$ 0.0 -- 0.5 
dex. Few exceptions are listed in Table~\ref{table5_sindicator}. 
The three CEMP-$s/r$ HD 209621, HE 1305+0007 and 
LP 625--44 have [La/Eu] = 1.06, 0.56 and 0.74, respectively
(column~14). For HD 209621 and LP 625--44, [La/Eu] is higher than the other 
 elements belonging to the second $s$-peak on which we based the $r$-enhancement. 
HE 1305+0007 has a high [La/Eu] because of its very high [La/Fe] $\sim$ 2.6.
Despite the low [La/Eu], the four stars CS 22942--019, CS 22964--161, 
CS 29513--032, HK II 17435--00532, are classified as CEMP-$s$I, 
because the low [La/Eu] is a consequence of the low $s$-process contribution
to [La/Fe], instead of a high $r$-process contribution to [Eu/Fe].
Among the seven CEMP-$s$ with Eu detected in Table~\ref{tablestellemancanti},
only three stars are CEMP-$s/r$ (CS 22891--171 by \citealt{masseron10}, 
CS 30338--089 by \citealt{lucatello04PhD} and HE 0131--3953 by \citealt{barklem05}).


\section{{\scriptsize r}-process enhancement in metal-poor stars}

\subsection{r-enhanced stars} 
\label{r}

In the past few decades, several spectroscopic efforts have been 
dedicated to the study of low metallicity stars showing high 
$r$-process enhancements. 
\citet{sneden94,sneden03a} firstly analysed the spectrum
of a giant with [Eu/Fe] = 1.64 and [La/Eu] = $-$0.55.
This star is the prototype of a new class of peculiar stars, the 
``pure $r$-process-enhanced metal-poor'' stars.
Following the standard classification given by \citet{christlieb04} 
and \citet{beers05}, $r$-II stars have [Eu/Fe] 
$\geq$ 1 and $r$-I show 0.3 $\leq$ [Eu/Fe] $\leq$ 1. In both cases
[Ba/Eu] $<$ 0 (or [La/Eu] $\leq$ 0) to exclude any $s$-process contribution. 
Up to date, several $r$-II stars have been analysed: most noteworthy are
 CS 31082--001 by \citep{hill02,plez04} ([Eu/Fe] $\sim$ 1.6), HD 115444 
 by \citet{westin00} ([Eu/Fe] = 0.85), HE 1523--0901 by \citet{frebel07}
([Eu/Fe] $\sim$ 1.8), CS 29497--004 
by \citet{christlieb04} ([Eu/Fe] = 1.64), CS 31078--018  by \citet{lai08} 
([Eu/Fe] = 1.23), CS 29491--069 ([Eu/Fe] = 1.0) and HE 1219-0312 
([Eu/Fe] = 1.4) by \citet{hayek09},
CS 22183--031 by \citet{honda04} ([Eu/Fe] = 1.2), CS 22953--003 by \citet{francois07}
([Eu/Fe] = 1.05). Recently, \citet{mashonkina10} discovered the new
star HE 2327-5642 ([Eu/Fe] = 0.98; [r/Fe] = 1.5) and \citet{aoki10} found
the $r$-II stars with the highest [Eu/Fe] detected, SDSS J2357--0052 (with 
[Fe/H] = $-$3.4 and [Eu/Fe] = 1.92 dex).
Identified $r$-I stars are CS 30306--132 by \citet{honda04,aoki05}, 
HD 221170 by \citet{ivans06}, and BD +173248 by \citet{cowan02,roederer10BD}
([Eu/Fe] $\sim$ 0.8 -- 0.9).

\subsection{r-process residual} \label{residual}

The $r$-process is associated with explosive conditions in massive
stars, although the astrophysical site is still unknown.
From the theoretical point of view, several models have been advanced
\citep{farouqi10,QW08,qian07,kratz07,wanajo06}, but an exhaustive interpretation
is still lacking.
The mostly adopted estimate of the solar $r$-process 
contribution to each isotope 
were first evaluated by \citet{kaeppeler82} with the \textit{residual method}, 
$N_{\rm r}$ = $N_\odot$ - $N_{\rm s}$. 
It was successfully adopted to derive the solar $s$-process
abundances as well as to acquire information on the physical conditions during 
the $s$-process far from the branching points. 
Subsequently, this classical approach was replaced 
by a first generation of stellar $s$-process models (\citealt{gallino98,arlandini99},
as anticipated in Section~\ref{intro}),
avoiding inconsistencies encountered close to the magic neutron numbers due to
the use of more accurate cross section measurements.
In Table~\ref{ns} the solar $s$-process contributions of \citet{arlandini99} 
(column~4 in $N_{\rm s}$; column~6 in percentages) are compared to the updated results 
of 2010 (column~5 in $N_{\rm s}$; column~7 in percentages) for isotopes from $^{63}$Cu up 
to $^{209}$Bi.
$N_{\rm r}$ values of each isotope obtained with 
the residual method are listed in column~8.
\\
However, the solar $s$ abundances should include the contributions of all 
stellar generations over the Galactic history. 
In particular, low metallicity stars of low initial mass produce a huge 
amount of Pb and Bi (the so-called strong $s$-process component;
Section~\ref{intro}), while only about 50\% of solar Pb and 6\% of solar Bi
are synthesised by the main component (see columns~6 and~8).
For this reason, an appropriate evaluation of the $s$-contributions
at the end of the $s$-path is provided by a Galactic Chemical Evolution
model as described by \citet{travaglio99,travaglio04} (updated by
\citealt{serminato09pasa}):
$N_{\rm s}^{\rm GCE}$(Pb) = 87\%, $N_{\rm s}^{\rm GCE}$(Bi) = 26\%.
These values are reported between brackets in column~6; the
resulting $r$-process percentages for Pb and Bi are listed in column~8. 
\\
Comparison between observations of elements from Ba to Bi

\onecolumn
\input{table5_revised2.tex}
\twocolumn

\input{table6.tex}
\input{table7.tex}

\begin{figure}
\includegraphics[angle=-90,width=8.5cm]{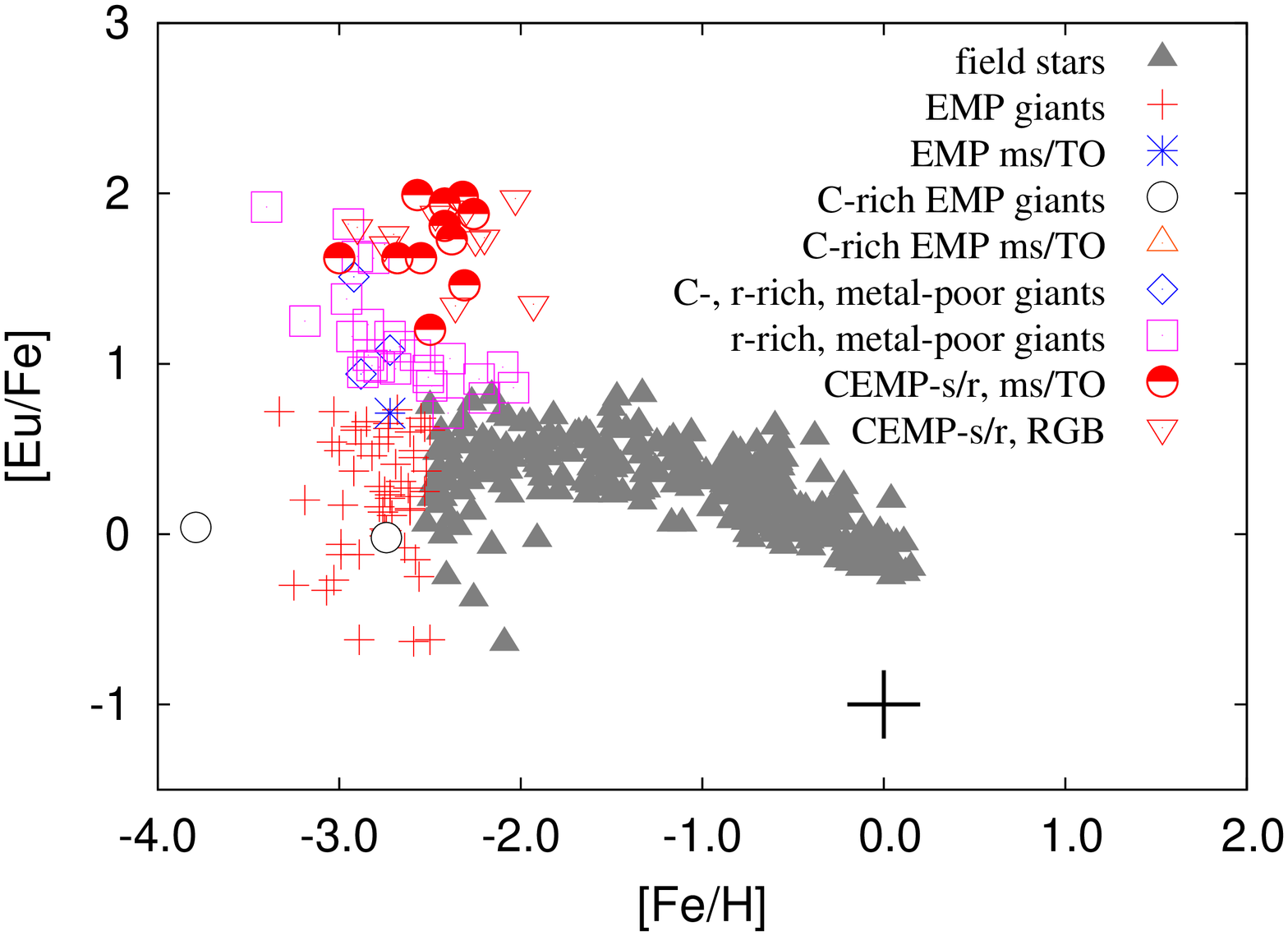}
\caption{[Eu/Fe] abundances collected from the literature 
using the SAGA database \citep{suda08,suda11}, including rII and rI stars listed in 
Section~\ref{r}.  For comparison, CEMP-$s/r$ stars discussed 
here are also shown (see text).
Filled triangles are field stars with [Fe/H] $\ga$ 2.5; plus and asterisks are Extremely 
Metal-Poor (EMP; [Fe/H] $\la$ 2.5) giants and ms/TO stars, respectively; filled circles and 
empty triangles are C-rich EMP giants and ms/TO stars, respectively.
Stars showing an $r$-enhancement are: empty diamonds (C-rich, EMP), empty squares (EMP),
half empty circles (CEMP-$s/r$ ms/TO), empty triangles (CEMP-$s/r$ giants).
Typical error bars are indicated.}
\label{mnras_EuvsFe_noUL3_nosigma1_ingrandito_gallinoRIO}
\end{figure}

\begin{figure}
   \centering
\includegraphics[angle=-90,width=8cm]{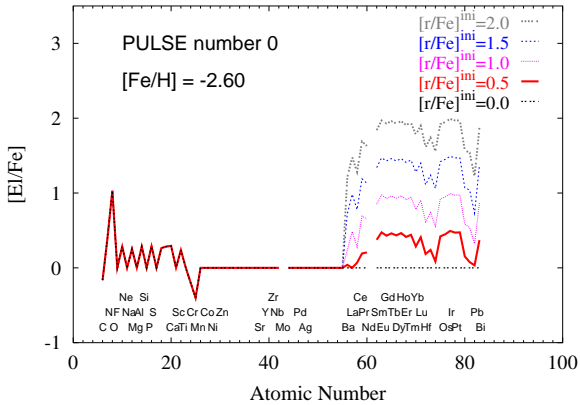}
\caption{Initial abundances for different initial $r$-process enrichments
([r/Fe]$^{\rm ini}$ = 0.0, 0.5, 1.0, 1.5, and 2.0) for elements from Ba to Bi.
For neutron capture elements below Ba, a mild initial $r$-process enrichment 
[r/Fe]$^{\rm ini}$ $\la$ 1.0 could be introduced (see discussion in Section~\ref{r}).
Note that for elements lighter than A = 30, we assumed the initial
abundances described in Paper I, Section~2.1.}
\label{rini}
\end{figure}

\begin{figure}
\includegraphics[angle=-90,width=8.5cm]{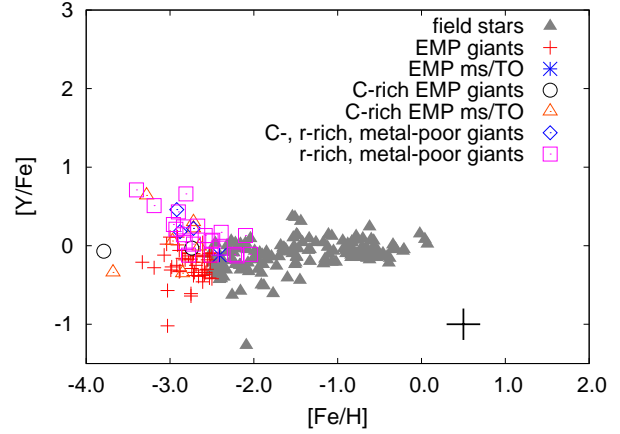}
\includegraphics[angle=-90,width=8.5cm]{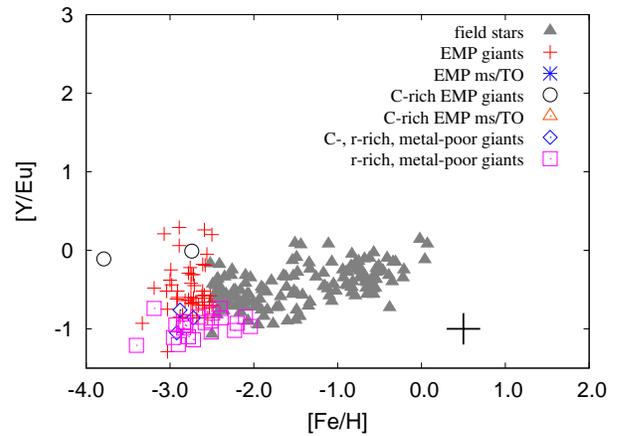}
\caption{\textit{Top panel}: The same as 
Fig.~\ref{mnras_EuvsFe_noUL3_nosigma1_ingrandito_gallinoRIO}, 
but for [Y/Fe]. Note that CEMP-$s/r$ stars have been excluded from this 
figure because, in these stars, Y is mainly produced by the $s$-process. 
\textit{Bottom panel}: the same as top panel but for [Y/Eu].
Typical error bars are indicated.}
\label{mnras_Y_noUL3_nosigma1_ingrandito_gallinoRIO}
\end{figure}

\noindent and the scaled solar-system $r$-element abundance distribution 
shows remarkable agreement in some $r$-rich stars (CS 22892-052, 
BD +173248, HD 221170, HD 115444), suggesting a perhaps unique 
$r$-process component in this range\footnote{For these four 
stars the authors provided Th (and U) measurements, also
in agreement with the solar-system $r$-element distribution.
Other r-rich stars (CS 30306-132, CS 31078-018, CS 31082-001 and HE 1219-0312) 
show an enhanced ratio in the actinide region (Z $\geq$ 90) with 
respect to solar, while elements from Ba to Bi follow the scaled 
Solar-system $r$-element abundances \citep{schatz02,roederer09PbTh}.}.
Instead, neutron capture elements below Ba (40 $<$ Z $<$ 56) 
show deviations from the solar $r$-process curve (e.g, \citealt{sneden03a}),
in agreement with the hypothesis of multiple $r$-process components. 
\\
These considerations make use of the residual method
to estimate the $r$-process contribution disputable in the region below Ba.
Despite that, the residual method remains
a valid approximation in the region between Ba and Bi (see 
Section~\ref{CEMPs+r}) because of the limited 
knowledge of the primary $r$-process nucleosynthesis, but it does 
not exclude different approaches.

\subsection{CEMP-s/r stars} 
\label{CEMPs+r}

Among the 45 CEMP-$s$ stars with Eu detection listed in 
Tables~\ref{table5_sindicator} and~\ref{tablestellemancanti}, half are 
CEMP-$s/r$, with [Eu/Fe] and [La/Eu] incompatible with a pure $s$-process
contribution. In some cases, Eu is strongly enhanced
([Eu/Fe] $\sim$ 2 and [La/Eu] $\sim$ 0, e.g. CS 29497--030 by 
\citealt{ivans05}, HE 0338--3945 by \citealt{jonsell06}, HE 1305+0007 
by \citealt{goswami06} and HE 2148--1247 by \citealt{cohen03}), 
(Section~\ref{r}).
\\
Starting from the pioneering work of \citet{BBF57}, it is clear that 
$s$- and $r$-process have to be ascribed to separate astrophysical sites.
As introduced in Section~\ref{intro}, the [La/Eu] ratio is a good 
indicator of the competition between the two processes.
We remember that 70$\%$ of solar La is synthesised by the 
$s$-process and 94$\%$ of solar Eu is provided by the $r$-process
(see Table~\ref{ns}, column~6).
The large enhancement of typical elements of both processes in the envelope 
of these stars is highly debated.
A pure $s$-process predicts
[La/Eu]$_{\rm s}$ $\sim$ 0.8 -- 1.1, depending on the AGB initial mass and
metallicity. All isotopes between Ba and Eu have well 
determined neutron-capture cross sections (\citealt*{winckler06,kaeppeler89}, 
http://www.kadonis.org).
In order to explain a [La/Eu] ratio close to 0 together with a high $s$-process
enhancement ([La/Fe] $\sim$ 2) as observed in some CEMP-$s/r$ stars, 
different scenarios have been proposed in the literature 
(\citealt{jonsell06}; \citealt{cohen03} and references therein; 
\citealt{zijlstra04}; \citealt{barbuy05}).
\\
We discuss here our hypothesis, see also \citet{SCG08} and 
\citet{bisterzo09pasa}, based on the spread observed in [Eu/Fe] in field
halo stars shown in Fig.~\ref{mnras_EuvsFe_noUL3_nosigma1_ingrandito_gallinoRIO}.
Spectroscopic data are from
\citet{MW95,hill00,JB01,mishenina01,mashonkina03,sneden03a,christlieb04,simmerer04,honda04,barklem05,francois07,frebel07,aokihonda08,hayek09,roederer10r}.
For comparison, we added also observations of the CEMP-$s/r$ stars listed in 
Tables~\ref{table5_sindicator} and~\ref{tablestellemancanti}.
High [Eu/Fe] enrichments are limited at [Fe/H] $<$ $-$2, sustaining 
the hypothesis that a range of progenitor massive stars 
contribute to the $r$-process.
The spread may be attributed to 
inhomogeneous mixing of the interstellar medium in the Galaxy 
\citep{ishimaru99,travaglio01inh,wanajo06}\footnote{Recent studies by 
\citet{carollo07,carollo10,carollo11astroph} show evidence of
dichotomy of the Galactic halo, formed from two individual 
(broadly overlapping) stellar components with different chemical
compositions and kinematics.}.
Note that the spread is not present at disc metallicities; 
[Eu/Fe] follows the behaviour of [O/Fe], linearly decreasing at
[Fe/H] $>$ $-$1 owing to a progressively higher contribution to 
Fe in the interstellar medium from long-lived SNe~Ia \citep{travaglio99}. 
\\
Starting with the spread observed in [Eu/Fe], we can hypothesise
different initial $r$-process enhancements within the 
observed range ($\sim$ 2 dex), averaged around $\sim$ 0.5 dex. 
The $r$-enhancement detected in peculiar stars with very low metallicities
may be due to a local SNe~II explosions, leading to an $r$-enrichment of molecular 
clouds from which CEMP-$s/r$ stars may have formed.
Both stars belonging to the binary system have the same initial $r$-enhancement.
The more massive star is supposed to evolve through the TP-AGB 
phase, synthesising $s$-elements and polluting the observed companion
through stellar winds. 
 \\
\citet{vanhalacameron98} showed through numerical simulations that 
supernova ejecta may interact with a molecular cloud, polluting 
it with freshly synthesised material, likely triggering the formation 
of binary systems consisting of stars with low mass.
This may explain the high frequency of CEMP-$s/r$ among very metal-poor stars.
 However, we do not exclude other hypotheses supporting an initial
$r$-process enrichment of the molecular cloud.\\
The choice of the initial $r$-enhancement (scaled to Eu) is made 
adopting the solar isotopic $r$-process contributions obtained with
the residual method.
As discussed in Section~\ref{r}, the residual method remains a valid 
approximation to estimate the $r$-process contributions for elements 
from Ba to Bi.
We consider as CEMP-$s/r$ those stars that need an initial $r$-process 
enhancement in the range between 1 $\leq$ [r/Fe]$^{\rm ini}$ $\sim$ 2. 
Note that stars with [r/Fe]$^{\rm ini}$ $\sim$ 1 lie at the 
limit between CEMP-$s$ and CEMP-$s/r$. 
As introduced in Section~\ref{observations}, following our 
definition of CEMP-$s/r$ stars most of the CEMP-sI stars 
with low $r$-enhancement are not considered $r$-rich.\\
Fig.~\ref{rini} shows the initial abundances (scaled to europium) 
adopted for different choices of the initial $r$-process enrichment 
[r/Fe]$^{\rm ini}$ = 0.0, 0.5, 1.0, 1.5, and 2.0, for elements from
Ba to Bi.\\
For neutron capture elements lighter than Ba, different initial $r$-process 
enrichment could be introduced under the assumption of a multiplicity
of the $r$-process components, as discussed in Section~\ref{residual}.
Particularly debated is the understanding of the origin of Sr, Y and
Zr, for which the hypothesis of an additional primary contribution 
of unknown origin was advanced by \citet{travaglio04} (called LEPP, 
lighter element primary process, see also \citealt{montes07}).
In particular, the same $r$-process by SNe~II that contributes to 
elements heavier than Ba synthesised only $\sim$ 10\% of solar Sr, Y and 
Zr \citep{travaglio04}.
This is sustained by the spread ($>$ 1 dex) shown by [Sr,Y,Zr/Fe] and
[Sr,Y,Zr/Eu] at [Fe/H] $<$ $-$2. 
As example, we show in Fig.~\ref{mnras_Y_noUL3_nosigma1_ingrandito_gallinoRIO},
 observations of [Y/Fe] and [Y/Eu] versus [Fe/H] 
for unevolved Galactic stars (top and bottom panels, respectively). 
Under the hypothesis that about 30\% of iron is produced by SNe~II, 
in first approximation, we may assume initial solar-scaled values
for the ls elements.
Indeed, the [Sr,Y,Zr/Fe] ratios observed in unevolved halo stars 
reach maximum values of about 0.5 dex, which has little impact on 
the [ls/Fe] in CEMP-$s$. 
This behaviour is confirmed by the recent study by
\citet{andr11}, who detected Sr in halo stars and accounted of NLTE 
corrections.
\\ 
In Table~\ref{riniLaEu}, column~3, we reported theoretical results 
of a pure $s$-process contribution ([r/Fe]$^{\rm ini}$ = 0) 
for [La/Fe], [Eu/Fe] and [La/Eu]. AGB models with initial masses 
$M$ = 1.3 and 1.5
$M_{\odot}$, [Fe/H] = $-$2.6 and three $^{13}$C-pockets 
(ST, ST/12, ST/75) are adopted. 
Low [La/Eu]$_{\rm s}$ ratios ($\sim$ 0.5 dex) may be a consequence 
of the low $s$-process contribution to [La/Fe] 
(e.g., case ST that overcomes La mainly producing Pb;
AGBs with low initial mass, which undergo a limited number of TDUs).
Different initial $r$-enrichments are adopted in columns~4 to~7 
([r/Fe]$^{\rm ini}$ = 0.5, 1.0, 1.5, and 2.0), reaching 
[La/Eu]$_{\rm s+r}$ values close to 0.
In case of a low $s$-process contribution to [La/Fe] (e.g., an AGB
model with initial mass $M$ = 1.3 $M_\odot$, case ST), the initial 
$r$-process enhancement predominates, providing [La/Eu]$_{\rm s+r}$ 
down to $-$0.3 $\div$ $-$0.4 for [r/Fe]$^{\rm ini}$ = 1.5 or 2. 
We include these values in the frame of a complete theoretical analysis. 
However, none of these models can provide theoretical interpretations for 
CEMP-$s/r$ stars: indeed, by definition stars with [La/Eu] $<$ 0 belong to a 
different category of stars, the CEMP-$r$ stars \citep{beers05}.

\begin{figure}
   \centering
\includegraphics[angle=-90,width=8.2cm]{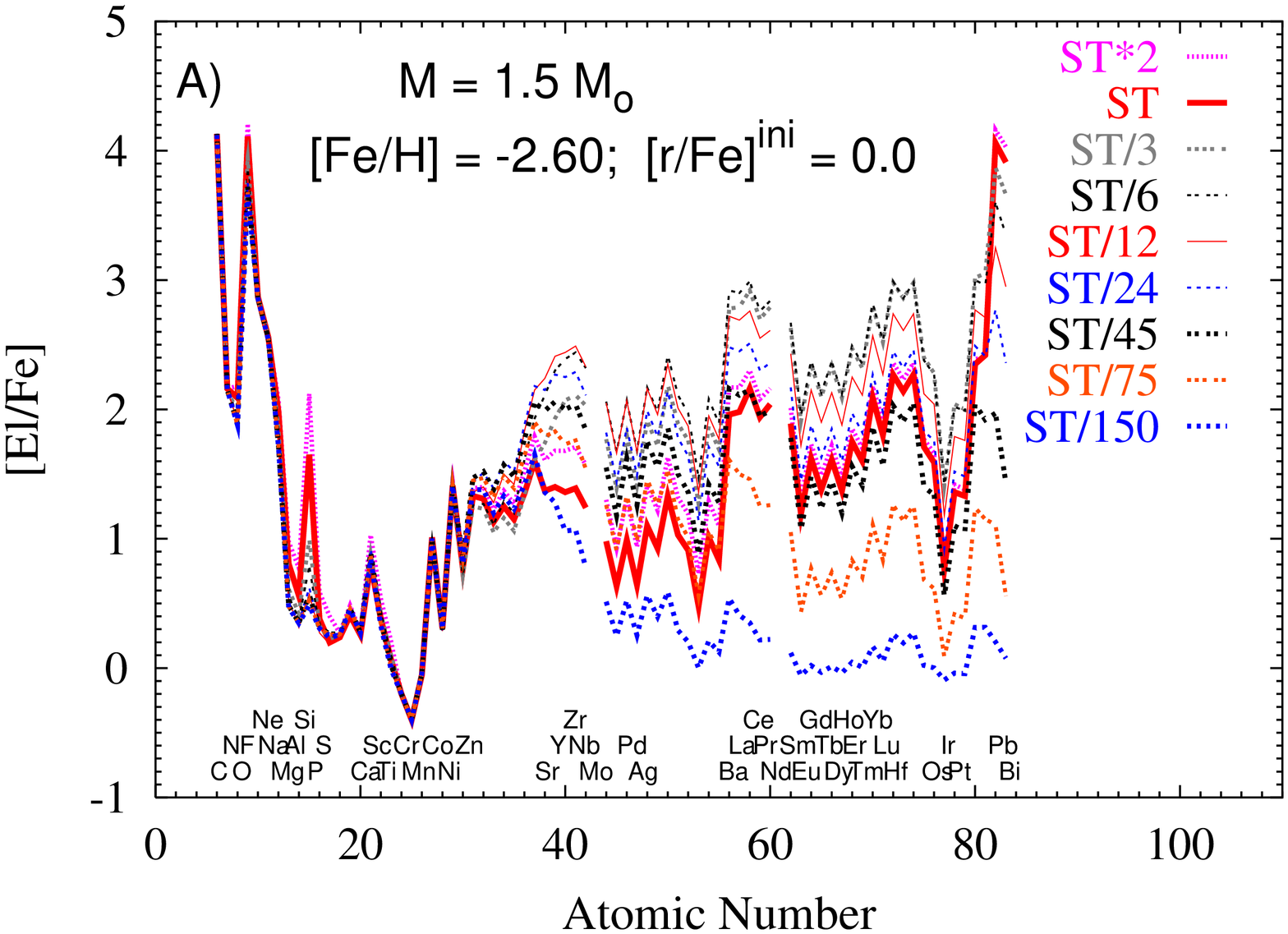}
\includegraphics[angle=-90,width=8.2cm]{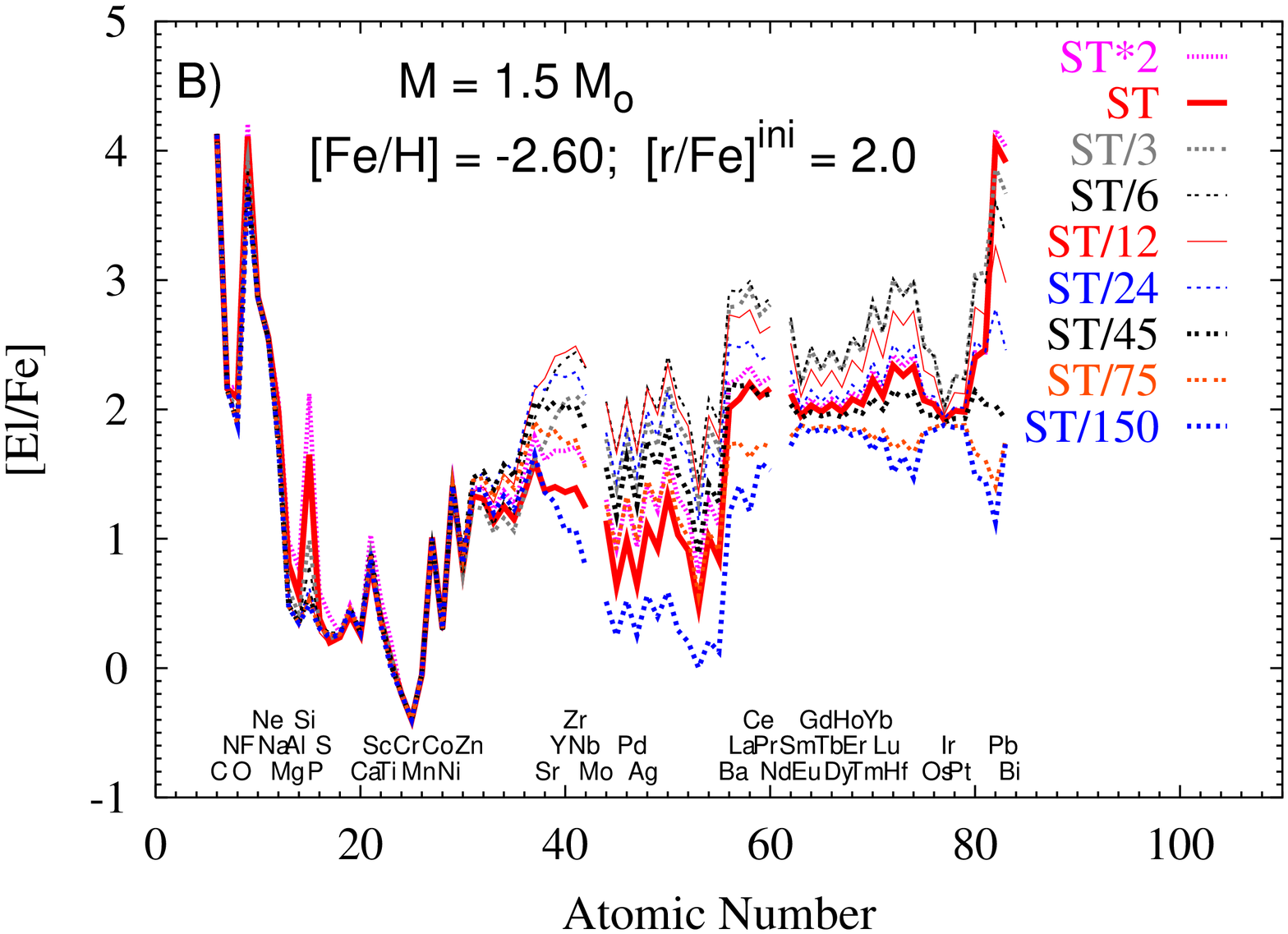}
\includegraphics[angle=-90,width=8.2cm]{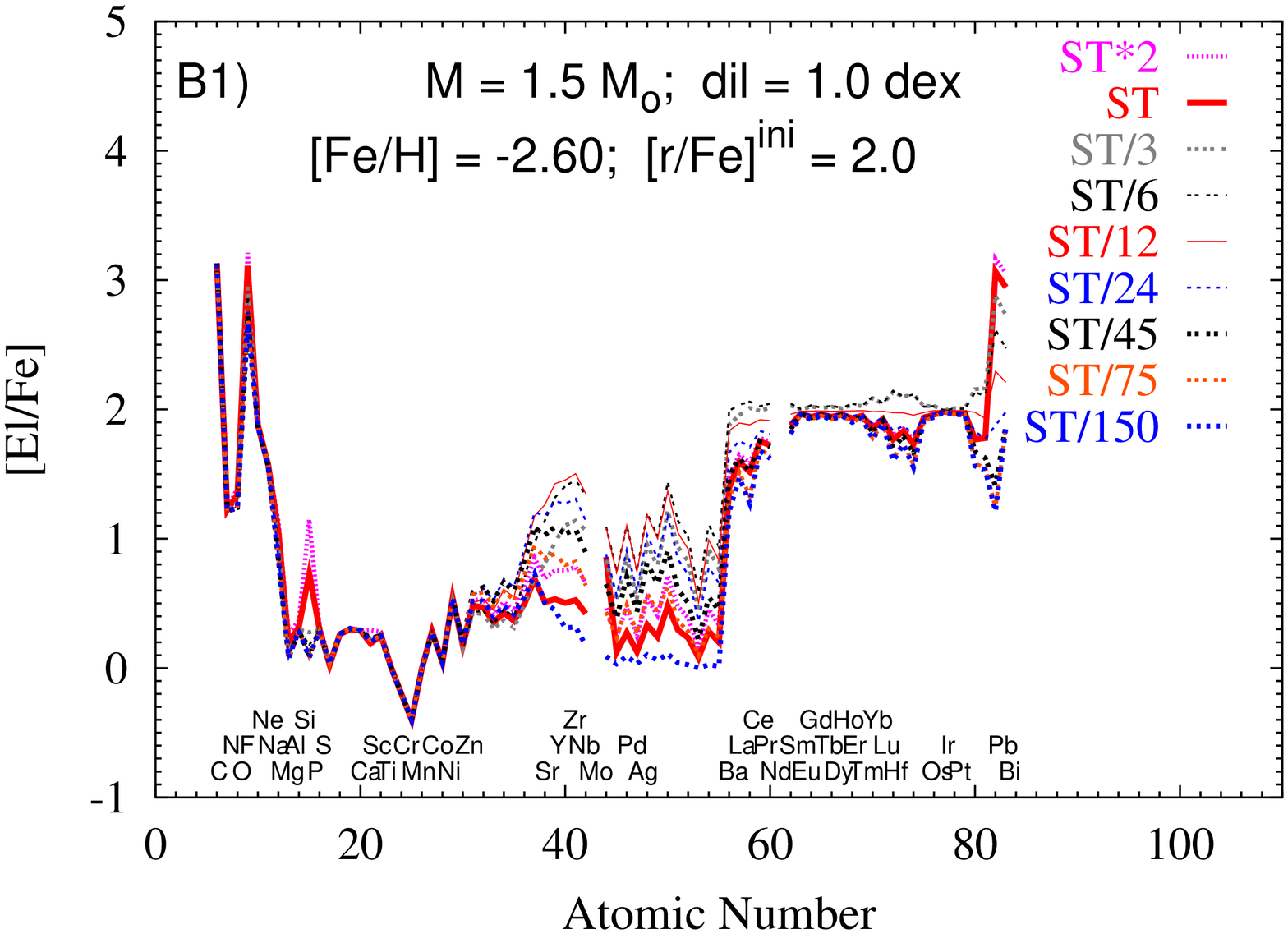}
\caption{Theoretical predictions for AGB models of $M$ = 1.5 $M_{\odot}$, [Fe/H] 
= $-$2.6, and a range of  $^{13}$C-pockets.
While in {\it panel A} no initial $r$-enhancement is assumed,
in {\it panel B} we adopt [r/Fe]$^{\rm ini}$ = 2.0. 
{\it Panel B1: } the same as panel B but with $dil$ = 1 dex.}
\label{bab10m1p5z5m5_rp2_alcuneST_n20}
\end{figure}

\begin{figure*}
   \centering
\includegraphics[angle=-90,width=5.5cm]{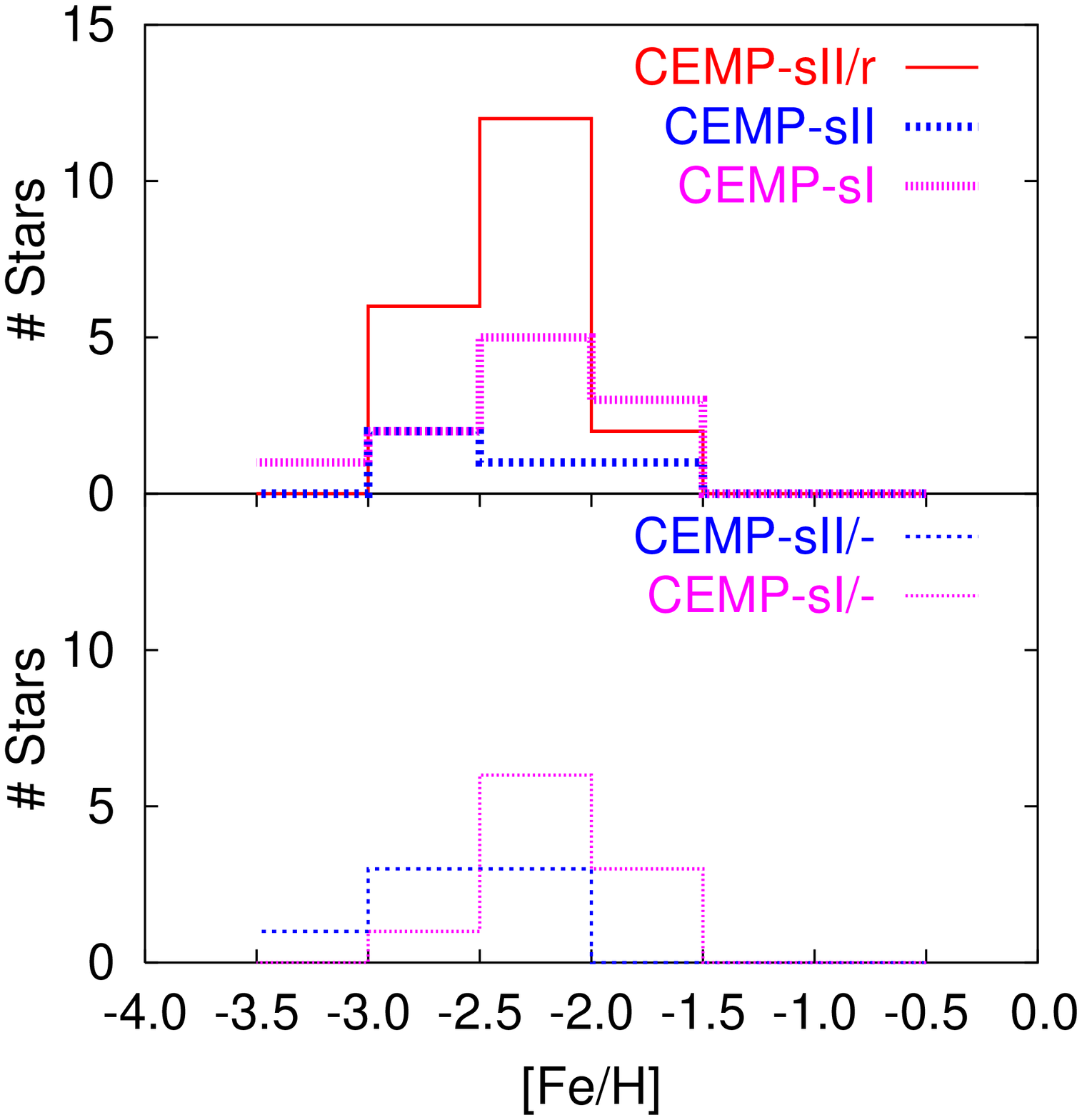}
\includegraphics[angle=-90,width=5.5cm]{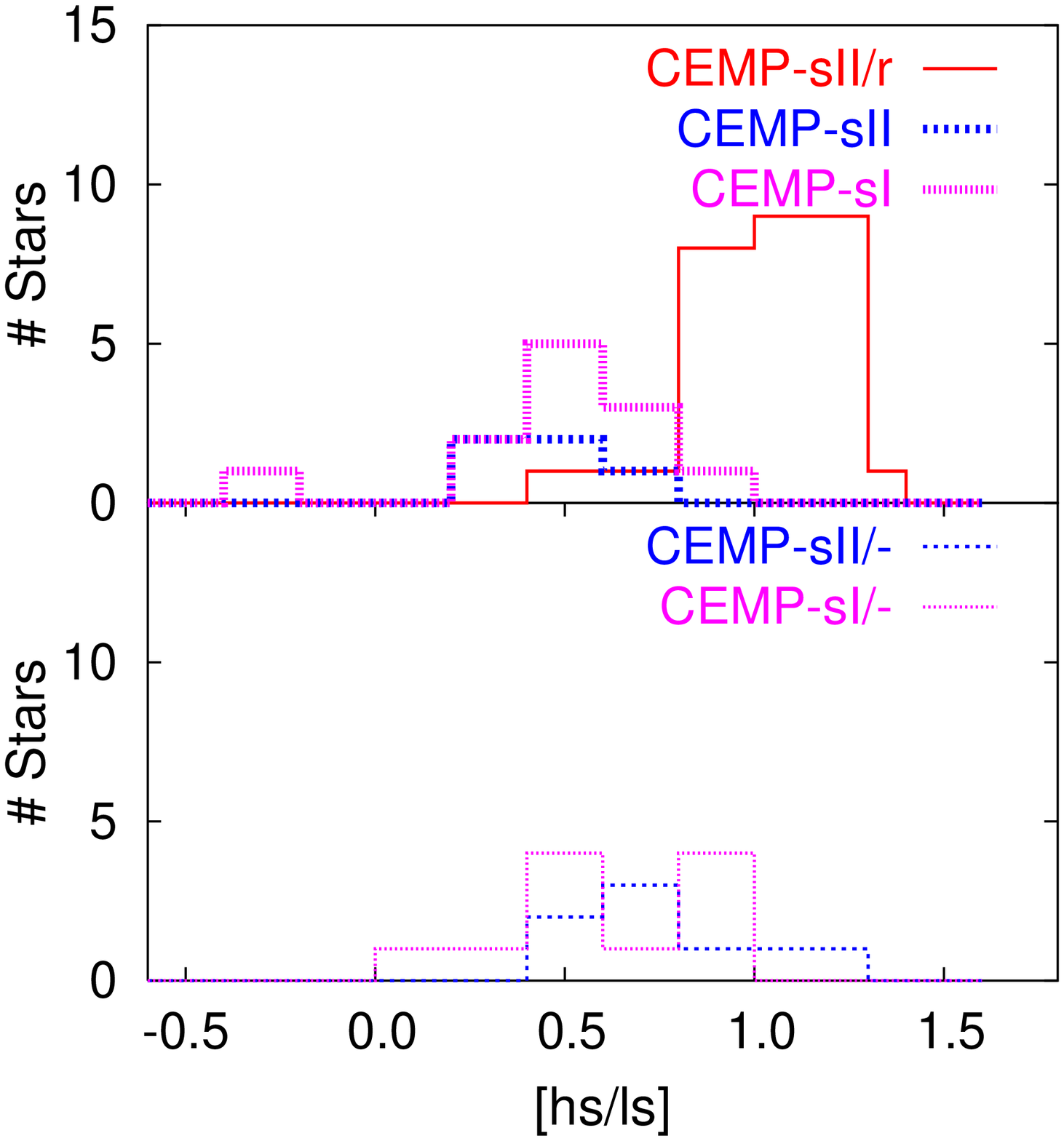}
\includegraphics[angle=-90,width=5.5cm]{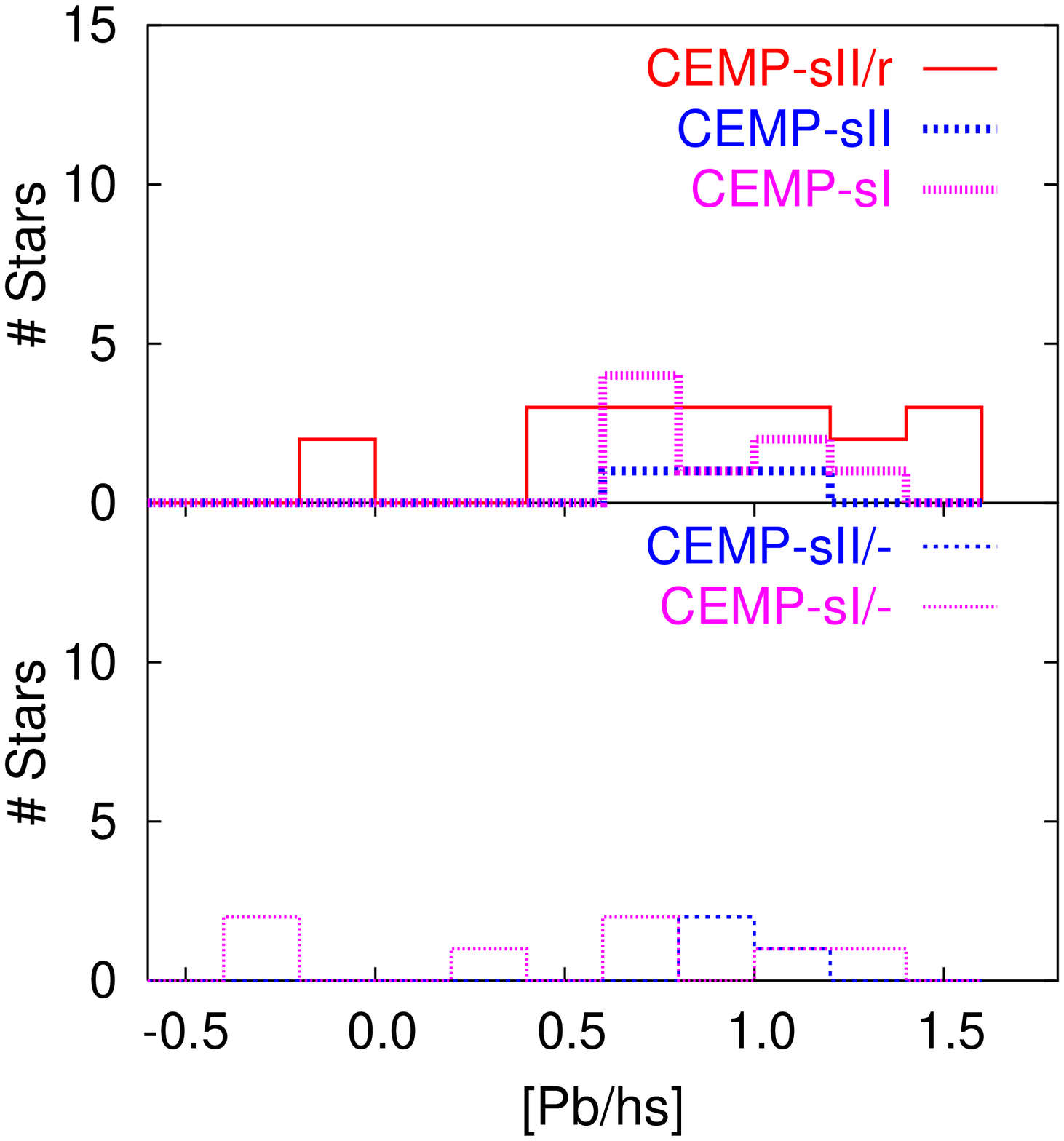}
\caption{Number of stars versus metallicity (\textit{left panel}).
Number of stars versus [hs/ls] (\textit{middle panel}) and 
[Pb/hs] (\textit{right panel}).
We present only the stars in Table~\ref{table5_sindicator} following  
the classification indicated in column~15: CEMP-$s$II/$r$, CEMP-$s$II
and CEMP-$s$I are shown in the \textit{top panels}, while stars with no Eu
detection (CEMP-$s$II/$-$ and CEMP-$s$I/$-$) in the \textit{bottom panels}.}
\label{isto1}
\end{figure*}

\noindent In Table~\ref{riniSrBi} [El/Fe] abundances (for Sr, Y, and Zr and for
elements from Ba to Bi) are reported
for AGB models with initial masses $M$ = 1.3 and 1.5 $M_{\odot}$ at [Fe/H] = $-$2.6, 
two $^{13}$C-pocket choices (ST in columns~3 to~6 and ST/12 in columns~7 to~10), 
and two initial $r$-process enrichments: [r/Fe]$^{\rm ini}$ = 0 and [r/Fe]$^{\rm ini}$ = 2.
The ratios [ls/Fe], [hs/Fe], [hs/ls] and [Pb/hs] are also listed at the 
end of the Table.
From Section~\ref{r}, Table~\ref{ns}, we derive that $\sim$70\% of solar La,
$\sim$60\% of solar Nd, and $\sim$30\% of solar Sm are 
synthesised by the $s$-process. 
Consequently, in presence of a very high $r$-process enrichment, the 
actual value of [hs/Fe] has also to account of an initial $r$-process 
contribution: $\sim$30\% of solar La, $\sim$40\% of solar Nd, and $\sim$70\% 
of solar Sm. Note that the $s$-process nucleosynthesis is not affected by 
different choices of initial $r$-enrichment. The changes in [hs/Fe]
are a consequence of the initial [hs/Fe]$_{\rm r}^{\rm ini}$ enhancement 
assumed molecular cloud by $r$-process contributions.
Then, in case of [r/Fe]$^{\rm ini}$ = 2, the initial [hs/Fe] 
of the molecular cloud is averaged among [La/Fe]$_{\rm r}^{\rm ini}$ 
$\sim$ 1.5, [Nd/Fe]$_{\rm r}^{\rm ini}$ $\sim$ 1.6 and 
[Sm/Fe]$_{\rm r}^{\rm ini}$ $\sim$ 1.8. 
In first approximation we assumed a solar-scaled Y and Zr, because 
the [Y,Zr/Fe] ratios observed in unevolved halo stars reach maximum values
of about 0.5 dex, which does not affect the [ls/Fe] ratios much.
By comparing AGB models with and without initial $r$-enhancement, the 
largest differences are shown for low [hs/Fe]$_{\rm s}$ in the primary AGB, 
because the final [hs/Fe]$_{\rm s+r}$ value is more sensible to the initial 
$r$-process enrichment of the molecular cloud.
These differences are reduced by increasing the number of thermal pulses.
Indeed, $M^{\rm ini}_{\rm AGB}$ = 1.5 $M_\odot$ models show negligible 
differences in [hs/ls].
The solar $r$-process contribution to Pb is 15\% ($\pm$ 5), 
with negligible effects on [Pb/Fe] for all initial mass models,
due to the high $s$-process production of $^{208}$Pb 
at low metallicities.  
In Fig.~\ref{bab10m1p5z5m5_rp2_alcuneST_n20} {\it top and middle panels}, 
we show the theoretical predictions of AGB models with initial mass
$M$ = 1.5 $M_{\odot}$, a wide range of $^{13}$C-pockets and [r/Fe]$^{\rm ini}$ 
= 0 and 2, respectively.
\\
To simulate large mixing between the $s$-rich AGB material accreted onto
the convective envelope of an observed giant (after the FDU, 
Section~\ref{observations}), we show
in Fig.~\ref{bab10m1p5z5m5_rp2_alcuneST_n20} {\it bottom panel},
AGB models with $M^{\rm ini}_{\rm AGB}$ = 1.5 $M_{\odot}$, 
[r/Fe]$^{\rm ini}$ = 2 and a dilution of 1 dex. 
The dilution does not affect the initial [El/Fe] enhancement assumed in 
the molecular cloud by $r$-process contributions, because
both stars of the binary system are supposed 
to have the same initial chemical composition.
Instead, the material transferred from the AGB is 
sensibly modified by large dilutions.
Therefore, the initial $r$-contribution dominates the 
$s$-process material, thus affecting the [hs/Fe] ratio.

\section{General comparison between
theory and observations in CEMP-{\scriptsize s} and CEMP-{\scriptsize s/r} stars:
 [L{\scriptsize a}/E{\scriptsize u}], [{\scriptsize hs}/{\scriptsize ls}] and 
 [P{\scriptsize b}/{\scriptsize hs}]} 
\label{generalcomparison}

\begin{figure}
   \centering
\includegraphics[angle=-90,width=8cm]{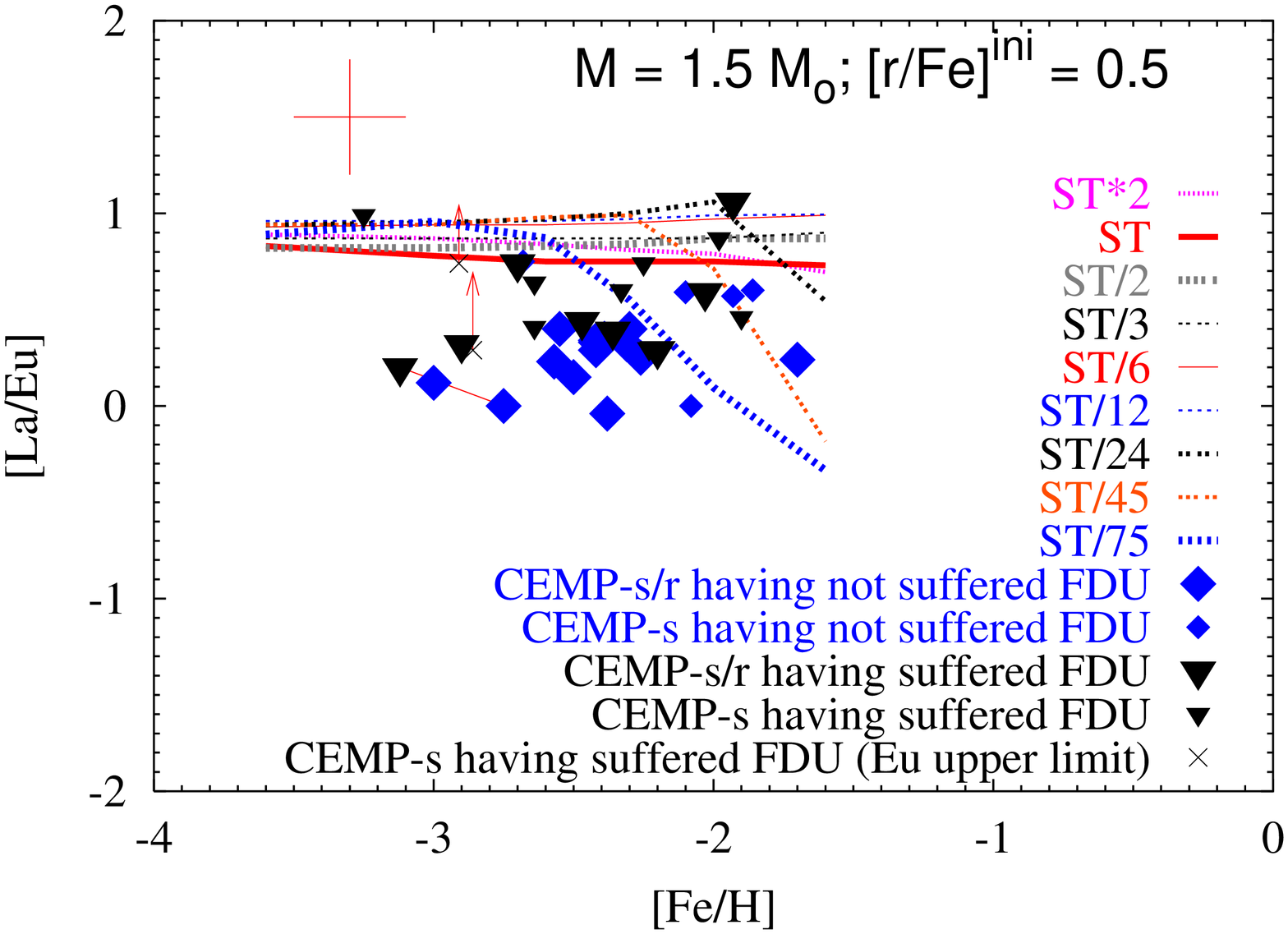}
\includegraphics[angle=-90,width=8cm]{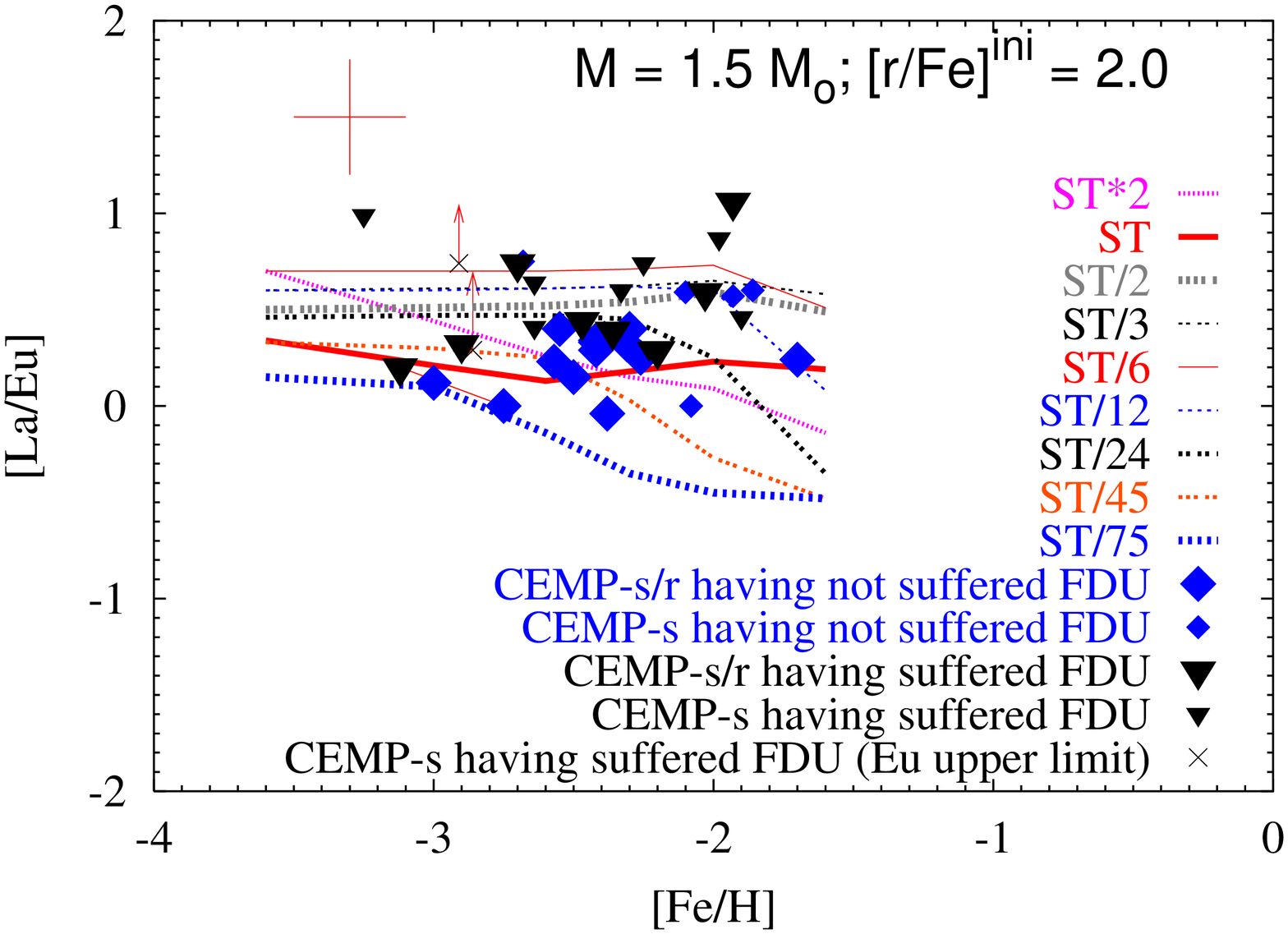}
\caption{Theoretical predictions of [La/Eu] versus 
metallicity for AGB models with initial mass $M$ = 1.5 $M_{\odot}$ 
and a range of $^{13}$C-pockets compared with observations of
CEMP-$s$ and CEMP-$s/r$ stars listed in Table~\ref{table5_sindicator}.
In the \textit{top panel} we adopt [r/Fe]$^{\rm ini}$ = 0.5, corresponding to an 
average of [Eu/Fe] in unevolved halo stars (Section~\ref{r}). 
\textit{Bottom panel}: same as top panel but for [r/Fe]$^{\rm ini}$ = 2, taken 
as the highest value adopted to interpret CEMP-$s/r$ stars.
Diamonds are stars before the FDU, while triangles are stars having 
suffered the FDU. CEMP-$s/r$ and CEMP-$s$ stars are 
represented by big and little symbols, respectively.
For the star CS 22183--015, having uncertain atmospheric parameters
\citep{cohen06,aoki07,JB02,lai07}, the values obtained 
by different authors are connected by a (red) line.
Upper limits for Eu are represented by cross symbols.
Note that HD 209621 ([Fe/H] = $-$1.93) is considered a 
CEMP-$s/r$ despite its high [La/Eu] $\sim$ 1, because an [r/Fe]$^{\rm ini}$ 
= 1 is needed to interpret [hs/Eu];
on the other side, CS 29513--032 ([Fe/H] = $-$2.08) is considered a CEMP-$s$ 
despite the observed [La/Eu] $\sim$ 0, because it has a low [La/Fe] and not
an enhanced [Eu/Fe] ratio.}
\label{mnras_AAbab9ltHT_laeuvsfe}
\end{figure}

\begin{figure}
   \centering
\includegraphics[angle=-90,width=8cm]{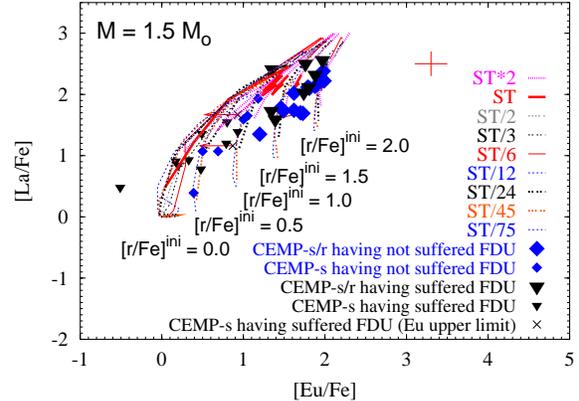}
\caption{Theoretical predictions of [La/Fe] versus [Eu/Fe] for AGB models with initial mass 
$M$ = 1.5 $M_{\odot}$ and a range of $^{13}$C-pockets
compared with observations of CEMP-$s$ and CEMP-$s/r$ stars.
The different theoretical ranges correspond to the initial $r$-process enrichments
[r/Fe]$^{\rm ini}$ = 0.0, 0.5, 1.0, 1.5 and 2.0, adopted to interpret the spread 
observed in CEMP-$s$ and CEMP-$s/r$ stars. Symbols are the same as in
Fig.~\ref{mnras_AAbab9ltHT_laeuvsfe}.
The star CS 30322--023 agrees with [r/Fe]$^{\rm ini}$ = $-$1 (see Paper III).}
\label{mnras_AAbab9ltHT_lafevseufe}
\end{figure}

\begin{figure}
   \centering
\includegraphics[angle=-90,width=8cm]{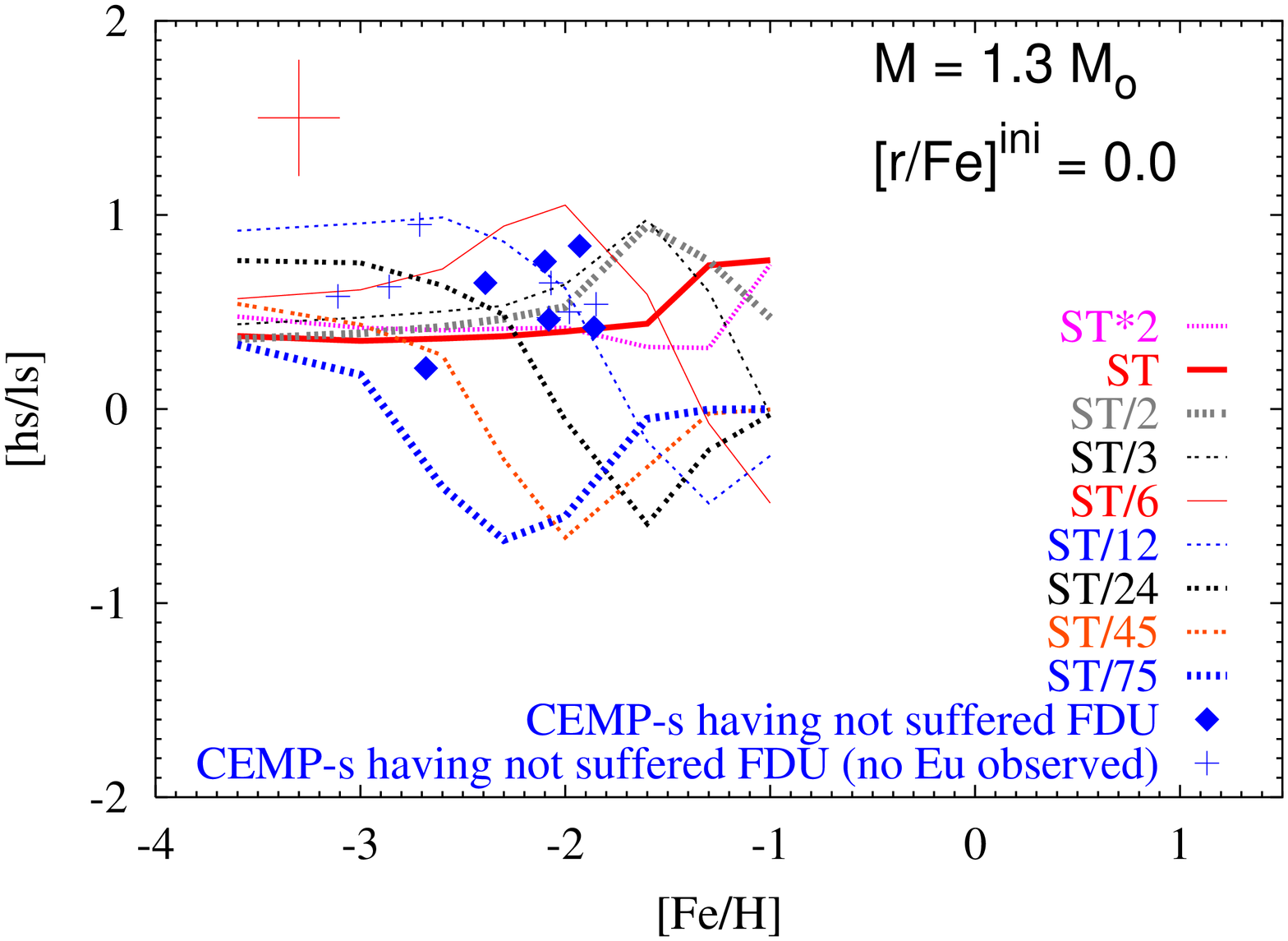}
\includegraphics[angle=-90,width=8cm]{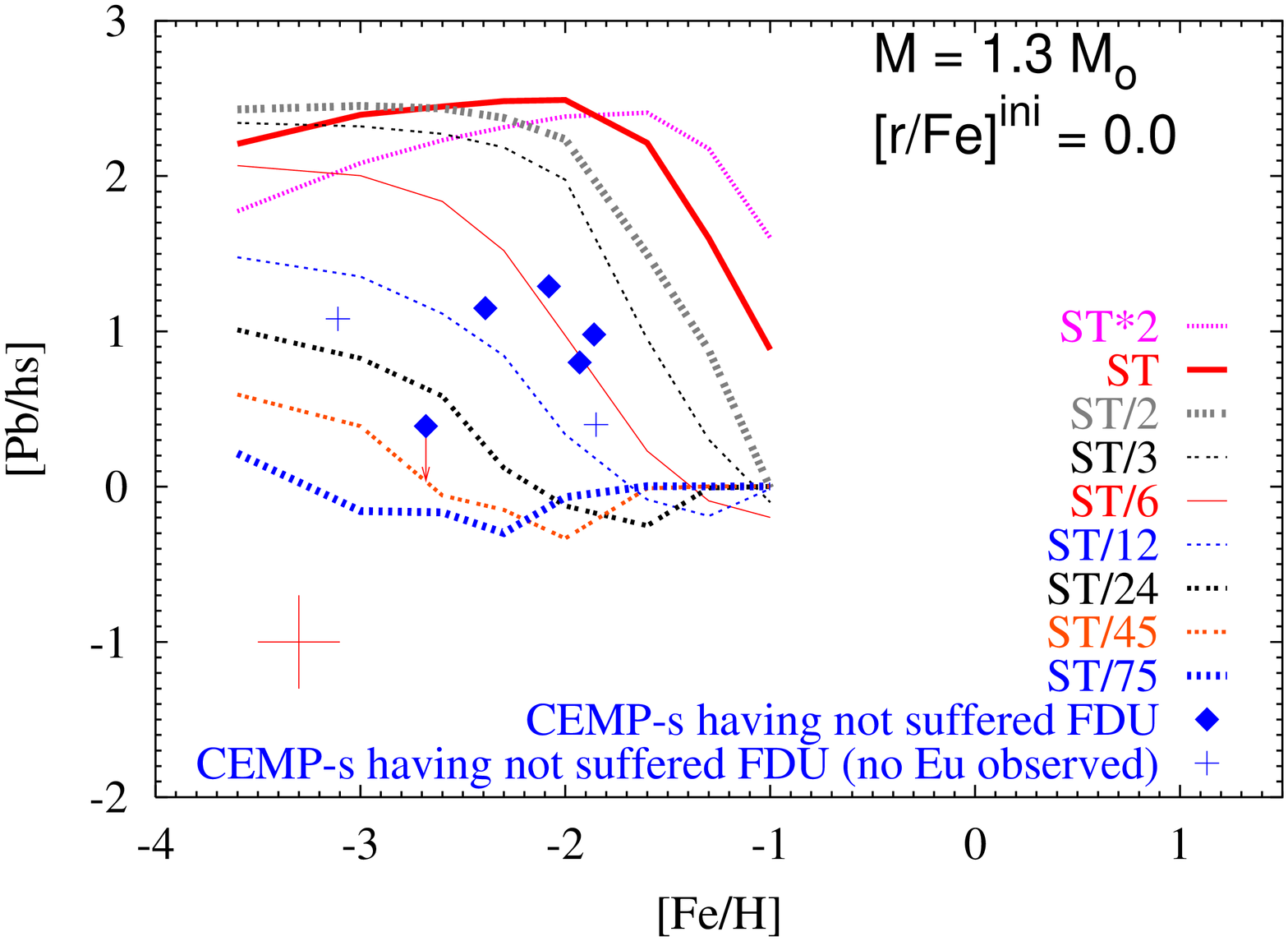}
\caption{\textit{Top Panel:} theoretical predictions of [hs/ls] versus 
metallicity for AGB models with initial mass 
$M$ = 1.3 $M_{\odot}$ and a range of $^{13}$C-pockets
compared with CEMP-$s$ observations (little diamonds). 
No initial $r$-process enhancement is assumed.
To simplify the discussion, we represent only 
main-sequence/turnoff/subgiant 
stars having not suffered the FDU (see text).
CEMP-$s$ without europium detection are indicated 
by plus symbols. 
\textit{Bottom Panel:} the same as the top panel, 
but for [Pb/hs] versus metallicity.
Typical error bars are $\Delta$[hs/ls] = $\Delta$[Pb/hs]
 = 0.3; $\Delta$[Fe/H] = 0.2.}
\label{mnras1_AAbab9ltHT_m1p5_WobsCEMP_n5_EunoEu_MSTO}
\end{figure}

\begin{figure}
   \centering
\includegraphics[angle=-90,width=8cm]{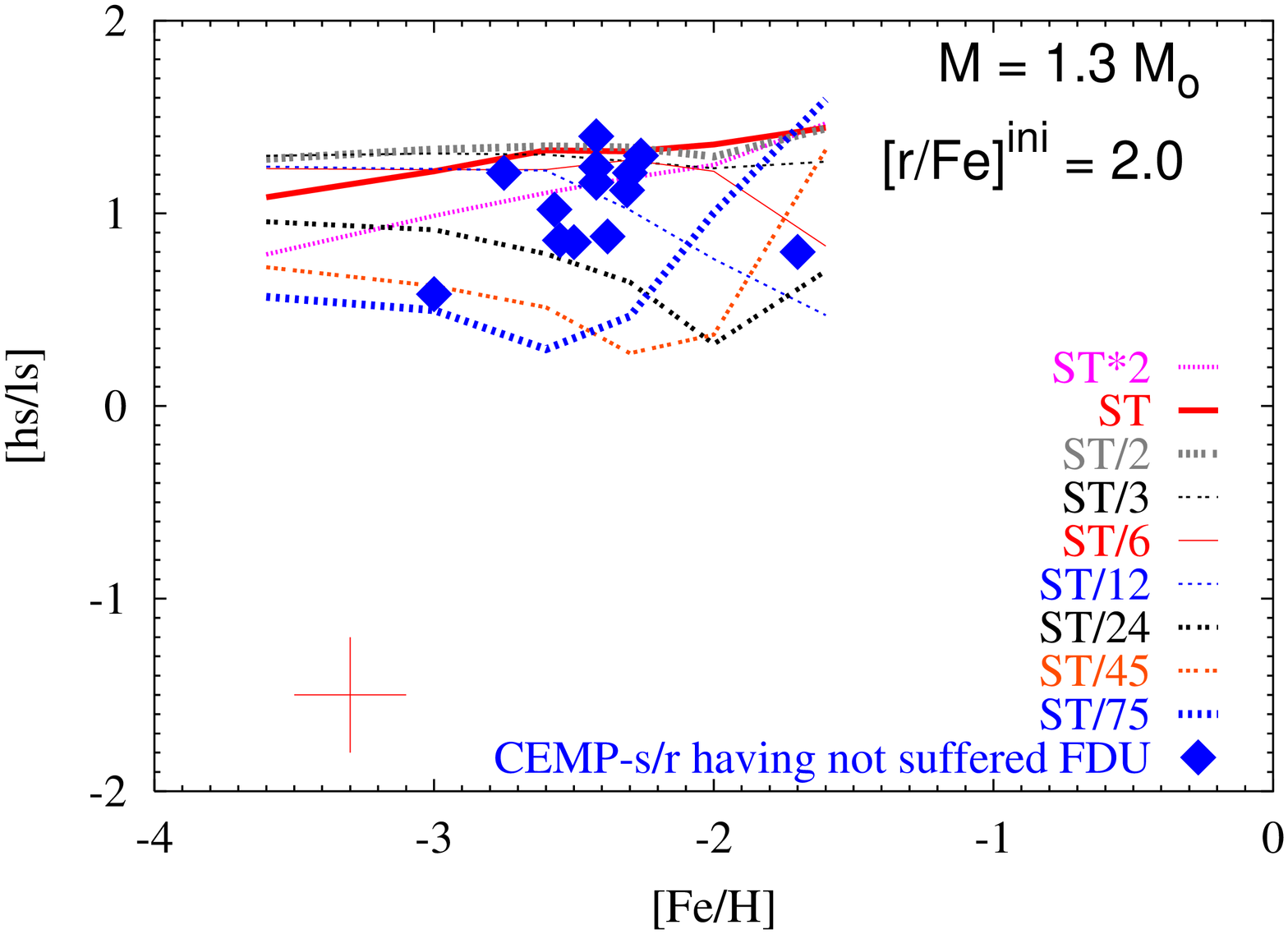}
\includegraphics[angle=-90,width=8cm]{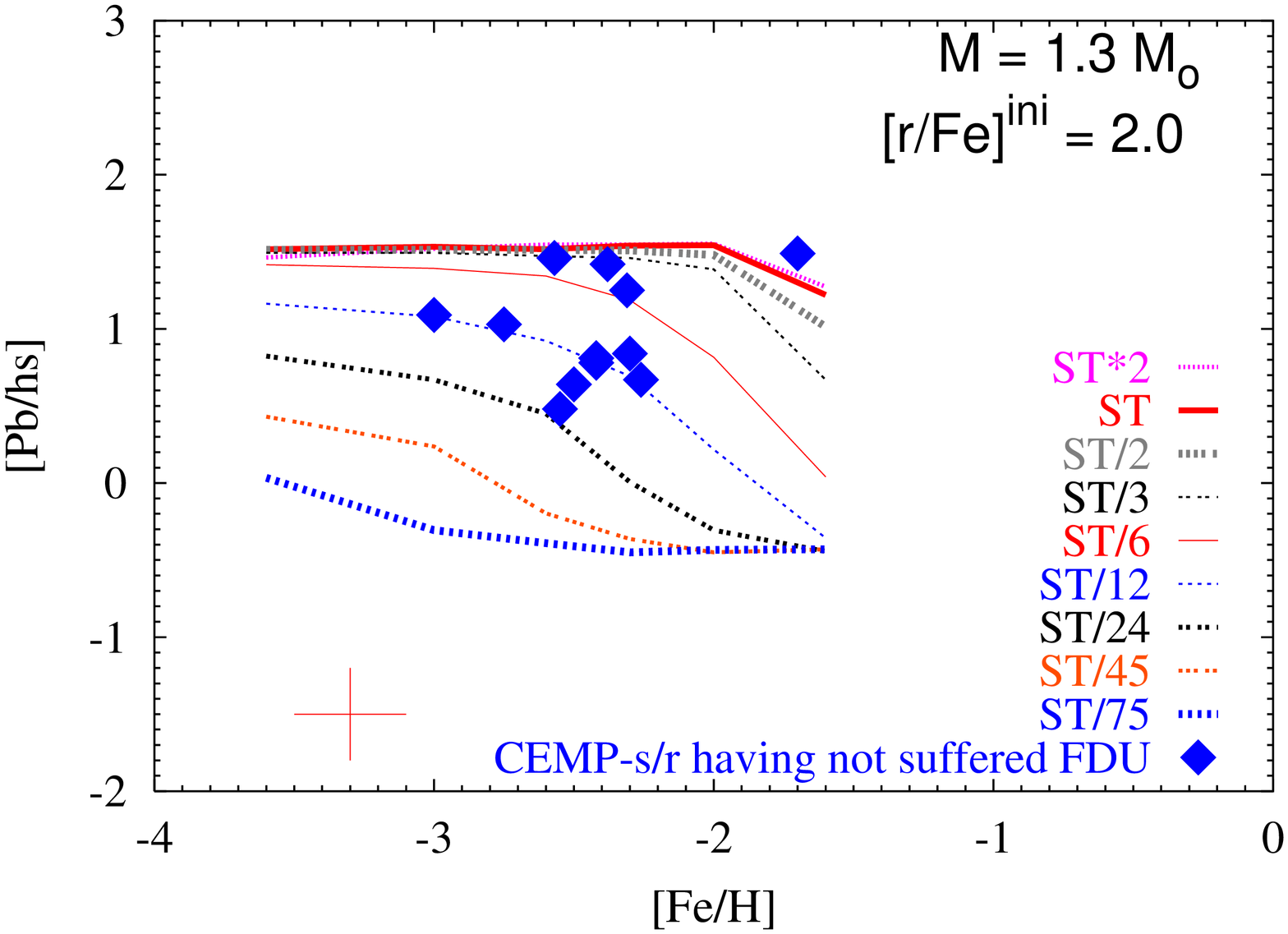}
\caption{The same as Fig~\ref{mnras1_AAbab9ltHT_m1p5_WobsCEMP_n5_EunoEu_MSTO}, 
but for AGB models with [r/Fe]$^{\rm ini}$ = 2,
compared with observations of main-sequence/turnoff/subgiant 
CEMP-$s/r$ having not suffered the FDU (big diamonds).}
\label{mnras1_AAbab10_m1p5_rp2_WobsCEMPs+r_n5_EunoEu_MSTO}
\end{figure}

\begin{figure}
   \centering
\includegraphics[angle=-90,width=8cm]{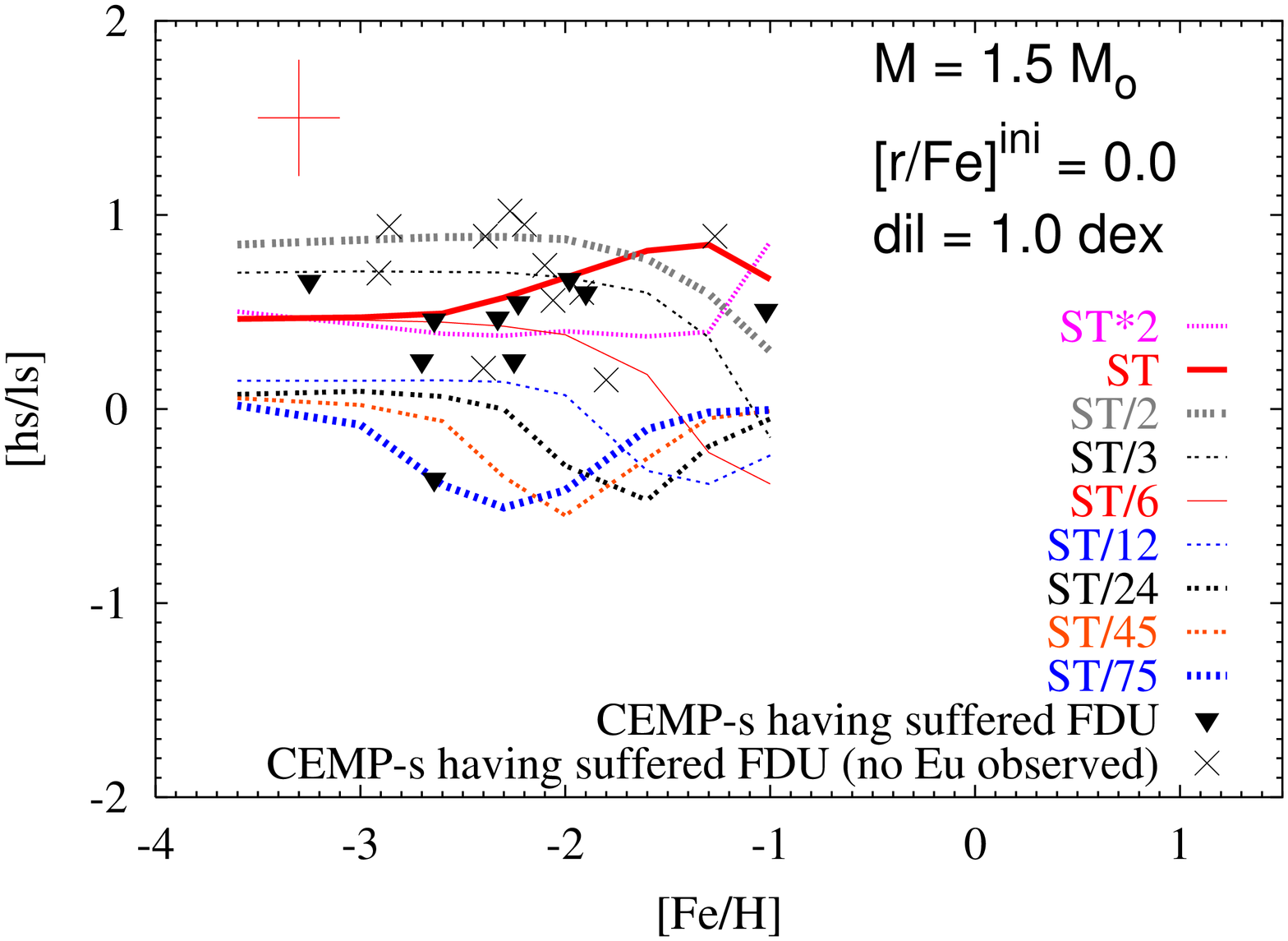}
\includegraphics[angle=-90,width=8cm]{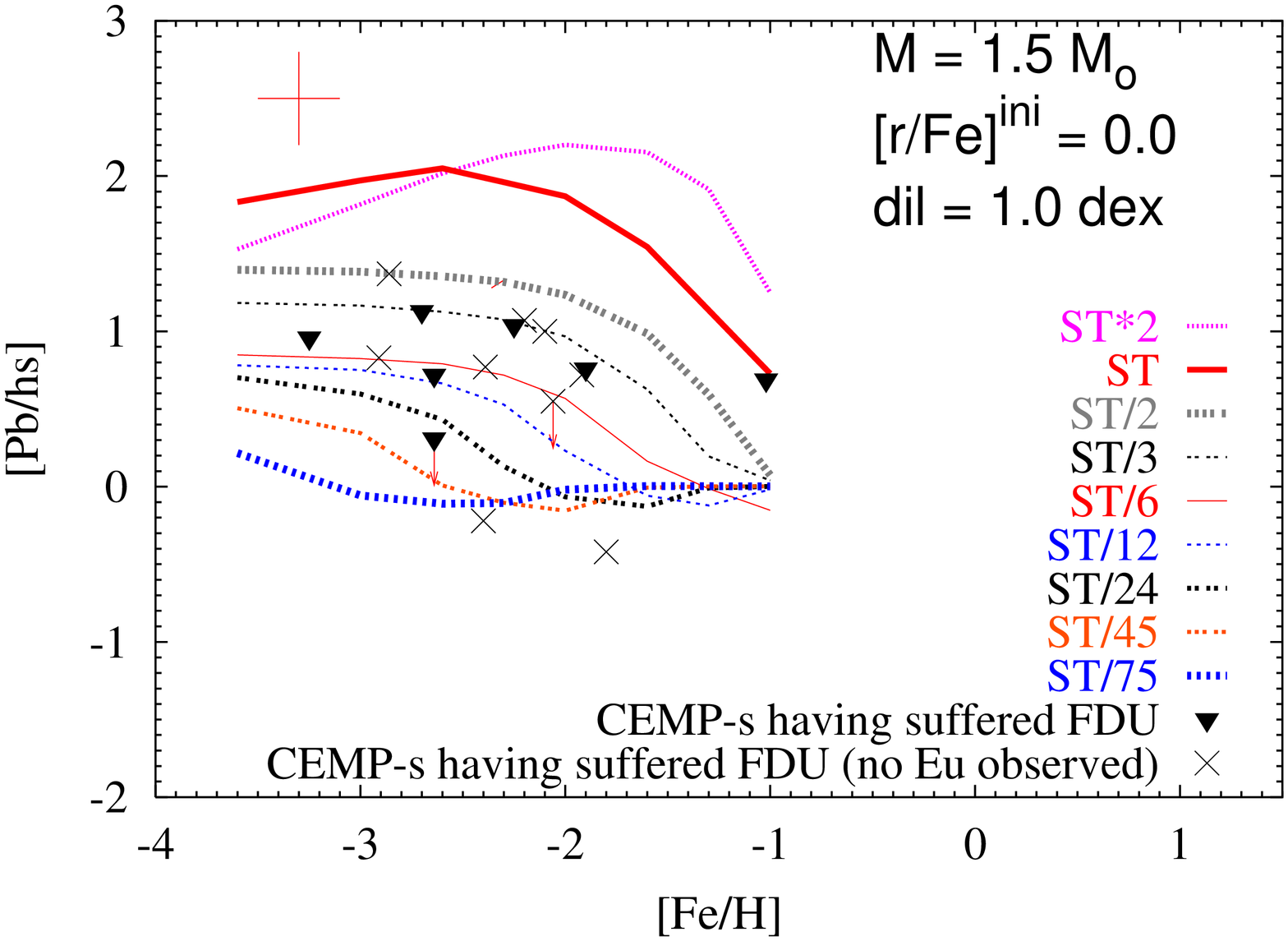}
\caption{\textit{Top Panel:} theoretical predictions of [hs/ls] versus 
metallicity for AGB models with initial mass 
$M$ = 1.5 $M_{\odot}$, a range of $^{13}$C-pockets and $dil$ = 1.0 dex,
compared with CEMP-$s$ observations (triangles). 
No initial $r$-process enhancement is assumed.
To simplify the discussion, we represent only subgiants and giants 
having suffered the FDU (see text).
CEMP-$s$ stars without europium detection are indicated by crosses. 
\textit{Bottom Panel:} the same as the top panel, but for 
[Pb/hs] versus metallicity.
Typical error bars are $\Delta$[hs/ls] = $\Delta$[Pb/hs] = 
0.3; $\Delta$[Fe/H] = 0.2.}
\label{mnras1_AAbab9ltHT_m1p5_WobsCEMP_n20_EunoEu_Giants}
\end{figure}

\begin{figure}
   \centering
\includegraphics[angle=-90,width=8cm]{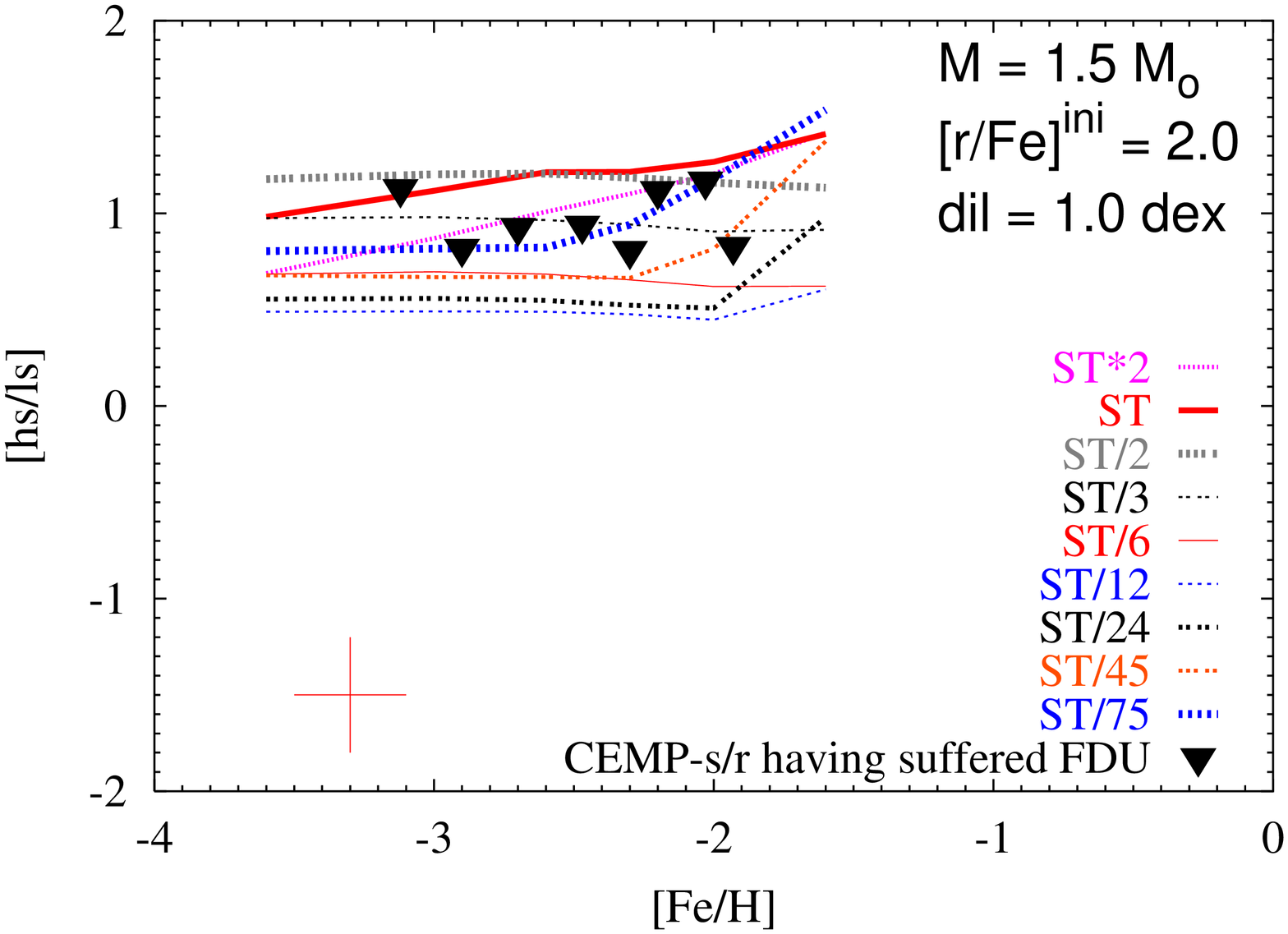}
\includegraphics[angle=-90,width=8cm]{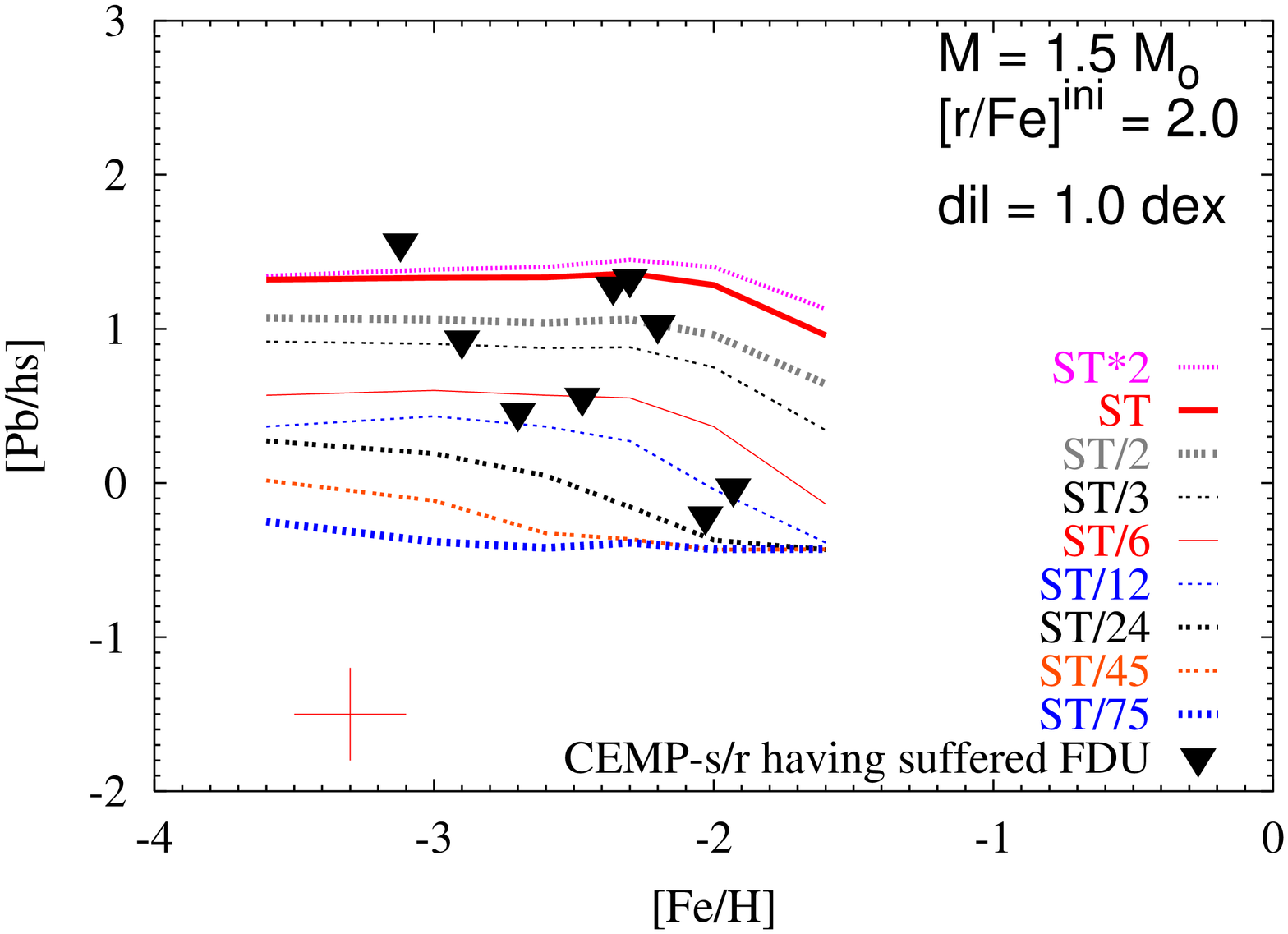}
\caption{The same as Fig.~\ref{mnras1_AAbab9ltHT_m1p5_WobsCEMP_n20_EunoEu_Giants}, 
but for AGB models with [r/Fe]$^{\rm ini}$ = 2,
as compared with CEMP-$s/r$  
stars having suffered the FDU
(triangles).}
\label{mnras1_AAbab10_m1p5_rp2_WobsCEMPs+r_n20_EunoEu_Giants}
\end{figure}

In this Section, we compare spectroscopic observations of 
CEMP-$s$ and CEMP-$s/r$ stars listed in Table~\ref{table5_sindicator} 
with AGB predictions. 
Stars from Table~\ref{tablestellemancanti} are excluded from this 
analysis owing to the limited number of detected $s$-process elements.
\\
To examine the characteristics of the classes of stars
indicated in Section~\ref{observations}, we show in Fig.~\ref{isto1}
three histograms, with the number of stars versus [Fe/H], [hs/ls]
and [Pb/hs], respectively. 
CEMP-$s$II$/r$, CEMP-$s$II and CEMP-$s$I are displayed in the top panels,
while stars without Eu measurement (CEMP-$s$II/$-$ and CEMP-$s$I/$-$) are shown 
in the bottom panels.
Note that no CEMP-$s$I$/r$ are present in the sample.
Most stars have metallicities between $-$2.5 $\la$ [Fe/H] 
$\la$ $-$2.0 (Fig.~\ref{isto1}, left panel). 
On average, the same ranges are covered by CEMP-$s/r$ and CEMP-$s$ stars.
A different behaviour appears in the middle panel for [hs/ls]:
the majority of the stars lay between 0.5 $\leq$ [hs/ls] $\leq$ 1.2, but
CEMP-$s$II$/r$ stars exhibit higher [hs/ls] values than CEMP-$s$II and 
CEMP-$s$I stars.
Instead, the [Pb/hs] observed in CEMP-$s$ and CEMP-$s/r$ stars is equally 
distributed within the range 0.5 $\leq$ [Pb/hs] $\leq$ 1.3 (right panel).
To improve the analysis, we then exclude CEMP-$s$II/$-$ and CEMP-$s$I/$-$ from
the following discussion to avoid possible medley between CEMP-$s/r$ and CEMP-$s$ stars.
We start with the analysis of [La/Eu] versus [Fe/H] and [La/Fe] 
versus [Eu/Fe]; then we discuss the behaviour of the two $s$-process 
indicators [hs/ls] and [Pb/hs] versus metallicity.
\\
In Fig.~\ref{mnras_AAbab9ltHT_laeuvsfe}, \textit{top panel}, theoretical
predictions of [La/Eu] versus metallicity for AGB models with initial mass 
$M$ = 1.5 $M_{\odot}$  and a range of $^{13}$C-pockets are
compared with observations of CEMP-$s$ (little symbols) and CEMP-$s/r$ (big
symbols) stars. 
Upper limits for Eu are represented by crosses with arrows.
Diamonds represent stars before their FDU, while triangles denote
giants having suffered the FDU (respectively `no' and `yes' in 
column~3 of Table~\ref{table5_sindicator}). 
For simplicity, the five stars for which the occurrence of the FDU is
uncertain (CS 31062--050, CS 22880--074, CS 29513--032, HE 0036+0113, 
HE 2232--0603, `no' in column~3 of Table~\ref{table5_sindicator}), 
are here included among the subgiants, which do not show FDU.
For CS 22183--015, with uncertain atmospheric parameters
\citep{cohen06,aoki07,JB02,lai07}, the values obtained 
by different authors are connected by a (red) line.
Typical error bars of $\Delta$[La/Eu] = $\pm$ 0.3 dex and $\Delta$[Fe/H] = $\pm$
0.2 dex are shown.
Theoretical predictions are normalised to the solar meteoritic abundances by
\citet{AG89}.
The differences between the solar normalisation adopted by different authors
are negligible if compared with the typical error bars that are shown in the figures. 
For the most accurate analysis provided in Section~\ref{method} and in Paper III, 
both predictions and observations are normalised to the solar photospheric 
values by \citet*{lodders09}. 
As anticipated in Section~\ref{observations}, in general CEMP-$s/r$ have
0.0 $\leq$ [La/Eu] $\leq$ 0.5. AGB models with [r/Fe]$^{\rm ini}$ = 0.5 are shown in the
\textit{top panel}, in agreement with an average of the 
[Eu/Fe] observed in field halo stars (Section~\ref{r}).
[La/Eu]$_{\rm th}$ is definitely overestimated in CEMP-$s/r$.
With an initial $r$-enrichment of [r/Fe]$^{\rm ini}$ = 2, \textit{bottom panel},
an agreement between CEMP-$s/r$ and theoretical predictions is reached 
except for two cases: the giant HD 209621 by \citet{GA10} ([Fe/H] = $-$1.93 
and [La/Eu] $\sim$ 1) and the CEMP-$s$ CS 29513--032 by \citet{roederer10} 
([Fe/H] = $-$2.08 and [La/Eu] $\sim$ 0).
HD 209621 has been classified as a CEMP-$s/r$ because [r/Fe]$^{\rm ini}$ 
= 1 is needed to interpret the observed [hs/Eu] ratio, even if [La/Fe] 
is $\sim$ 0.5 dex higher than the average for the hs elements.  
CS 29513--032 shows a mild $s$-process enhancement; then, the low [La/Eu] 
observed has to be attributed to a low $s$-process contribution to [La/Fe] 
instead of an enhanced [Eu/Fe] produced by the $r$-process.  
\\
In
Fig.~\ref{mnras_AAbab9ltHT_lafevseufe}, [La/Fe] is shown versus [Eu/Fe].
Symbols are the same as in Fig.~\ref{mnras_AAbab9ltHT_laeuvsfe}.
Spectroscopic observations 
are compared with AGB theoretical distributions assuming different initial 
$r$-process enrichments, [r/Fe]$^{\rm ini}$ = 0.0, 0.5, 1.0, 1.5 and 2.0.
The giant CS 30322--023 by \citet{masseron06} with a negative [Eu/Fe], 
is matched with a subsolar initial enrichment of the molecular 
cloud (see Paper III).

In Figs.~\ref{mnras1_AAbab9ltHT_m1p5_WobsCEMP_n5_EunoEu_MSTO} --~\ref{mnras1_AAbab10_m1p5_rp2_WobsCEMPs+r_n20_EunoEu_Giants}
we show the behaviour of [hs/ls] (\textit{top panels}) 
and [Pb/hs] (\textit{bottom panels}) versus metallicity for
AGB models with $M^{\rm ini}_{\rm AGB}$ = 1.3 and 1.5 $M_{\odot}$
and a range of $^{13}$C-pockets, compared with the observations of
stars listed in Table~\ref{table5_sindicator}. 
Main-sequence/turnoff and subgiants having not suffered the
FDU are compared with $M^{\rm ini}_{\rm AGB}$ = 1.3 $M_{\odot}$ models
(Figs.~\ref{mnras1_AAbab9ltHT_m1p5_WobsCEMP_n5_EunoEu_MSTO}
--~\ref{mnras1_AAbab10_m1p5_rp2_WobsCEMPs+r_n5_EunoEu_MSTO}); 
subgiants/giants having suffered the FDU are compared with 
$M^{\rm ini}_{\rm AGB}$ = 1.5 $M_{\odot}$ models 
(Figs.~\ref{mnras1_AAbab9ltHT_m1p5_WobsCEMP_n20_EunoEu_Giants}
--~\ref{mnras1_AAbab10_m1p5_rp2_WobsCEMPs+r_n20_EunoEu_Giants})
and a $dil$ = 1.0 dex to simulate the FDU mixing (Section~\ref{observations}).
This distinction is made to simplify the discussion: indeed, 
the large dilution adopted for giants is still compatible with 
an observed high $s$-enhancement
(e.g., [hs/Fe] $\sim$ 2) if $M^{\rm ini}_{\rm AGB}$ = 1.5 or
2 $M_\odot$ models are adopted. $M^{\rm ini}_{\rm AGB}$ 
= 1.3 $M_\odot$ models undergo a limited number of TDUs (at the 5th TDU
the maximum [hs/Fe] predicted is $\sim$ 2.1 for cases close to ST/12,
as shown in Fig.~8 of Paper I), and, in first
approximation may be excluded during the analysis of CEMP-$s$II giants. 
However, this distinction has to be considered with caution: indeed, AGB 
models with different initial masses may plausibly interpret
the spectroscopic observations of a given star (see e.g., Table~\ref{summary1}).
We defer to Paper III for a detailed analysis of individual 
stars.
The star CS 22183--015, having uncertain atmospheric parameters
is represented in both figures.
Typical uncertainties are $\Delta$[hs/ls] = $\pm$ 0.3 dex, 
$\Delta$[Pb/hs] = $\pm$ 0.3 dex, and $\Delta$[Fe/H] = $\pm$ 0.2 dex.
\\
While no initial $r$-enhancement is assumed 
for CEMP-$s$ stars (Figs.~\ref{mnras1_AAbab9ltHT_m1p5_WobsCEMP_n5_EunoEu_MSTO}
and~\ref{mnras1_AAbab9ltHT_m1p5_WobsCEMP_n20_EunoEu_Giants}),
CEMP-$s/r$ stars are compared with AGB models with [r/Fe]$^{ini}$ =
2.0 (Figs.~\ref{mnras1_AAbab10_m1p5_rp2_WobsCEMPs+r_n5_EunoEu_MSTO}
and~\ref{mnras1_AAbab10_m1p5_rp2_WobsCEMPs+r_n20_EunoEu_Giants}).
On average we confirm higher [hs/ls] ratios for CEMP-$s/r$ stars 
(big symbols) than for CEMP-$s$ stars (little symbols) as shown in Fig.~\ref{isto1},
middle panel. Preliminary results have been presented 
by \citealt{kaeppeler10rmp}.
This agrees with AGB model predictions if a very high initial 
$r$-process enhancement of the molecular cloud is adopted.
As discussed in Section~\ref{CEMPs+r}, an initial $r$-enhancement of 
[r/Fe]$^{ini}$ = 2.0 may affect the final [hs/Fe], because  
$\sim$30\% for solar La, $\sim$40\% for solar Nd, and 
$\sim$70\% for solar Sm are synthesised by the $r$-process.
Indeed, with [r/Fe]$^{ini}$ = 2.0, AGB models predict a maximum [hs/ls] 
ratio $\sim$ 0.3 dex higher than pure $s$-process nucleosynthesis models.
In first approximation, this allows to reproduce the [hs/ls] range 
covered by CEMP-$s/r$ stars, which are otherwise underestimated by pure 
$s$-process predictions. 
Note that, if lower [r/Fe]$^{ini}$ are assumed, the initial 
$r$-enhancement is dominated by the $s$-process contribution,
and the maximum [hs/ls] value is only marginally affected.
No particular distinction between CEMP-$s/r$ and CEMP-$s$ appears 
for [Pb/hs], which lies within the theoretical predictions 
for a range of the $^{13}$C-pockets between ST/2 and ST/45.
Lead is largely produced at low metallicity (as $^{208}$Pb) by the
$s$-process, with a low $r$-process contribution ($\sim$ 15\% 
to solar Pb). Then, [Pb/Fe]$_{\rm s+r}$ is not affected by high initial 
$r$-enrichments and possible variations in [Pb/hs] are the consequence of
the [hs/Fe]$_{\rm s+r}$.


\section{Method adopted to interpret the spectroscopic data} 
\label{method}

In this Section we illustrate the method adopted
to analyse the spectroscopic abundances of elements from C to Bi
and to provide theoretical interpretations with AGB models.
For this accurate investigation, we normalise both theoretical 
and spectroscopic abundances to the solar photospheric values
by \citet{lodders09}.
Differences higher than 0.05 dex between meteoritic solar abundances 
by \citet{AG89} and photospheric solar abundances by \citet{lodders09}
are listed in the online material, Table~A2 of Appendix~A.
We discuss here three stars with different characteristics taken 
as example: the giant CEMP-$s$I HD 196944, a second giant 
CEMP-$s$I without lead detection, HE 1135+0139 (Section~\ref{how1}),
and the main-sequence CEMP-$s$II/$r$ HE 0338--3945 (Section~\ref{how2}). 
In Paper III, similar analyses will be provided for the stars 
listed in Tables~\ref{table5_sindicator} and~\ref{tablestellemancanti}.

\subsection{How to fit CEMP-s stars} \label{how1}

\vspace{2mm}
\textbf{HD 196944} (Fig.~\ref{HD196944_es2},~\ref{HD196944_es3})

\begin{figure}
\includegraphics[angle=-90,width=9cm]{fig13.ps}
\caption{Spectroscopic [El/Fe] abundances of the CEMP-$s$I giant HD 196944 
compared with AGB models of initial mass $M$ = 1.5 $M_{\odot}$, three 
$^{13}$C-pockets (ST/3, ST/5, ST/9), and $dil$ $\sim$ 2.0 dex.
Observations are from \citet{aoki02c,aoki02d,aoki07}, 
\citet{vaneck03} and \citet{masseron10}.
The lower panel displays the differences between observations and 
theoretical predictions (case ST/5), [El/Fe]$_{\rm obs-th}$. The range 
between the two lines corresponds to an uncertainty of 0.2 dex.
Here and in the following Figures [C/Fe] and [N/Fe] are not represented
in bottom panel because they may be affected by extra-mixing
 (see Section~\ref{carbon}).} 
\label{HD196944_es2}
\end{figure}

\begin{figure}
\includegraphics[angle=-90,width=9cm]{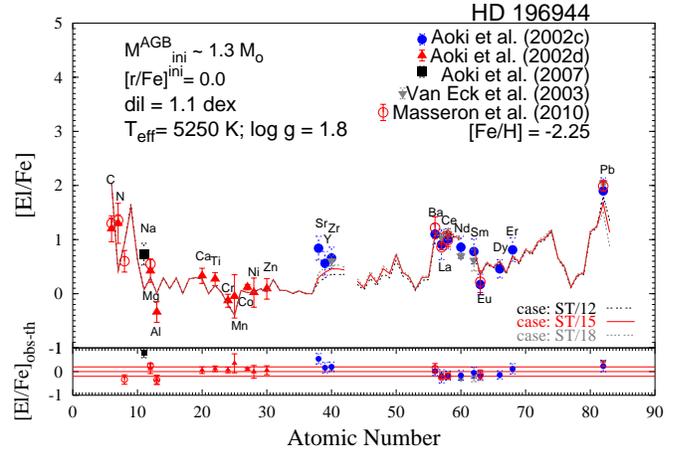}
\includegraphics[angle=-90,width=9cm]{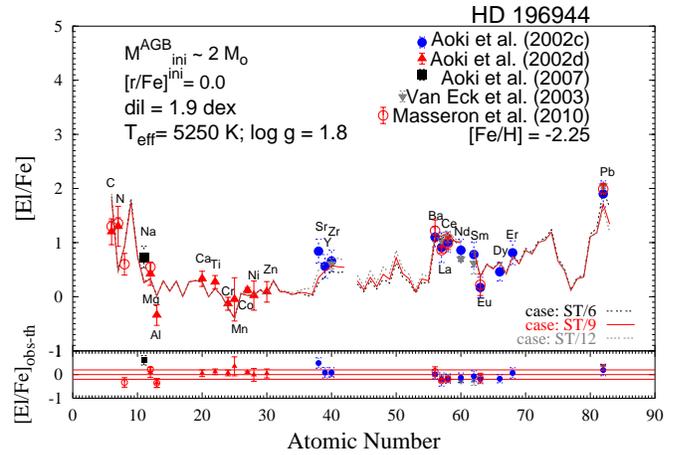}
\caption{The same as Fig.~\ref{HD196944_es2}, but for 
AGB models of initial mass $M$ = 1.3  $M_{\odot}$, cases ST/12, ST/15, ST/18, 
$dil$ = 1.1 dex (upper panel), and $M$ = 2  $M_{\odot}$, cases ST/6, ST/9, ST/12, 
$dil$ = 1.9 dex (bottom panel).
The observed [Na/Fe] is $\sim$ 0.7 and 0.5 dex higher than the theoretical
predictions, respectively.
Differences between observations and theoretical predictions, [El/Fe]$_{\rm obs-th}$,
are shown for cases ST/15 and ST/9 (upper and bottom panels, respectively).}
\label{HD196944_es3}
\end{figure}

\input{table8.tex}

HD 196944 is a CEMP-$s$I giant 
(\citealt{aoki02c,aoki02d,aoki07}, \citealt{vaneck03}, and \citealt{masseron10}),
with observed [hs/ls] = 0.3 and [Pb/hs] = 1.0. 
At [Fe/H] = $-$2.25, these values correspond to a limited range 
of $^{13}$C-pockets, depending on the initial AGB mass adopted
(see e.g., Fig.~\ref{mnras1_AAbab9ltHT_m1p5_WobsCEMP_n20_EunoEu_Giants}).
Once the $^{13}$C-pocket efficiency that reproduces the $s$-process 
indicators [hs/ls] and [Pb/hs] is established, a proper dilution factor
should be applied in order to fit the spectroscopic data.
As described in Section~\ref{intro}, the dilution factor $dil$ 
simulates the mixing between the $s$-rich material transferred from
the AGB and the convective envelope of the observed star.
HD 196944 is a giant having already suffered the FDU episode, 
requiring a dilution of the order of 1 dex or more.
The difference between our predicted [La/Fe] and the one observed in
HD 196944 gives the first assessment of the dilution, afterward 
optimized with a more careful analysis of the uncertainties of the 
single species. 
An example is given in Fig.~\ref{HD196944_es2} for $M^{\rm AGB}_{\rm ini}$ 
= 1.5 $M_{\odot}$, and cases from ST/3 down to ST/6. 
Here and in the following Figs.~\ref{HD196944_es2} 
to~\ref{HE0338-3945Jonsell06_bab10d6d7d10m1p5z1m4rp2alf0p5_n5},
the name of the star, its metallicity, the literature 
of the spectroscopic data, and the parameters of the AGB models 
adopted (i.e. initial mass, $^{13}$C-pocket, dilution factor and 
initial $r$-process enhancement) are given in the inset.
A solution is found for the case ST/5 (full line) with a dilution factor 
dil = 2.0 dex. This means that the material transferred from the AGB
is 1\%
of the convective envelope of the observed star.
The choice of the best fit is weighted on all the observed 
elements from carbon to lead, with particular attention to the
three $s$-process peaks. 
AGB models of different initial masses, with proper choice of the dilution 
factor, may provide plausible solutions for the $s$-process elements:
ST/15 and dil = 1.1 dex for $M^{\rm AGB}_{\rm ini}$ = 1.3 
$M_{\odot}$, ST/6 and dil = 1.6 dex for 
$M^{\rm AGB}_{\rm ini}$ = 1.4 $M_{\odot}$ and 
ST/9 and dil = 2.0 dex for $M^{\rm AGB}_{\rm ini}$ = 2 $M_{\odot}$ 
(Fig.~\ref{HD196944_es3}). 
Significant differences between observed and predicted Na are found
for $M^{\rm AGB}_{\rm ini}$ = 1.3 and 2 $M_{\odot}$ ($\sim$ 0.7 and 0.5 dex,
respectively).
The goodness of the fits is tested with a $\chi$$^2$ distribution,
by defining 
\begin{equation}
\chi^2_{\small N} = \sum^{N}_{i=1} \frac{(O_i - P_i )^2}{(\sigma_i)^2}, 
\end{equation}
where $N$ is the number of elements considered, 
$O_i$ is the observed abundance of the element $i$,
$P_i$ is the abundance predicted by the AGB model for the element $i$,
$\sigma_i$ is the uncertainty of the observed element $i$.
Taking the number of elements $N$ as degrees of freedom,
we calculate the confidence level of the $\chi^2_{\small N}$ test.
The distribution function of $\chi^2_{\small N}$ is $P$($\chi^2_{\small N}$),
obtained from the related tables. 
The results for models presented in Figs.~\ref{HD196944_es2} and~\ref{HD196944_es3}
are listed in Table~\ref{chi}. We considered different cases by changing the
number of elements involved. In general, the theoretical interpretations
are tested with the six $s$-elements Y, Zr, La, Nd, Sm and Pb (case A, columns~3
and~4).
Note that the observed Na and Mg are important initial mass
discriminators (case B, columns~5 and~6): levels of confidence higher than 
95\% are obtained only with AGB models of initial mass $M$ = 1.5 $M_\odot$ 
and $^{13}$C-pockets ST/5 - ST/6, even including other heavy elements as Ba, 
Ce, Eu, Dy, Er (cases C and D, last four columns).
\\
Disagreements between predictions and observations remain for
C, N, and Sr. A possibility to improve the C and N predictions is to
hypothesise the occurrence of the Cool Bottom Processing (\textbf{CBP}, see Section~\ref{carbon}),
\citep{nollett03,busso10}. 
\\
Concerning Sr, all AGB models underestimate the observed [Sr/Fe] ratio
by about 0.4 dex.
Note that the Sr abundance has been determined by using
three lines, while Y and Zr seems to be more reliable with 7 and 6 lines
detected, respectively. 
\\
For this star, the observed [La/Eu] ratio is in agreement with a pure 
$s$-process contribution ([La/Eu]$_{\rm th}$ = 0.74, obtained with
[r/Fe]$^{\rm ini}$ = 0).
\\
AGB model of initial mass $M$ = 1.5 $M_{\odot}$, case ST/5, $dil$ $\sim$ 2.0
dex and no initial $r$-process enhancement provide a theoretical
interpretation for this giant.

\begin{figure}
\includegraphics[angle=-90,width=9cm]{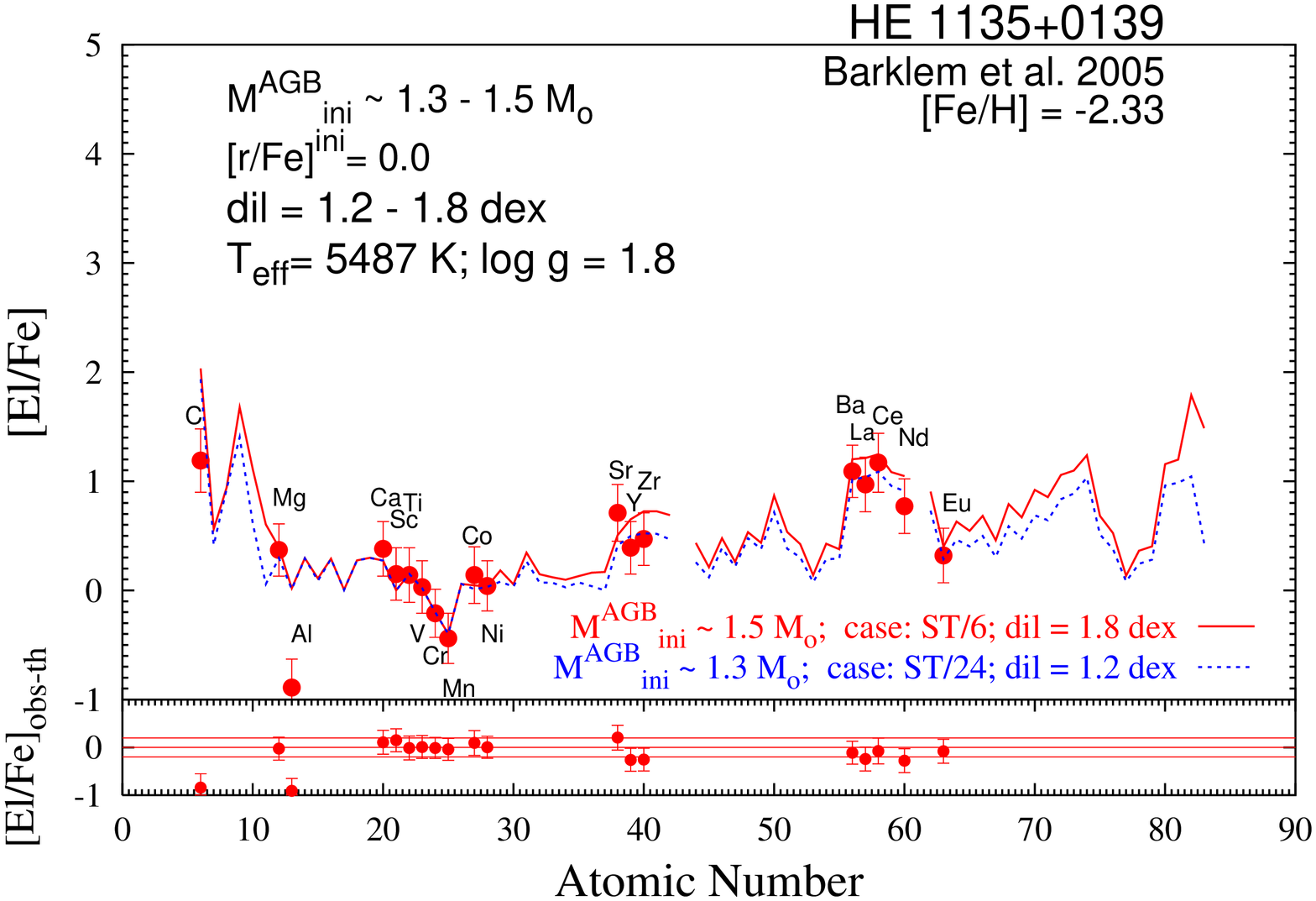}
\caption{Spectroscopic [El/Fe] abundances of the CEMP-$s$I giant HE 1135+0139
by \citet{barklem05}, compared with two AGB stellar
models of initial masses: $M^{\rm AGB}_{\rm ini}$ = 1.3 $M_{\odot}$,
case ST/24, $dil$ = 1.2 dex (dashed line) and $M^{\rm AGB}_{\rm ini}$ = 1.5 $M_{\odot}$,
case ST/6, $dil$ = 1.8 dex (solid line).
We predict a [Pb/Fe]$_{\rm th}$ $\sim$ 1.0 -- 1.8.
The observed [Al/Fe] is $\sim$ 1 dex lower than the solar value.
The lower panel displays the differences between observations and 
theoretical predictions [El/Fe]$_{\rm obs-th}$ for $M^{\rm AGB}_{\rm ini}$ 
= 1.5 $M_{\odot}$ model. The range between the two lines corresponds to an 
uncertainty of 0.2 dex.}
\label{HE1135+0139_es}
\end{figure}

\begin{figure}
\includegraphics[angle=-90,width=9cm]{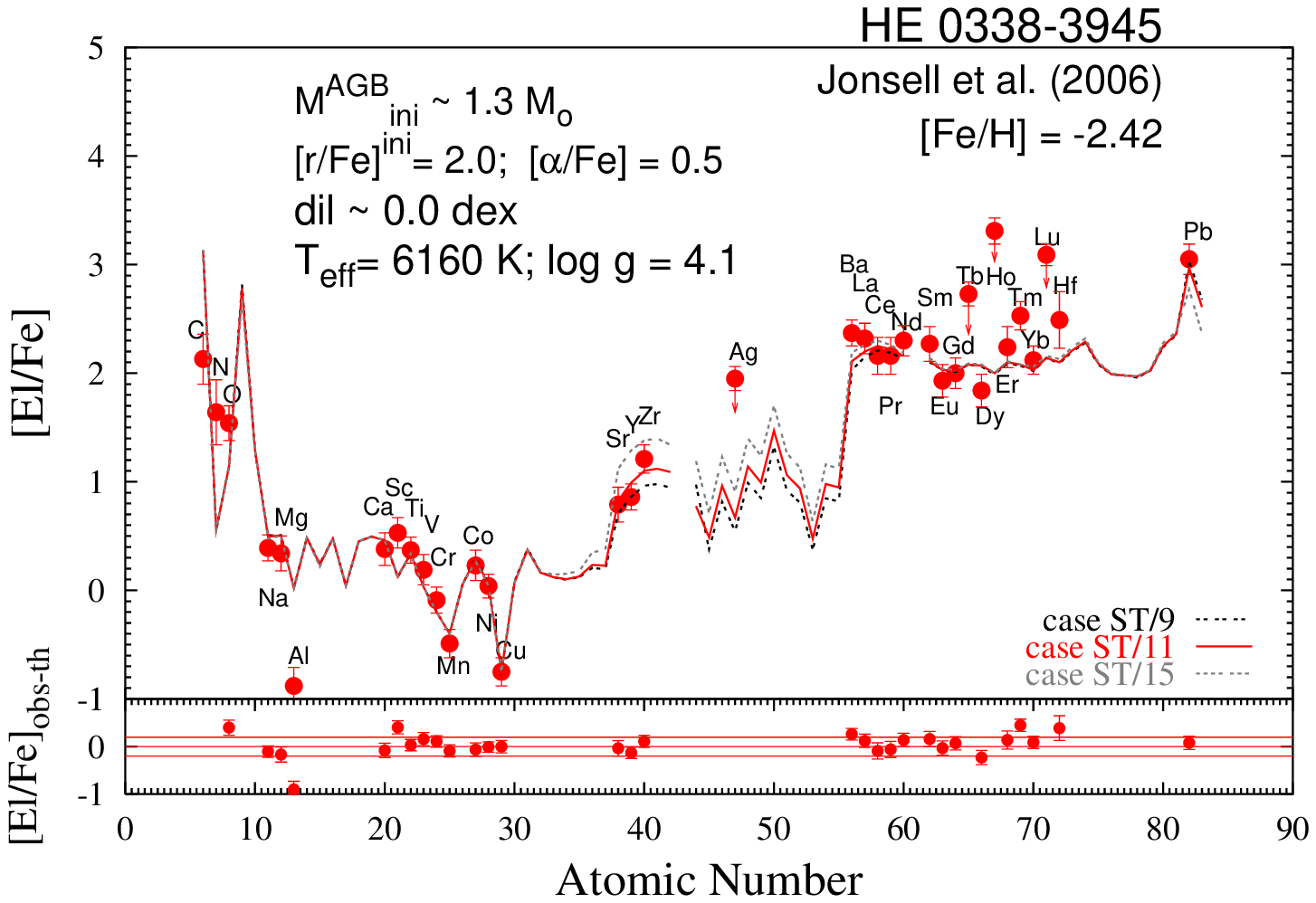}
\caption{Spectroscopic [El/Fe] abundances of the CEMP-$s$II/$r$ 
main-sequence HE 0338--3945 by \citet{jonsell06}, with AGB stellar models
of 1.3 $M_{\odot}$, cases ST/9, ST/11 and ST/15, and no dilution factor.
Pure $s$-process AGB models predict [Eu/Fe]$_{\rm s}$ = 1.2, about
0.8 dex lower than that observed: an initial $r$-process 
enrichment of [r/Fe]$^{\rm ini}$ = 2.0 is needed.
The lower panel displays the differences between observations and 
theoretical predictions (case ST/11), [El/Fe]$_{\rm obs-th}$. The range 
between the two lines corresponds to an uncertainty of 0.2 dex.
Four elements are not in agreement with the theoretical predictions (O, Al, Sc, Tm),
while Ba, Dy, Hf lie within the uncertainties of the AGB models (see text).}
\label{HE0338-3945Jonsell06_bab10d6d7d10m1p5z1m4rp2alf0p5_n5}
\end{figure}

\vspace{2mm}
\textbf{HE 1135+0139} (Fig.~\ref{HE1135+0139_es})

\citet{barklem05} analysed a sample of 253 metal-poor halo stars
during the HERES Survey (VLT/UVES), with a spectral resolution of
$R$ $\sim$ 20\,000 ($S/N$ $\sim$ 50). One of them is
the giant HE 1135+0139.
No Pb has been detected in this star; however, many $s$-process elements 
are available: Sr, Y, Zr, Ba, La, Ce and Nd. 
Therefore, we can give a robust Pb prediction, within the
range of the error bars. 
We evaluate the $^{13}$C-pocket starting with the observed [hs/ls] 
= 0.49.   
Similarly to HD 196944, this CEMP-$s$I star has a low hs peak 
([hs/Fe] = 0.92), requiring a high dilution factor depending on the
 AGB initial mass.
In Fig.~\ref{HE1135+0139_es}, we show two possible theoretical 
interpretations with $M^{\rm AGB}_{\rm ini}$ = 1.5 $M_{\odot}$, 
dil = 1.8 dex, case ST/5 (full line) and $M^{\rm AGB}_{\rm ini}$ = 1.3
$M_{\odot}$, dil = 1.2 dex, case ST/24 (dashed line). 
Solutions with a negligible dilution are discarded, 
in agreement with a giant after the FDU.
We predict [Pb/Fe]$_{\rm th}$ $\sim$ 1.0 -- 1.8.
The observed [Al/Fe] is $\sim$ 1 dex lower than the solar value,
as observed in field metal-poor stars \citep{barklem05}.

\subsection{How to fit CEMP-$s/r$ stars} \label{how2}

\vspace{2mm}
\textbf{HE 0338--3945} \label{HE0338} 
(Fig.~\ref{HE0338-3945Jonsell06_bab10d6d7d10m1p5z1m4rp2alf0p5_n5})

This CEMP-$s$II$/r$ star was first analysed by \citet{barklem05}.
Subsequently, \citet{jonsell06} detected 33 species and provided upper 
limits for 6 elements, using a high-quality VLT-UVES spectra ($R$ = 30\,000 
-- 40\,000; $S/N$ $\sim$ 74). This is a main-sequence star, with 
{\textit T}$_{\rm eff}$ = 6160 $\pm$ 100 K, log $g$ = 4.13 $\pm$ 0.33 dex.
\\
The [hs/ls] and [Pb/hs] ratios determine the $s$-process distribution
efficiency as described in Section~\ref{how1} for the giant HD 196944. 
In this case, no dilution is applied for low AGB initial masses,
because of the high $s$-process enhancement observed in HE 0338--3945 
([hs/Fe] $\sim$ 2.3).
An interpretation of the observed abundances is obtained with AGB
models of initial mass 1.3 $M_{\odot}$, cases ST/9, ST/11, ST/15 
and dil = 0.0 dex (see Fig.~\ref{HE0338-3945Jonsell06_bab10d6d7d10m1p5z1m4rp2alf0p5_n5}). 
Main-sequence/turnoff/subgiant stars have not suffered the
FDU. 
Moreover, old main sequence stars of low mass ($M$ $\sim$ 0.8 $M_\odot$)
have a very thin convective envelope and no large mixing
between the $s$-rich material transferred from the AGB and the envelope
of the observed star are expected contrary to the case of giants.
As discussed in Section~\ref{observations}, the efficiency of mixing processes 
as thermohaline, gravitational settling and radiative levitation
acting during the main-sequence phase is difficult to estimate.
Then, for stars having not suffered the FDU we do not assume initial
constraints on the dilution.
If present, information about possible mixing provided by the dilution factor
in main-sequence stars will be discussed in Section~\ref{summary} and Paper III.
\\
For a pure $s$-process prediction ([r/Fe]$^{\rm ini}$ = 0), we found 
a difference of about 0.7 dex between observed and predicted [Eu/Fe]. 
This is solved by adopting an initial $r$-enrichment of the interstellar 
cloud of [r/Fe]$^{\rm ini}$ = 2.0, which explains the observed  
[La/Eu]. The upper limit of Ag does not provide constraints for a possible
initial $r$-enhancement of the elements lighter than Ba (see Section~\ref{r}).
\\
As discussed by \citet{jonsell06}, C is uncertain due to the 
strong temperature sensitivity of CH molecule formation and 3D model 
calculations may add further uncertainties of about $-$0.5 to $-$0.3 dex 
\citep{asplund05}. For N, a reasonable estimate of the NLTE and 3D effects may 
amount to $-$0.5 dex. 
The observed [O/Fe] is slightly underestimated by AGB models. 
[Al/Fe] is strongly subsolar, but NLTE effects may be important for Al 
\citep{asplund05,baumuller97,andr08}.
[Sc/Fe] may be overestimated \citep{jonsell06}, indeed it appears 
enhanced compared to field halo stars.
Note that AGBs do not synthesise copper: indeed
[Cu/Fe] is negative, in accord with the observations of unevolved
stars in the same range of metallicities \citep{bisterzo04,romano07}.
[Ba/Fe] is slightly higher than the AGB prediction; however, 
NLTE effects together with 3D models can decrease the abundances 
by about 0.3 dex \citet{asplund04,asplund05,andr09}.
La could be underestimated, due to the incomplete hyperfine 
splitting (hfs) data.
Dy and Tm are lower and higher by about 0.2 and
0.4 dex with respect to the case ST/11, respectively.
For Pb an overionisation is possible, which would lead to 
a decrease of the observed abundance.
\\
By increasing the number of TDUs (e.g, for AGB models of initial mass 1.5
$M_{\odot}$; dil $\sim$ 0.8 dex), the observed [Na/Fe]\footnote{Note that 
NLTE effects further decrease the observed [Na/Fe] 
\citep{baumuller98,andr07}.} and [ls/Fe]
are overestimated.
Then, solution with AGB models of initial mass higher than 1.3 $M_{\odot}$ 
are excluded for this star.
\\
AGB model of initial mass $M$ = 1.3 $M_{\odot}$, case ST/11, $dil$ = 0.0
dex and [r/Fe]$^{\rm ini}$ = 2.0 provide a theoretical
interpretation for this star.


\subsection{Carbon and Nitrogen}\label{carbon}

\input{table9.tex}

For most CEMP-$s$ and CEMP-$s/r$ stars, the observed
[C/Fe] and [N/Fe], as well as the carbon isotopic ratio 
$^{12}$C/$^{13}$C, are reproduced by AGB models.
Spectroscopic observations of C and N in CEMP stars 
are affected by high uncertainties due to strong 
molecular bands, NLTE corrections or 3D hydrodynamical model atmospheres
(\citealt{asplund05}, \citealt{collet07}; 
\citealt{GAS07}; \citealt{caffau09}).
However, measurements in pre-solar grains of AGB origin \citep{zinner06}, 
which are more reliable than observations in CEMP-$s$ stars, also show 
a disagreement from AGB predictions.
\\
A large mixing, affecting C and N and decreasing the $^{12}$C/$^{13}$C 
ratio, occurs during the red giant branch by the FDU\footnote{As anticipated 
in Section~\ref{observations}, during the FDU, the material processed 
in the hydrogen shell is brought to the surface, changing the CNO isotopic 
composition of the envelope. 
The main isotopes involved are $^{13}$C, $^{14}$N and $^{17}$O which 
increase, while $^{12}$C, $^{15}$N and $^{18}$O decrease (the observed 
isotopic ratio $^{14}$N/$^{15}$N increases by a factor of 6).
From the point of view of the observations, since $^{12}$C decreases of 
30\% and $^{13}$C increases of a factor of 2 $\sim$ 3, the evidence of the FDU 
is confirmed by the reduction of the $^{12}$C/$^{13}$C ratio 
to about 20 \citep{boot88,busso99}.}.
However, it is not sufficient to explain the observed $^{12}$C/$^{13}$C 
(e.g, see \citealt{abia01} for disc C stars).
Further extra-mixing has to be hypothesised during the following
AGB phase to interpret the observations.
To solve this problem, it was proposed that some material could be 
transported from the fully convective envelope 
into the underlying radiative region, down to the outer zone of the
hydrogen shell, where partial H burning occurs.
CBP is predicted to occur in low-mass AGB stars (see \citealt{nollett03}; 
\citealt{dominguez04}; \citealt{wasserburg95,wasserburg06}; \citealt{busso10}),
and may decrease $^{12}$C in the envelope while $^{14}$N increases 
\citep{wasserburg95}; consequently, [C/Fe] 
and $^{12}$C/$^{13}$C also decrease, together with an increase of [N/Fe]. 
\\
CBP is the most plausible hypothesis to reproduce the observations, 
although its efficiency is difficult to estimate because it may be affected
by many physical processes (e.g., rotation, magnetic fields, thermohaline
mixing).

\subsubsection{$^{12}$C/$^{13}$C}

In Table~\ref{c12c13}, we collected the isotopic ratio $^{12}$C/$^{13}$C 
as well as [C/Fe], [N/Fe] and [O/Fe]
for the CEMP-$s$ and CEMP-$s/r$ stars listed in Tables~\ref{table5_sindicator}
(first group) and~\ref{tablestellemancanti} (second group).
The same references and labels are adopted here in columns~2,~5 and~11.
When AGB models predict higher [C/Fe] ratio than the   
observed value, we simulate the occurrence of CBP 
by applying the method used for HD 196944 and HE 0338--3945.
In column~12, comments about the hypothesis of extra-mixing are listed.
The label `CBP' means that CBP may explain the observed 
[C/Fe] and [N/Fe]. Most stars need the occurrence of a `CBP'.
Only in four stars the observed [C/Fe] ratios agree with AGB predictions 
without involving extra-mixing: CS 22880--074, CS 30315--91,
HE 0507--1653 and HE 1429--0551. The first two show also relatively high 
lower limits for $^{12}$C/$^{13}$C ($>$ 40 -- 60).
If a CBP overestimates the observed [N/Fe], we adopt the
label ``low [N/Fe]$_{\rm obs}$''.
Five stars have [N/Fe] higher than [C/Fe], but only two of them are incompatible 
with the hypothesis of a CBP (label ``high [N/Fe]$_{\rm obs}$''; 
CS 30322--023, HE 1031--0020).
\\
Several stars show $^{12}$C/$^{13}$C $\sim$ 10 in agreement with
Early-Type CH stars (see e.g., \citealt{vaneck03}). 
An exception is V Ari, a giant for which \citet{beers07} 
detected $^{12}$C/$^{13}$C = 90 $\pm$ 10.
$^{12}$C/$^{13}$C lower limits are measured for CS 29497--030 ($>$ 10), 
HE 0143--0441, HE 0012--1441 and HE 1410--0004 ($\ga$ 3), 
HE 2232--0603 ($>$ 6), CS 22967--07 ($>$ 60), as well as 
 CS 22880--074 and CS 30315--91. 
Theoretical AGB predictions are strongly enhanced: 
($^{12}$C/$^{13}$C)$_{\rm th}$ $\sim$ 2 -- 3 $\times$ 10$^{4}$ for 
AGB models of initial mass $M$ = 1.3 $M_\odot$ and 
2 $\times$ 10$^{5}$ for $M$ = 1.5 $M_\odot$.
This confirms that an extra-mixing must be assumed to reduce the 
$^{12}$C observed in the envelope of CEMP-$s$ stars.
\\
Another mechanism may affect the C and N ratios during the AGB phase.
Low mass stars with [Fe/H] $<$ $-$2.5
may experience a huge thermal pulse 
\citep{hollowell90,iwamoto04,cristallo09pasa,campbell08,lau09,suda10,straniero10}. 
When the CNO abundances in the envelope
are low, the entropy barrier provided by the H-burning shell may vanish and 
protons are ingested from the envelope down to the He-intershell during 
the first fully developed thermal pulse resulting in a violent H-burning. 
During this episode large amount of $^{13}$C and $^{14}$N are produced, 
which are mixed to the surface by an extremely deep TDU, thus decreasing the 
$^{12}$C/$^{13}$C ratio sensibly, while [N/Fe] increases.
Note that the mass and metallicity limits for the occurrence of this deep TP 
would be decreased if an initial O enhancement is adopted.
However, different models may provide contrasting results because different 
prescriptions are adopted. For instance, \citet{lau09} find an enhancement in C,
 but only a marginal enrichment in N. 
Similar behaviours are obtained for low mass models by \citet{campbell08} and
\citet{suda10}.

In intermediate mass stars (3 $<$ $M/M_\odot$ $\la$ 8), the Hot Bottom Burning 
(HBB) also modifies the final C and N abundances 
(\citealt{sugimoto71}, \citealt{iben73},\citealt{karakas03},\citealt{ventura05}), destroying
C and producing a large amount of N (NEMPs, nitrogen-enhanced metal-poor
stars \citealt{johnson07}).
Moreover, AGB models of intermediate mass and halo metallicity seem to be affected
by extreme mixing, as hot TDU (when the envelope penetrates
burning protons at its base), which may modify the stellar structure
of the star and its evolution (\citealt{herwig04,goriely04,campbell08,lau09}). 
However, the mass of the He-intershell is smaller than low mass AGBs
and also the $^{13}$C-pocket and the TDU efficiency are reduced.
Moreover, the maximum temperature reached during the TP is high enough 
to efficiently activate the $^{22}$Ne($\alpha$, n)$^{25}$Mg reaction. Consequently,
the production of the first $s$-peak should be favoured, while the contribution 
of the $^{13}$C-pocket to the $s$-process elements is marginal in these stars.
Because there are no information about the consequences of hot TDU on the 
$s$-process distribution so far, these models are not considered in the present
discussion.

In this paper, AGB models with specific range of initial masses ($M$ $\sim$
1.3 $\div$ 2 $M_\odot$) were considered. 
 Further studies would be necessary to establish the contribution
by AGB stars with initial mass out of this range.
Low metallicity AGB models with initial mass below the adopted range would help
to understand the effects of the proton ingestion episode on C and N, as well
as on the $s$-process elements distribution. 
The need of stars with low initial mass has also been invoked by \citet{izzard09} 
to explain the high fraction of carbon enhanced stars among very metal-poor stars.
However, additional sources of carbon, besides the nucleosynthesis of AGB stars in
 binary systems, may be hypothesised (see e.g., \citealt{carollo11astroph}).

\input{table10.tex}

\input{table11.tex}


\section{Summary of the results}\label{summary}

\begin{figure*}
\includegraphics[angle=-90,width=7.5cm]{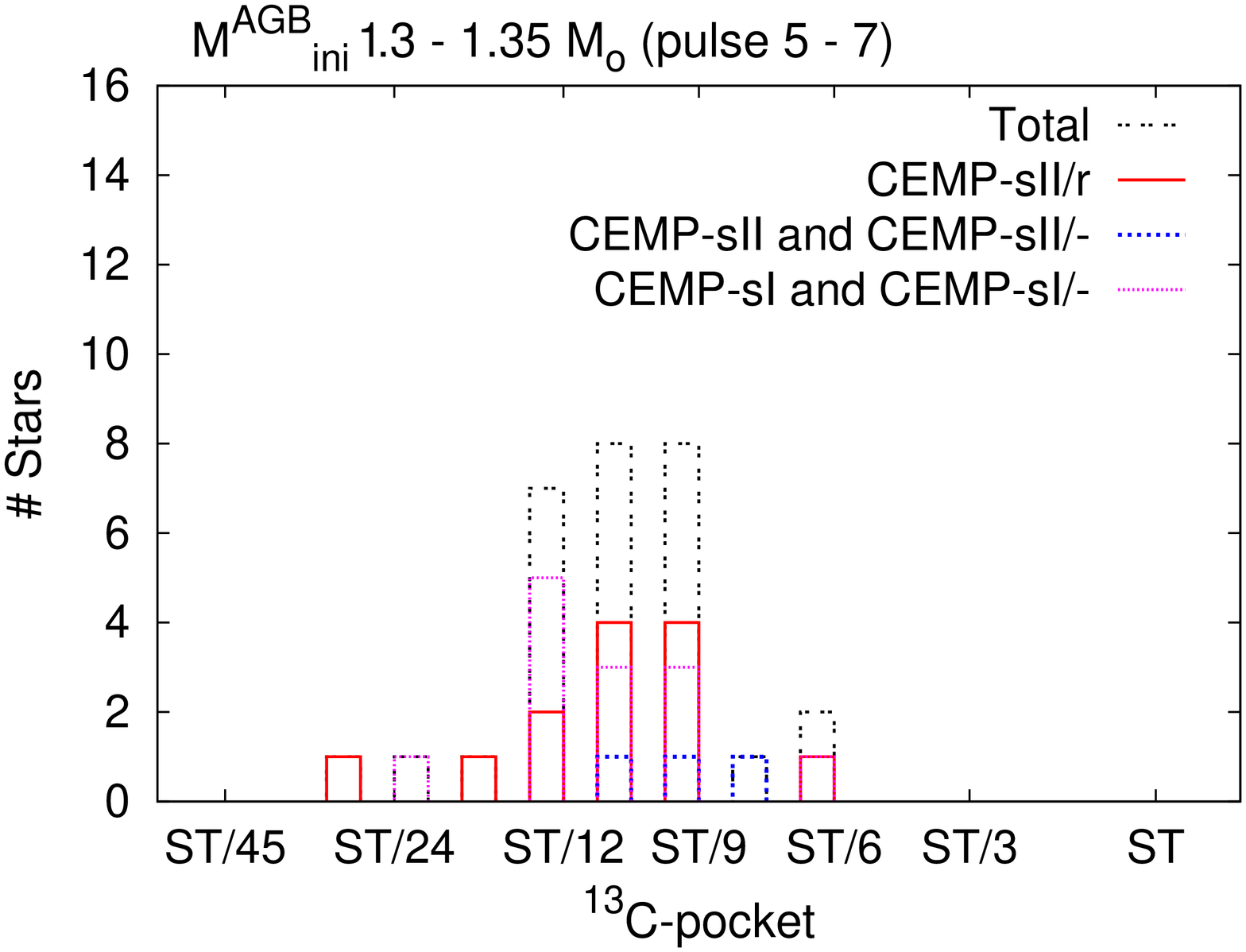}
\includegraphics[angle=-90,width=7.5cm]{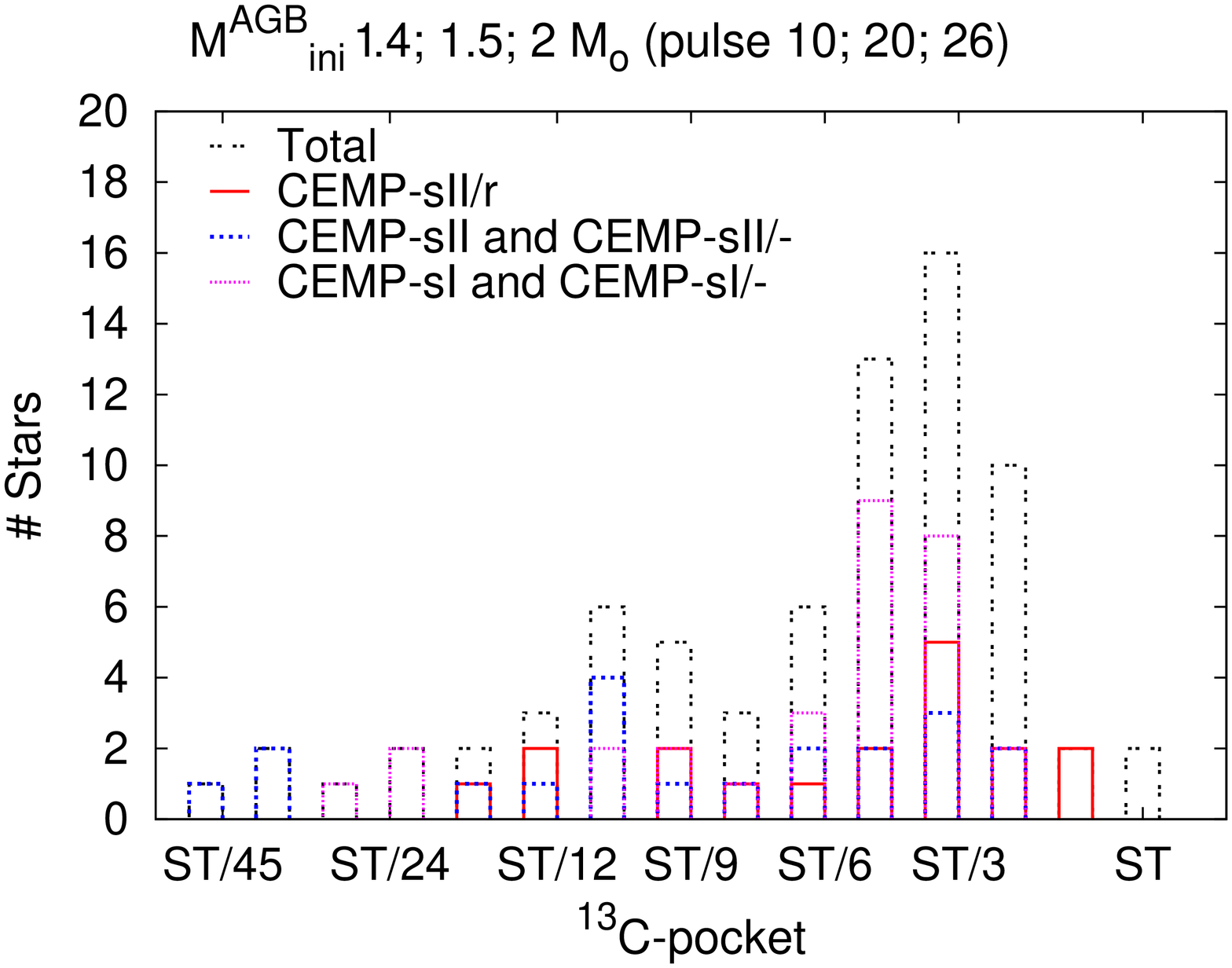}
\vspace{10mm}
\caption{Number of stars versus the size of the $^{13}$C-pocket,
for AGB models with initial masses $M$ = 1.3 and 1.35 $M_\odot$ (left
panel), $M$ = 1.4, 1.5 and 2 $M_\odot$ (right panel).
We only consider stars from Table~\ref{summary1}, following  
the classification indicated in column~5.
All solutions listed in Table~\ref{summary1} are shown.}
\label{istotasche}
\end{figure*}

The results are summarised in Tables~\ref{summary1} and~\ref{summary2}.
In Table~\ref{summary1}, 55 CEMP-$s$ and CEMP-$s/r$ stars with a great number of
observations among the $s$-elements are listed, and 
in Table~\ref{summary2} we report 36 CEMP-$s$ and CEMP-$s/r$ stars
for which a limited number of ls and hs elements is detected (in several
cases only Sr and Ba).
All the stars lie in a metallicity range $-$3 $\la$ [Fe/H] $\la$ $-$1.7.
The only exceptions are five CH stars with [Fe/H] $\sim$ $-$1.2 (CS 29503--010, 
HD 26, HD 206983, HE 0507--1653 and HE 1152--0355), which may be consided as a 
link between CEMP-$s$ and Ba stars of nearly solar metallicity, because their C 
and $s$-enrichment are due to the same physical reasons (Paper III).
In columns~1 to~3 in Tables~\ref{summary1} and~\ref{summary2}, the name of the 
stars, their references and metallicities are listed.
In column~4 we distinguish between main-sequence/turnoff or subgiants
before and giants after the FDU (labeled `no' and `yes', respectively).
The class of each star is specified in column~5.
In columns~6 to~9, AGB models that best reproduce the spectroscopic
observations are listed: the AGB initial mass (in solar masses), 
the $^{13}$C-pocket, the dilution factor, and the initial $r$-process 
enhancement. 
\\
Robust predictions for $s$-process elements are provided
for stars in Table~\ref{summary1}, while a degeneracy of solutions 
may equally interpret the limited number of observations for stars in 
Table~\ref{summary2}. 
In column~10 of Table~\ref{summary1}, [Pb/Fe]$_{\rm th}$ predictions
are given within $\pm$ 0.3 dex of uncertainties.
In general, reasonable solutions are obtained by using AGB models with 
initial masses in the range 1.3 $\leq$ $M/M_{\odot}$ $\leq$ 2. 
Otherwise, in column~11 of Table~\ref{summary1} and in column~10 of 
Table~\ref{summary2}, we report the elements that provide a
constraint on AGB models. The label `FDU' means that the only constraint
is given by the occurrence of the FDU, in agreement with a high dilution.
For instance, CEMP-$s$II giants need high AGB initial mass, because
high $s$-process enhancements ([hs/Fe] = 2) are obtained together with
a large dilution factor (dil $\sim$ 1 dex) only if $M^{\rm AGB}_{\rm ini}$ 
= 1.5 -- 2 $M_{\odot}$ are adopted. Indeed, AGB models with 
$M^{\rm AGB}_{\rm ini}$ = 1.3 $M_{\odot}$ underwent 5 TDUs, with a maximum 
[hs/Fe] $\sim$ 2.
\\
A range of $s$-process efficiencies was adopted by \citet{busso01} 
in order to interpret the observed [hs/ls] and [Pb/hs] ratios of disc stars;
\citet{kaeppeler10rmp} included in their analysis recent spectroscopic
observations of different stellar populations (Ba stars, CH stars, Post-AGB, 
CEMP-$s$ and CEMP-$s/r$ stars) sustaining this hypothesis. 
We confirm the need of a spread of $^{13}$C-pocket efficiencies. 
In Fig.~\ref{istotasche}, we display the number of stars (taken from 
Table~\ref{summary1}) as a function of the strength of the $^{13}$C-pocket, for AGB models 
of initial mass $M$ = 1.3 and 1.35 $M_{\odot}$ (left panel) and $M$ = 1.4, 1.5 
and 2 $M_{\odot}$ (right panel). 
The most common $^{13}$C-pockets are close to ST/12 for $M^{\rm AGB}_{\rm ini}$ 
= 1.3 -- 1.35 $M_{\odot}$ and ST/4 for $M^{\rm AGB}_{\rm ini}$ = 1.4 -- 2 
$M_{\odot}$. 
One finds that $s$-process efficiencies below $\sim$ ST/24 are 
rare (CS 22942--019, CS 31062--012, HE 0336+0113, V Ari,
HE 1135+0139, HD 189711, as well as CS 22891--171, CS 22956--28, 
for which [Pb/hs] $\sim$ 0 are observed).
\\
In Table~\ref{summary1}, seventeen stars lie on the main-sequence,
and ten of them can be only interpreted with AGB models of low initial mass
and negligible dilutions. Note that, in several stars, the main constraint 
of the initial mass model is given by Na.
This seems to indicate that no efficient mixing takes place in main-sequence stars, 
in agreement with model calculations by \citet{thompson08}, (see also 
\citealt{vauclair04}; \citealt{richard02}), who showed that gravitational 
settling can confine the efficiency of thermohaline mixing in these low mass 
metal-poor stars. 
Two stars showing a low $s$-process enhancement, BS 16080--175 and 
CS 22964--161, may only be interpreted with AGB models with dil = 0.5 
-- 1.0 dex. Three other stars do not show any dilution constraints.


\section{Conclusions}\label{conclusions}

We have compared spectroscopic observations of 94 CEMP-$s$ and CEMP-$s/r$
stars collected from the literature with AGB models of different initial
masses and metallicities presented in Paper I.
\\
All CEMP-$s$ stars are old halo main-sequence/turnoff or giants of low initial 
mass ($M$ $<$ 0.9 $M_\odot$). 
The most plausible explanation of their $s$-enhancement is the pollution by stellar 
winds diffused by a more massive AGB companion during their thermally 
pulsing phase, which evolved to a white dwarf afterwards.
\\
Particularly debated is the interpretation of those CEMP-$s$ stars showing an enhancement
in $r$-processes elements, incompatible with a pure $s$-process nucleosynthesis: 
among 45 stars with measured Eu, 23 are CEMP-$s/r$.
Even if the astrophysical site of the $r$-process is not well known, it is 
associated to explosive conditions in massive stars, in environments separated 
by the $s$-process. 
Several hypotheses have been advanced to interpret the CEMP-$s/r$ stars
(\citealt{jonsell06} and references therein).
We suggest that a supernova 
exploded in the neighborhood of the molecular cloud from which 
the binary system was formed, polluting it with $r$-process elements. 
The more massive companion of the binary system evolved 
as AGB star, synthesising $s$-elements, which are detected in the 
 envelope of the observed star after the mass transfer. 
The initial $r$-enhancement [r/Fe]$^{\rm ini}$ is evaluated by using the 
residual method, $N_{\rm r}$ = $N_\odot$ - $N_{\rm s}$, where the $s$-process 
solar contributions are obtained as in \citet{arlandini99} (updated in 
Table~\ref{ns}).
This is an approximation, adopted because of the poor knowledge of 
the primary $r$-process nucleosyntheses. 
We apply an initial $r$-process distribution [r/Fe]$^{\rm ini}$ scaled to 
the Eu observed in the CEMP-$s/r$ stars, to the isotopes between Ba and Bi.
Indeed, observations of $r$-II stars sustain the hypothesis 
of the existence of multiple $r$-process components: a light-$r$ component 
for Z $<$ 56, a heavy-$r$ component for 56 $<$ Z $<$ 83 (see review by 
\citealt{SCG08}), as well as an additional third component invoked to 
explain a subsolar Th and U component \citep{roederer09PbTh}.
For neutron capture elements below Ba, we did not apply any initial
$r$-process enhancement. Spectroscopic observations in this region
are extremely limited: only one star has Pd detected, CS 31062--050
\citep{JB04}, and upper limit for Ag have been obtained for CS 29497--030
\citep{ivans05} and HE 0338--3945 \citep{jonsell06}.
\\
In general, the [La/Eu] ratio (where La and Eu are typical $s$ and $r$
elements) provides an important discriminator between 
CEMP-$s/r$ and CEMP-$s$ stars. 
For [Fe/H] $<$ $-$2, an observed [La/Eu] $\sim$ 0.0 -- 0.5 dex (together with
a high $s$-enhancement) can not be explained by a pure $s$-process contribution. 
The range of the initial $r$-enhancement adopted to interpret the
observed Eu is a consequence of the inhomogeneity of the Galactic interstellar 
medium, and it is independent of AGB models. 
Starting from the spread observed in the [Eu/Fe] ratios in field stars with [Fe/H] 
$\la$ $-$2, we hypothesise a similar range of initial 
$r$-process enrichments in order to interpret observations in CEMP-$s/r$ stars
([r/Fe]$^{\rm ini}$ up to 2.0).
We classify as CEMP-$s/r$ those stars that need an [r/Fe]$^{\rm ini}$ 
from 1.0 to 2.0. 
Five stars show an observed [La/Eu] $\sim$ 0, together with [La/Fe] $\sim$ 2:
CS 22898--027, CS 29497--030, HE 0338--3945, HE 1305+0007, HE 2148--1247.
Their interpretation requires the highest $r$-enhancement [r/Fe]$^{\rm ini}$ = 2.0.
Note that stars with [r/Fe]$^{\rm ini}$ = 1.0 lie at the limit between
normal CEMP-$s$ and CEMP-$s/r$ (CS 22887--048, HD 209621, HE 0143--0441,
SDSS J1349--0229). 
\\
The initial $r$-process enhancements do not affect in any way the
nucleosynthesis of the $s$-process.
However, we have to consider that also the isotopes mainly produced
by the $s$-process receive a partial $r$ contribution for 
the hs elements, e.g. $\sim$ 30\% of solar La, $\sim$ 40\% of solar Nd, 
and $\sim$ 70\% of solar Sm. In case of high initial $r$-enhancements
as [r/Fe]$^{\rm ini}$ = 2.0, this implies an increase of the maximum
[hs/ls] up to 0.3 dex. 
In this regard, we discuss the behaviour of the two $s$-process indicators, 
[hs/ls] and [Pb/hs] versus metallicity. 
On average, the [hs/ls] observed in CEMP-$s/r$ stars shows higher values
than that in CEMP-$s$.
From a general analysis, [hs/ls] in CEMP-$s/r$ stars seems to be better 
reproduced by models with very high initial $r$-enhancement.
However, a detailed analysis of individual stars is needed (Paper III).
No particular distinction between CEMP-$s/r$ and CEMP-$s$ 
appears for [Pb/hs].

We have focused our attention on three stars taken as example to 
explain the method adopted to interpret the observations.
All the elements are considered, with particular attention to C, 
N, Na, Mg, ls, hs, Pb, as well as Eu, which is fundamental to evaluate possible
$r$-process contributions.
The [hs/ls] and [Pb/hs] ratios provide informations about the $^{13}$C-pocket,
while the first assessment of the dilution factor is based on the observed [hs/Fe].
The occurrence of the FDU provides the first constraint on the AGB models: 
during the FDU subgiants or giants undergo a large mixing 
involving about 80\% of the mass of the star, which dilutes the C and $s$-rich 
material previously transferred from the AGB companion, implying 
solutions with $dil$ $\ga$ 1 dex.
The choice of the AGB models reproducing the observations are based on the 
analysis of single species including their uncertainties and the number of detected 
lines, in particular within the three $s$-peaks.

We have presented a general description of the sample, and the main results
obtained by this study.
However, discussions of individual stars are necessary to underline
possible discrepancies between AGB models and observations and to suggest
possible points of debate for unsolved problems. This will be the topic of
Paper III. 
\\
To simplify the analysis we divided the stars in different classes,
according to the observed abundance pattern of the $s$-elements and of Eu.
Stars with [hs/Fe] $\ga$ 1.5 are called CEMP-$s$II (or CEMP-$s$II$/r$
if they show also an $r$-enhancement), 
while stars with [hs/Fe] $<$ 1.5 are CEMP-$s$I (or CEMP-$s$I$/r$).
The level of $s$-enhancement, `II' or `I', depends on different factors:
the $s$-process nucleosynthesis of the primary AGB, the efficiency of
mass transfer by stellar winds, the distance between the two stars, 
and the mixing of the transferred material with the envelope of the observed star. 
\\
The range covered by the observed [hs/ls] and [Pb/hs] requires the assumption of
different $^{13}$C-pocket efficiencies. 
This may be related to the uncertainty affecting the formation of the $^{13}$C-pocket,
the hydrogen profile and, then, the amount of $^{13}$C and $^{14}$N in the pocket.
A clear answer to the properties of the mixing processes at radiative/convective 
interfaces during TDU episodes, which lead to the formation of the $^{13}$C-pocket, 
has not been reached yet.
Moreover, models including rotation, gravity waves or magnetic fields, 
may influence the formation of the $^{13}$C-pocket \citep{langer99,herwig03,siess04,denissenkov03}.
This translates into different $s$-process distributions.
Further investigations are desirable.
\\
AGB models with $M^{\rm AGB}_{\rm ini}$ = 1.4 -- 2 $M_{\odot}$ 
an asymptotic trend beyond the 10th TDU (Paper I). Consequently, negligible
differences are observed in the $s$-process distribution at the last TDUs
when the envelope is affected by efficient stellar winds. 
Instead, $M^{\rm AGB}_{\rm ini}$ = 1.3 -- 1.35 $M_{\odot}$ models undergo
5 to 7 thermal pulses with TDU, without reaching an asymptotic trend.
Deeper investigations are in project for $M^{\rm AGB}_{\rm ini}$ $\sim$ 1.3 $M_{\odot}$
models, which provide theoretical interpretations for several CEMP-$s$ and CEMP-$s/r$
stars.
\\
In general, AGB models with different initial masses in the range 1.3 $\leq$ 
$M/M_{\odot}$ $\leq$ 2 and a proper choice of $^{13}$C-pocket
 may equally interpret the $s$-process observations. 
On average, possible fits with $M^{\rm AGB}_{\rm ini}$ $\sim$ 1.3 --
1.35 $M_{\odot}$ models require $^{13}$C-pockets close to case ST/12,
while case ST/4 are achieved for $M^{\rm AGB}_{\rm ini}$ = 1.4 -- 2 $M_{\odot}$ 
models. A restricted number of stars need 
 $s$-process efficiencies below $\sim$ ST/24 
(CS 22942--019, CS 31062--012, HE 0336+0113, V Ari,
HE 1135+0139, HD 189711, as well as CS 22891--171, CS 22956--28).
The $^{13}$C-pocket spread observed seems larger than that
suggested by \citet{bonacic07} in their population synthesis study. 
A detailed investigation of individual stars will be presented in Paper III.
\\
{[Na/Fe]} may provide indications on the AGB initial mass.
In CEMP-$s$II or CEMP-$s$II$/r$ before the FDU, an observed ratio
[Na/Fe] $\leq$ 0.5 is only interpreted by $M^{\rm AGB}_{\rm ini}$ $\sim$ 1.3 
$M_{\odot}$ models, because a large $s$-enhancement together with
a low [Na/Fe] may be reached only after a limited number of thermal pulses.
Nine stars among the sample show a high Na abundances
([Na/Fe] $\geq$ 1): CS 22942--019 \citep{PS01}, CS 29497--34, 
CS 29528--028 and CS 30301--015 \citep{aoki07,aoki08}, CS 30322--023
\citep{masseron06,aoki07}, LP 625--44 \citep{aoki02a},
SDSS J1349--0229 \citep{behara10}, 
SDSS 0924+40, and SDSS 1707+58 \citep{aoki08}.
The maximum values observed are for CS 29528--028 with [Na/Fe] = 2.33
and SDSS 1707+58 with [Na/Fe] = 2.71.
For these stars, $M^{\rm AGB}_{\rm ini}$ $\sim$ 1.5 $M_{\odot}$ models
are adopted. 
\\ 
Apart from the information on the $s$-process efficiency, the ls peak
may also represent an indicator for the AGB initial mass. In 
particular, in CEMP-$s$II stars with low Sr-Y-Zr (with respect to [hs/Fe]), AGB 
models with low initial mass are adopted ($M^{\rm AGB}_{\rm ini}$ $\leq$ 
1.4 $M_{\odot}$, e.g., CS 22183--015, CS 22880--074, CS 22898--027, 
CS 31062--012, HE 0338--3945,
HE 2232--0603).

Despite the uncertainty of C and N in very metal-poor
stars (due to strong molecular bands, 3D model atmospheres and non-LTE
corrections), the observed [C/Fe] and [N/Fe] values and the low 
$^{12}$C/$^{13}$C ratio in most stars indicate that the CBP
is needed during the AGB phase. However, its efficiency 
is difficult to estimate because several physical processes
may concur in this mixing (e.g., magnetic fields, rotation, thermohaline).
We are planning a detailed study about the discrepancy
between observed and predicted carbon in a forthcoming paper.

Several mixing processes may occur in the envelope of the star during the
main-sequence phase (e.g., thermohaline, gravitational settlings, radiative
levitation).
Their efficiency is different from star to star, depending on its age,  
initial metallicity, initial mass, on the amount of 
material accreted from the AGB companion, and the time at which
this material was accreted. 
These additional effects have not been included in the present study, but the 
comparison between models and observations may provide important clues for
these aspects.

\section*{Acknowledgments}

We are deeply grateful to W. Aoki, T. C. Beers, J. J. Cowan, I. I. Ivans, 
C. Pereira, G. W. Preston, I. U. Roederer, C. Sneden, I. B. Thompson, S. 
Van Eck, for enlightening discussions about 
CEMP-$s$ and CEMP-$s/r$ stars. 
This work has been supported by MIUR
and by KIT (Karlsruhe Institute of Technology, Karlsruhe).


 



\appendix





\section*{ {\bf Appendix A.} See Supplementary Material} \label{onlinematerial}


\bsp

\label{lastpage}

\end{document}

%% file: table1_revised2.tex
\begin{center}
\begin{deluxetable}{lllllll}
\tabletypesize{\scriptsize}
\tablewidth{0pt}
\tablecaption{Metallicity, atmospheric parameters and evolutionary state of CEMP-$s$ and CEMP-$s/r$.
Labels: `ms' means main-sequence, `TO' turnoff, `SG' subgiant 
and `G' giant; in column~7, the label `no' indicates stars having not suffered the FDU,
`yes' is for stars having already suffered the
FDU,  and `no?' means that the occurrence 
of the FDU is uncertain (see text).
Authors providing data with resolution spectra $R$ $\sim$ 2\,000 -- 3\,000 are indicated with (*).
In bold are marked the references considered for further discussions.\label{tuttestelleparam}}
\tablecolumns{7}
\tablehead{
\colhead{Stars}                  &
\colhead{Ref.s\tablenotemark{(a)}}                  &
\colhead{[Fe/H]}                 &   
\colhead{{\textit T}$_{\rm eff}$}&
\colhead{log $g$}               &
\colhead{Phase}                 &
\colhead{FDU}                  \\
\colhead{(1)}                  &
\colhead{(2)}                  &
\colhead{(3)}                  &
\colhead{(4)}                  &
\colhead{(5)}                  &
\colhead{(6)}                  &
\colhead{(7)}                  \\
}       
\startdata
BD +04$^{\circ}$2466 &  \textbf{P09}     & -1.92   & 5100   &   1.8   & G      & yes    \\
 "                   &  \textbf{I10}    & -2.10   & 5065   &   1.8   & "      & "    \\
 "                   &  \textbf{Z09}     & -1.92   & 5115   &   1.9   & "      & "    \\
BS 16080--175         & \textbf{T05}      & -1.86   & 6240   &   3.7   & ms/TO  & no    \\
BS 17436--058         & \textbf{T05}      & -1.90   & 5390   &   2.2   & G      & yes    \\
CS 22183--015         &  \textbf{JB02}    & -3.12   & 5200   &   2.5   & G      & yes    \\
 "                   &  \textbf{C06,A07} & -2.75   & 5620   &   3.4   & SG     & no?   \\
 "                   &Lai04(*),Lai07(*)  & -3.17   & 5178   &   2.7   & "      & yes    \\
 "                   &  \textbf{T05}     & -3.00   & 5470   &   2.9   & "      & "    \\
CS 22880--074         &  \textbf{A02,A07} & -1.93   & 5850   &   3.8   & SG     & no?   \\
 "                   &   PS01            & -1.76   & 6050   &   4.0   & "      &  "   \\
CS 22881--036         & \textbf{PS01}     & -2.06   & 6200   &   4.0   & ms/TO  & no    \\
CS 22887--048         & \textbf{T05}      & -1.70   & 6500   &   3.4   & ms/TO  & no   \\
 "                   &  J07(*)           & -2.79   & 6455   &   4.0   & "      &  "    \\
CS 22891--171         &  \textbf{M10}     & -2.25   & 5100   &   1.6   & G      & yes    \\
CS 22898--027         &  \textbf{A02,A07} & -2.26   & 6250   &   3.7   & ms/TO  & no    \\
 "                   &  \textbf{T05}     & -2.61   & 6240   &   3.7   & "      &  "  \\
 "                   &  PS01             & -2.15   & 6300   &   4.0   & "      &  "  \\
 "                   &  Lai07(*)         & -2.29   & 5750   &   3.6   & "      &  "  \\
CS 22942--019         &  \textbf{A02}     & -2.64   & 5000   &   2.4   & G      & yes    \\
 "                   &  \textbf{Sch08}   & -2.66   &  "      &  "     & "      &  "   \\
 "                   &  \textbf{PS01}    & -2.67   & 4900   &   1.8   & "      &  "   \\
 "                   &  \textbf{M10}    & -2.43   & 5100   &   2.5   & "      &  "   \\
CS 22948--27          &  \textbf{BB05}    & -2.47   & 4800   &   1.8   & G      & yes    \\
 "                   &  \textbf{A07}     & -2.21   & 5000   &   1.9   & "      &  "  \\
 "                   &  PS01             & -2.57   & 4600   &   0.8   & "      &  "  \\
 "                   &  \textbf{A02}     & -2.57   & 4600   &   1.0   & "      &  "  \\
CS 22956--28          &  L04 & -1.91   & 7038   &   4.3   & ms     & no    \\
 "                   &  \textbf{S03}   & -2.08   & 6900   &   3.9   & "      &  "  \\
  "                   &  \textbf{M10}    & -2.33   & 6700   &   3.5   & ms/TO  & "   \\
CS 22960--053         &  \textbf{A07}     & -3.14   & 5200   &   2.1   & G      & yes  \\
"                    &  J07(*)           & -3.08   & 5061   &   2.4   &  "      &  "   \\
CS 22964--161A/B      &  \textbf{T08}     & -2.39   & 6050   &   3.7   & ms/TO  & no    \\
CS 22967--07          &  \textbf{L04}     & -1.81   & 6479   &   4.2   & ms     & no    \\
CS 29495--42          &  \textbf{L04}     & -1.88   & 5544   &   3.4   & SG     & no?   \\
"                    &   J07(*)          & -2.30   & 5400   &   3.3   & "      & yes    \\
CS 29497--030         &  \textbf{I05}     & -2.57   & 7000   &   4.1   & ms     & no    \\
"                     &   S03           & -2.16   & 7050   &   4.2   & "      & "  \\
"                     &   S04             & -2.77   & 6650   &   3.5   & "      & "  \\
"                     &   J07(*)          & -2.20   & 7163   &   4.2   & "      & "  \\
CS 29497--34          &  \textbf{BB05}    & -2.90   & 4800   &   1.8   & G      & yes    \\
"                     &  \textbf{A07}     & -2.91   & 4900   &   1.5   & "      & "   \\
"                     &   L04             & -2.57   & 4983   &   2.1   & "      & "   \\
CS 29503--010         &  \textbf{A07}     & -1.06   & 6500   &   4.5   & ms     & no    \\
CS 29509--027         &  \textbf{S03}   & -2.02   & 7050   &   4.2   & ms     & no    \\
CS 29513--032         &  \textbf{R10}     & -2.08   & 5810   &   3.3   & SG     & no?   \\
CS 29526--110         &  \textbf{A02,A07} & -2.38   & 6500   &   3.2   & ms/TO  & no    \\
"                     &  \textbf{A08}     & -2.06   & 6800   &   4.1   & "      & "   \\
CS 29528--028         &  \textbf{A07}     & -2.86   & 6800   &   4.0   & ms     & no    \\
CS 30301--015         &  \textbf{A02,A07} & -2.64   & 4750   &   0.8   & G      & yes    \\
CS 30315--91          &  \textbf{L04}     & -1.68   & 5536   &   3.4   & SG     & no?   \\
CS 30322--023         &  \textbf{M06}     & -3.50   & 4100   &   -0.3  & G      & yes    \\
"                     &  \textbf{A07}     & -3.25   & 4300   &   1.0   & "      & "   \\
"                     &  \textbf{M10}     & -3.39   & 4100   &  -0.3  & "      & "   \\  
CS 30323--107         &  \textbf{L04}     & -1.75   & 6126   &   4.4   & ms     &  no   \\
CS 30338--089         &  \textbf{A07}     & -2.45   & 5000   &   2.1   & G      & yes    \\
"                     &  \textbf{L04}     & -1.75   & 5202   &   2.6   & "      & "   \\
CS 31062--012         &  \textbf{A02,A07} & -2.55   & 6250   &   4.5   & ms     & no    \\
"                     &  \textbf{A08}     & -2.53   & 6200   &   4.3   & "      & "  \\
"                     &  \textbf{I01}     & -2.81   & 6090   &   3.9   & ms/TO   & "  \\
CS 31062--050         &  \textbf{JB04}    & -2.42   & 5500   &   2.7   & SG/G   & yes    \\
"                     &  \textbf{A02,A06,A07}     & -2.31   & 5600   &   3.0   & SG     & no?   \\
"                     &   Lai07(*)        & -2.65   & 5313   &   3.1   & SG/G   & yes  \\
HD 26                 &  \textbf{VE03}    & -1.25   & 5170   &   2.2   & G      & yes    \\
"                     &  \textbf{M10}     & -1.02   & 4900   &   1.5   & "      & "   \\
HD 5223               &  \textbf{G06}     & -2.06   & 4500   &   1.0   & G      & yes  \\    
HD 187861             &  \textbf{VE03}    & -2.30   & 5320   &   2.4   & G      & yes    \\      
"                     &  \textbf{M10}     & -2.36   & 4600   &   1.7   & "      & "   \\
HD 189711             &  \textbf{VE03}    & -1.80   & 3500   &   0.5   & G      & yes    \\
HD 196944             &  \textbf{A02,A07} & -2.25   & 5250   &   1.8   & G      & yes    \\
"                     &  J05              & -2.23   & 5250   &   1.7   & "      & "   \\
"                     &  \textbf{VE03}    & -2.40   & 5250   &   1.7   & "      & "   \\
"                     &  \textbf{M10}     & -2.19   & 5250   &   1.7   & "      & "   \\
"                     &  R08B             & -2.46   & 5170   &  1.8    & "      & "    \\
  G 18--24             &  \textbf{I10}     & -1.62   & 5447   &   4.2   & ms    & no    \\  
  HD 198269           &  \textbf{VE03}    & -2.20   & 4800  &   1.3   & G      & yes  \\ 
  HD 201626           &  \textbf{VE03}    & -2.10   & 5190  &   2.3   & G      & yes  \\ 
  HD 206983           &  \textbf{M10}     & -0.99   & 4200   &  0.6   & G      & yes \\
     "                &  \textbf{JP01}    & -1.43   & 4200   &  1.4   & "      & "   \\
     "                &  \textbf{DP08}    & "       & "      &  "     & "      & "   \\
  HD 209621           &  \textbf{GA10}    & -1.93   & 4500  &   2.0   & G      & yes  \\      
  HD 224959           &  \textbf{VE03}    & -2.20   & 5200  &   1.9   & G      & yes  \\ 
"                     &  \textbf{M10}     & -2.06   & 4900   &  2.0   & "      & "   \\  
  HE 0012--1441       &  \textbf{C06}     & -2.52   & 5730  &   3.5   & SG     & no? \\
  HE 0024--2523       &  \textbf{L03,C04} & -2.72   & 6625  &   4.3   & ms     & no  \\    
  HE 0131--3953       &  \textbf{B05}     & -2.71   & 5928  &   3.8   & TO/SG  & no  \\        
  HE 0143--0441       &  \textbf{C06}     & -2.31   & 6240  &   3.7   & ms/TO  & no  \\     
  HE 0202--2204       &  \textbf{B05}     & -1.98   & 5280  &   1.7   & G      & yes  \\     
  HE 0206--1916       &  \textbf{A07}     & -2.09   & 5200  &   2.7   & SG     & yes  \\     
  HE 0212--0557       &  \textbf{C06}     & -2.27   & 5075  &   2.2   & G      & yes  \\     
  HE 0231--4016       &  \textbf{B05}     & -2.08   & 5972  &   3.6   & SG     & no  \\     
  HE 0322--1504       &   Beers07(*)      & -2.00   & 4460  &   0.8   & G/HB?  & yes  \\     
  HE 0336+0113        &  \textbf{C06}     & -2.68   & 5700  &   3.5   & SG     & no? \\     
  "                   &   L04             & -2.41   & 5947  &   3.7   & "      & "  \\    
  HE 0338--3945       &  \textbf{J06}     & -2.42   & 6160  &   4.1   & ms/TO  & no  \\     
  "                   &   B05             & -2.41   & 6162  &   4.1   & "      & "  \\    
  HE 0400--2030       &  \textbf{A07}     & -1.73   & 5600  &   3.5   & SG     & no? \\     
  HE 0430--4404       &  \textbf{B05}     & -2.07   & 6214  &   4.3   & ms     & no  \\     
  HE 0441--0652       &  \textbf{A07}     & -2.47   & 4900  &   1.4   & G      & yes  \\     
  HE 0507--1430       &   Beers07(*)      & -2.40   & 4560  &   1.2   & G      & yes  \\     
  HE 0507--1653       &  \textbf{A07}     & -1.38   & 5000  &   2.4   & G      & yes  \\     
  "                   &  \textbf{Sch08}   & -1.42   &  "    &    "    &  "     & "  \\    
  HE 0534--4548       &   Beers07(*)      & -1.80   & 4250  &   1.5   & G      & yes  \\   
  HE 1001--0243       &  \textbf{M10}     & -2.88   & 5000  &   2.0   & G      & yes  \\     
  HE 1005--1439       &  \textbf{A07}     & -3.17   & 5000  &   1.9   & G      &  yes \\     
  "                   &  \textbf{Sch08}   & -3.08   & "     &   "     & "      & "   \\    
  HE 1031--0020       &  \textbf{C06}     & -2.86   & 5080  &   2.2   & G      & yes  \\     
  HE 1045--1434       &   Beers07(*)      & -2.50   & 4950  &   1.8   & G      & yes  \\     
  HE 1105+0027        &  \textbf{B05}     & -2.42   & 6132  &   3.5   & ms/TO  & no  \\     
  HE 1135+0139        &  \textbf{B05}     & -2.33   & 5487  &   1.8   & G      & yes  \\     
  HE 1152--0355       &  \textbf{G06}     & -1.27   & 4000  &   1.0   & G      & yes  \\     
  HE 1157--0518       &  \textbf{A07}     & -2.34   & 4900  &   2.0   & G      & yes  \\     
  HE 1305+0007        &  \textbf{G06}     & -2.03   & 4750  &   2.0   & G      & yes  \\     
  "                   &   Beers07(*)      & -2.50   & 4560  &   1.0   & "      & "  \\    
  HE 1305+0132        &   Sch07           & -2.50   & 4462  &   0.8   & G/HB?  & yes  \\
  "                   &  \textbf{Sch08}   & -1.92   &  "   &    "     &  "     & "  \\         
  HE 1319--1935       &  \textbf{A07}      & -1.74   & 4600  &   1.1   & G     & yes  \\     
  HE 1410--0004       &  \textbf{C06}      & -3.02   & 5605  &   3.5   & SG    & no? \\   
  HE 1419--1324       &  \textbf{M10}      & -3.05   & 4900  &   1.8   & G      & yes  \\   
  HE 1429--0551       &  \textbf{A07}      & -2.47   & 4700  &   1.5   & G      & yes  \\     
  HE 1430--1123       &  \textbf{B05}      & -2.71   & 5915  &   3.8   & SG     & no  \\     
  HE 1434--1442       &  \textbf{C06}      & -2.39   & 5420  &   3.2   & SG     & yes  \\     
  HE 1443+0113        &  \textbf{C06}      & -2.07   & 4945  &   2.0   & G      & yes  \\     
  HE 1447+0102        &  \textbf{A07}      & -2.47   & 5100  &   1.7   & G      & yes  \\     
  HE 1509--0806       &  \textbf{C06}      & -2.91   & 5185  &   2.5   & G      & yes  \\     
  HE 1523--1155       &  \textbf{A07}      & -2.15   & 4800  &   1.6   & G      & yes  \\     
  HE 1528--0409       &  \textbf{A07}      & -2.61   & 5000  &   1.8   & G      & yes  \\     
  HE 2148--1247       &  \textbf{C03},C04  & -2.30   & 6380  &   3.9   & ms/TO  & no  \\     
  HE 2150--0825       &  \textbf{B05}      & -1.98   & 5960  &   3.7   & SG     & no  \\         
  HE 2158--0348       &  \textbf{C06}      & -2.70   & 5215  &   2.5   & G      & yes  \\     
  HE 2221--0453       &  \textbf{A07}      & -2.22   & 4400  &   0.4   & G      & yes  \\     
  HE 2227--4044       &  \textbf{B05}      & -2.32   & 5811  &   3.9   & SG     & no? \\     
  HE 2228--0706       &  \textbf{A07}      & -2.41   & 5100  &   2.6   & G      & yes  \\     
  HE 2232--0603       &  \textbf{C06}      & -1.85   & 5750  &   3.5   & SG     & no? \\     
  HE 2240--0412       &  \textbf{B05}      & -2.20   & 5852  &   4.3   & SG     & no  \\     
  HE 2330--0555       &  \textbf{A07}      & -2.78   & 4900  &   1.7   & G      & yes  \\     
  HK II 17435--00532  &  \textbf{R08}      & -2.23   & 5200  &   2.2   & G      & yes  \\   
  LP 625--44          &  \textbf{A02,A06}  & -2.70   & 5500  &   2.5   &  G     & yes  \\      
  V Ari               &  \textbf{VE03}     & -2.40   & 3580  &   -0.2  & G      & yes  \\     
  "                   &   Beers07(*)       & -2.50   & 3500  &   0.5   & "      & "  \\    
  SDSS 0126+06        &  \textbf{A08}      & -3.11   & 6600  &   4.1   & ms     & no  \\
  SDSS 0817+26        &  \textbf{A08}      & -3.16   & 6300  &   4.0   & ms     & no  \\     
  SDSS 0924+40        &  \textbf{A08}      & -2.51   & 6200  &   4.0   & ms     & no  \\     
  SDSS 1707+58        &  \textbf{A08}      & -2.52   & 6700  &   4.2   & ms     & no  \\     
  SDSS 2047+00        &  \textbf{A08}      & -2.05   & 6600  &   4.5   & ms     & no  \\     
  SDSS J0912+0216     &  \textbf{B10}      & -2.50   & 6500  &   4.5   &   ms   & no  \\       
  SDSS J1349--0229    &  \textbf{B10}      & -3.00   & 6200  &   4.0   &   ms   & no  \\ 
\enddata
\tablenotetext{(a)}{References are \citet{aoki02a,aoki02c,aoki02d,aoki06,aoki07,aoki08}, A02a, A02c, A02d, A06, A07, A08; 
\citet{barbuy05}, BB05; 
\citet{barklem05}, B05;
\citet{beers07}, Beers07;
\citet{behara10}, B10;
\citet{cohen03,cohen04,cohen06}, C03, C04, C06;
\citet{drakepereira08}, DP08;
\citet{goswami06}, G06;
\citet{GA10}, GA10;
\citet{ivans05}, I05;
\citet{ish10}, I10;
\citet{israelian01}, I01;
\citet{jonsell05,jonsell06}, J05,J06;
\citet{JB02,JB04}, JB02, JB04; 
\citet{johnson07}, J07;
\citet{lai07}, Lai07;
\citet{lai04}, Lai04;
\citet{lucatello03}, L03;
\citet{lucatello04PhD}, L04;
\citet{masseron06,masseron10}, M06, M10;
\citet{pereira09}, P09;
\citet{PS01}, PS01;
\citet{roederer08,roederer08a,roederer10}, R08, R08B, R10;
\citet{schuler07,schuler08}, Sch07, Sch08;
\citet{sivarani04}, S04;
\citet{sneden03b}, S03;
\citet{thompson08}, T08;
\citet{tsangarides05}, T05;
\citet{vaneck03}, VE03;
\citet{zhang09}, Z09.}
\end{deluxetable}
\end{center}      

%% file: table2.tex
\begin{center}  
\begin{table*}                  
\caption{Observed [Na/Fe], [Mg/Fe], [ls/Fe], [hs/Fe], [Pb/Fe], their $s$-process
indicators [hs/ls] and [Pb/hs], as well as [La/Fe], [Eu/Fe] and [La/Eu] are 
listed for CEMP-$s$ and CEMP-$s/r$ stars with a major number of 
observations. 
References and labels in columns~2 and~3 are the same given in
Table~\ref{tuttestelleparam}.
We distinguish between stars with high $s$-process enhancement, 
[hs/Fe] $\ga$ 1.5, called CEMP-$s$II (or CEMP-$s$II$/r$II and CEMP-$s$II$/r$I
depending on the $r$-process enhancement; see text and Section~\ref{CEMPs+r}),
and stars with a mild $s$ enrichment, [hs/Fe] $<$ 1.5, 
called CEMP-$s$I.
In column~15 the two classes are labeled with `sII' (or `sII/rII' and `sII/rI') and `sI'.
The labels `sI/$-$' and `sII/$-$' stay for stars without europium detection.}
\label{table5_sindicator}       
\resizebox{17.3cm}{!}{\begin{tabular}{lcccccccccccccc}                 
\hline                                   
Star       &  Ref.     & FDU & [Fe/H]  & [Na/Fe] & [Mg/Fe] &[ls/Fe] & [hs/Fe] &[hs/ls]&     [Pb/Fe]   &   [Pb/hs]   &  [La/Fe] & [Eu/Fe]    &   [La/Eu]    
&   Type  \\  
(1) & (2) & (3) &(4) & (5) & (6) & (7) & (8) & (9) & (10) & (11) & (12) &(13) & (14) & (15) \\      
\hline  
BD +04$^{\circ}$2466  &  P09,I10          & yes&-1.92,-2.10&-0.04&0.25,0.44&  0.60  &  1.20 &  0.60  &    1.92  &  0.72    & 1.20 &  -       &  -        & sI/$-$     \\
BS 16080$-$175        & T05               &  no   &  -1.86  &  -    &  -   &  1.20  &  1.62 &  0.42  &    2.60  &  0.98    & 1.65 &  1.05    &  0.60     & sII     \\ 
BS 17436$-$058        & T05               &  yes  &  -1.90  &  -    &  -   &  0.75  &  1.35 &  0.60  &    2.11  &  0.76    & 1.39 &  0.93    &  0.46     & sI      \\ 
CS 22183$-$015        & C06,A07           &  no?  &  -2.75  & 0.11  & 0.54 &  0.55  &  1.76 &  1.21  &    2.79  &  1.03    & 1.70 &  1.70    &  0.00     & sII/rII   \\
 "                    & JB02              &  yes  &  -3.12  &  -    &  -   &  0.48  &  1.61 &  1.13  &    3.17  &  1.56    & 1.59 &  1.39    &  0.20     & "        \\
CS 22880$-$074        & A02d,A07          &  no?  &  -1.93  & -0.09 & 0.46 &  0.26  &  1.10 &  0.84  &    1.90  &  0.80    & 1.07 &  0.50    &  0.57     & sI      \\
CS 22881$-$036        & PS01              &  no   &  -2.06  &  0.16 & 0.40 &  0.99  &  1.75 &  0.76  &    -     &  -       & 1.59 &  1.00    &  0.59     & sII     \\ 
CS 22887$-$048        & T05               &  no   &  -1.70  &  -    &  -   &  1.11  &  1.91 &  0.80  &    3.40  &  1.49    & 1.73 &  1.49    &  0.24     & sII/rI  \\ 
CS 22898$-$027        & A02d,A07          &  no   &  -2.26  &  0.33 & 0.41 &  0.87  &  2.17 &  1.30  &    2.84  &  0.67    & 2.13 &  1.88    &  0.25     & sII/rII  \\
CS 22942$-$019        & A02d,PS01         &  yes  &  -2.64  &  1.44 & 0.58 &  1.65  &  1.29 &  -0.36 &$\leq$1.6 &$\leq$0.31& 1.20 &  0.79    &  0.41     & sI      \\ 
CS 22948$-$27         & BB05,A07          &  yes  &  -2.47,-2.21 & 0.52 &0.31,0.55&1.23 &2.16 & 0.93 &    2.72  &  0.56    & 2.32 &  1.88    &  0.44     & sII/rII   \\ 
CS 22964$-$161        & T08               &  no   &  -2.39  &  0.00 & 0.36 &  0.39  &  1.04 &  0.65  &    2.19  &  1.15    & 1.07 &  0.69    &  0.38     & sI      \\  
CS 29497$-$030        & I05               &  no   &  -2.57  &  0.58 & 0.44 &  1.17  &  2.19 &  1.02  &    3.65  &  1.46    & 2.22 &  1.99    &  0.23     & sII/rII  \\ 
CS 29497$-$34         & BB05,A07          &  yes  &  -2.90  &  1.37 &0.72,1.31& 1.21&  2.02 &  0.81  &    2.95  &  0.93    & 2.12 &  1.80    &  0.32     & sII/rII   \\ 
CS 29513$-$032        & R10               &  no?  &  -2.08  &  0.15 & 0.52 &  0.07  &  0.52 &  0.45  &    1.81  &  1.29    & 0.39 &  0.39    &  0.00      & sI     \\
CS 29526$-$110        & A02c,A07          &  no   &  -2.38  & -0.07 & 0.30 &  1.00  &  1.88 &  0.88  &    3.30  &  1.42    & 1.69 &  1.73    &  -0.04    & sII/rII   \\ 
CS 29528$-$028        & A07               &  no   &  -2.86  & 2.33  & 1.69 &  2.23  &  2.85 &  0.62  &    -     &   -      & 2.93 &   -      &  -        & sII/$-$    \\  
CS 30301$-$015        & A02d,A07          &  yes  &  -2.64  & 1.09  & 0.86 &  0.53  &  0.98 &  0.45  &    1.70  &  0.72    & 0.84 &  :0.2    &  :0.64    & sI      \\ 
CS 30322$-$023        & M06,A07           &  yes  &  -3.25  & 1.04 &0.80,0.54&-0.13 &  0.53 &  0.66  &    1.49  &  0.96    & 0.48 &  -0.51   &  0.99     & sI      \\ 
CS 31062$-$012        & A02d,A07          &  no   &  -2.55  & 0.60  & 0.45 &  1.06  &  1.92 &  0.86  &    2.40  &  0.48    & 2.02 &  1.62    &  0.40     & sII/rII   \\ 
CS 31062$-$050        & JB04,A06,A07      &  no?  &  -2.42  & 0.34  &0.84,0.60& 0.62&  2.02 &  1.40  &    2.81  &  0.79    & 2.12 &  1.79    &  0.33     & sII/rII   \\ 
HD 26$^{a}$           & V03,M10           &  yes&-1.25,-1.02& -     & 0.93 &  0.83  &  1.33 &  0.50  &    2.02  &  0.69    & 1.39 &  0.75    &  0.64     & sII     \\ 
HD 5223               & G06               &  yes  &  -2.06  & 0.46  & 0.58 &  1.10  &  1.66 &  0.56  &$\leq$2.21&$\leq$0.55& 1.76 &   -      &  -        & sII/$-$    \\
HD 187861             & V03               &  yes  &  -2.30  &  -    & -    &  1.16  &  1.97 &  0.81  &   3.30   &  1.33    & 2.0  &   -      &  -        & sII/$-$    \\ 
"                     & M10               &  yes  &  -2.36  &  -    & 0.37 &   -    &  1.58 &   -    &   2.86   &  1.28    & 1.73 &  1.34    &  0.39     & sII/rI  \\
HD 189711             & V03               &  yes  &  -1.80  &  -    & -    &  1.18  &  1.32 &  0.15  &   0.90   &  -0.42   & 1.20 &   -      &  -        & sI/$-$     \\   
HD 196944             & A02d,A07          &  yes  &  -2.25  & 0.86  & 0.42 &  0.61  &  0.86 &  0.25  &   1.90   &  1.04    & 0.91 &  0.17    &  0.74     & sI      \\  
HD 198269             & V03               &  yes  &  -2.20  &  -    & -    &  0.39  &  1.33 &  0.94  &   2.40   &  1.07    & 1.60 &   -      &  -        & sI/$-$     \\ 
HD 201626             & V03               &  yes  &  -2.10  &  -    & -    &  0.87  &  1.60 &  0.73  &   2.60   &  1.00    & 1.90 &   -      &  -        & sII/$-$    \\  
HD 206983$^{a}$       & M10,JP01          &  yes&-0.99,-1.43&0.42,-&0.25,0.61& 0.44 &  0.82 &  0.38  &   1.49   &  0.67    & 1.04 &  0.73    &  0.31     & sI     \\
HD 209621             & GA10              &  yes  &  -1.93  & 0.01  & 0.17 &  1.08  &  1.91 &  0.83  &   1.88   &  -0.03   & 2.41 &  1.35    &  1.06     & sII/rI   \\ 
HD 224959             & V03,M10           &  yes  &  -2.20,-2.06  &  -    & 0.76 &  0.95  &  2.07 &  1.12  &   3.10   &  1.03    & 2.03 &  1.74    &  0.29     & sII/rII   \\ 
HE 0143$-$0441        & C06               &  no   &  -2.31  &  -    & 0.63 &  0.74  &  1.86 &  1.12  &   3.11   &  1.25    & 1.78 &  1.46    &  0.32     & sII/rI   \\
HE 0202$-$2204        & B05               &  yes  &  -1.98  &  -    & -0.01&  0.43  &  1.10 &  0.67  &   -      &  -       & 1.36 &  0.49    &  0.87     & sI      \\
HE 0212$-$0557        & C06               &  yes  &  -2.27  &  -    & 0.04 &  1.03  &  2.05 &  1.02  &   -      &  -       & 2.28 &   -      &  -        & sII/$-$    \\
HE 0231$-$4016        & B05               & no    &  -2.08  &  -    & 0.22 &  0.79  &  1.26 &  0.47  &   -      &  -       & 1.22 &   -      &  -        & sI/$-$     \\
HE 0336+0113          & C06               & no?   &  -2.68  &  -    & 1.04 &  1.68  &  1.89 &  0.21  &$\leq$2.28&$\leq$0.39& 1.93 &  1.18    &  0.75     & sII     \\
HE 0338$-$3945        & J06               & no    &  -2.42  &  0.36 & 0.30 &  1.05  &  2.29 &  1.24  &   3.10   &  0.81    & 2.28 &  1.94    &  0.34     & sII/rII   \\
HE 0430$-$4404        & B05               & no    &  -2.07  &  -    & 0.29 &  0.69  &  1.34 &  0.65  &   -      &  -       & 1.41 &   -      &  -        & sI/$-$     \\
HE 1031$-$0020        & C06               & yes   &  -2.86  &  -    & 0.50 &  0.35  &  1.29 &  0.94  &   2.66   &  1.37    & 1.16 &$\leq$0.87&$\geq$0.29 & sI/$-$     \\
HE 1105+0027          & B05               & no    &  -2.42  &  -    & 0.47 &  0.91  &  2.06 &  1.15  &   -      &  -       & 2.10 &  1.81    &  0.29     & sII/rII   \\
HE 1135+0139          & B05               & yes   &  -2.33  &  -    & 0.33 &  0.39  &  0.87 &  0.48  &   -      &  -       & 0.93 &  0.33    &  0.60     & sI      \\
HE 1152$-$0355$^{a}$  & G06               & yes   &  -1.27  &  -    & -0.01&  0.07  &  0.96 &  0.89  &   -      &  -       & 1.57 &   -      &  -        & sI/$-$    \\
HE 1305+0007          & G06               & yes   &  -2.03  & 0.26  & 0.25 &  1.41  &  2.58 &  1.17  &   2.37   &  -0.21   & 2.56 &  1.97    &  0.59     & sII/rII   \\
HE 1430$-$1123        & B05               & no    &  -2.71  &  -    & 0.35 &  0.73  &  1.67 &  0.94  &   -      &  -       & -    &   -      &  -        & sII/$-$    \\
HE 1434$-$1442        & C06               & yes   &  -2.39  &  0.03 & 0.30 &  0.52  &  1.41 &  0.89  &   2.18   &  0.77    & -    &   -      &  -        & sI/$-$     \\
HE 1509$-$0806        & C06               & yes   &  -2.91  &  -    & 0.64 &  1.08  &  1.78 &  0.70  &   2.61   &  0.83    & 1.67 &$\leq$0.93&$\geq$0.74 & sII/($-$)  \\
HE 2148$-$1247        & C03               & no    &  -2.30  &  -    & 0.50 &  1.07  &  2.28 &  1.21  &   3.12   &  0.84    & 2.38 &  1.98    &  0.40     & sII/rII   \\
HE 2150$-$0825        & B05               & no    &  -1.98  &  -    & 0.36 &  0.89  &  1.39 &  0.50  &   -      &  -       & 1.41 &   -      &  -        & sI/$-$     \\
HE 2158$-$0348        & C06               & yes   &  -2.70  &  -    & 0.68 &  1.22  &  1.47 &  0.25  &   2.60   &  1.13    & 1.55 &  0.80     &  0.75    & sII    \\
HE 2232$-$0603        & C06               & no?   &  -1.85  &  -    & 0.85 &  0.62  &  1.15 &  0.53  &   1.55   &  0.40    & 1.23 &   -      &  -        & sI/$-$     \\
HK II 17435--00532    & R08               & yes   &  -2.23  &  0.69 & 0.42 &  0.37  &  0.92 &  0.55  &    -     &  -       & 0.78 &  0.48    &  0.30     & sI      \\
LP 625$-$44           & A02a,A02d,A06     & yes   &  -2.70  &  1.75 & 1.12 &  1.28  &  2.21 &  0.93  &   2.67   &  0.46    & 2.50 &  1.76    &  0.74     & sII/rII  \\
V Ari                 & V03               & yes   &  -2.40  &   -   & -    &  1.21  &  1.42 &  0.21  &   1.20   &  -0.22   & 1.30 &   -      &  -        & sI/$-$     \\
SDSS 0126+06          & A08               & no    & -3.11   & 0.69  & 0.61 &  1.75  &  2.33 &  0.58  &   3.41   &   1.08   & 2.46 &   -      &  -        & sII/$-$    \\
SDSSJ 0912+0216       & B10               & no    & -2.50   & 0.38  & 0.21 &  0.85  &  1.69 &  0.84  &   2.33   &  0.64    & 1.35 &  1.20    &  0.15     & sII/rI   \\
SDSSJ 1349-0229       & B10               & no    & -3.00   & 1.49  & 0.57 &  1.43  &  2.00 &  0.57  &   3.09   &  1.09    & 1.74 &  1.62    &  0.12     & sII/rII   \\
\hline
Range   &-&-&-1.0$\div$-3.3&-0.1$\div$2.3&0.0$\div$1.7&-0.1$\div$2.2&0.5$\div$2.9&-0.4$\div$1.4&0.9$\div$3.7&-0.4$\div$1.6&0.4$\div$2.9&-0.5$\div$2.0&0.0$\div$1.1& - \\
\hline 
\multicolumn{15}{l}{$^{a}$ The three disc stars, HD 26 ([Fe/H] = $-$1.0), HD 206983 ([Fe/H] = $-$1.0) and 
HE 1152--0355 ([Fe/H] = $-$1.3), will be discussed in Paper III.} \\  
\end{tabular}} 
\end{table*}
\end{center}  

%% file: table3.tex
\begin{table*}                  
\caption{The same as Table~\ref{table5_sindicator}, but 
for CEMP-$s$ and CEMP-$s/r$ stars with a limited number
of spectroscopic observations among the characteristic $s$-process elements. 
Labels `sI/$-$' or `sII/$-$' between brackets mean a low upper limit detected for Eu.
[ls$\dag$/Fe] = [Sr/Fe] in all these stars, with only one exception,
SDSS 2047+00, for which Y and Zr are detected.
[hs$\dag$/Fe] = [Ba/Fe], with the exception of those stars with La
measurement (column~12), and for HE 0131--3953 for which both La and
Nd have been detected. 
The two disc stars mentioned here, CS 29503--010 ([Fe/H] = $-$1.1) and HE 0507--1653 
([Fe/H] = $-$1.4), will be discussed in a separate Section in Paper III.}
\label{tablestellemancanti}       
\begin{center}   
\resizebox{17cm}{!}{\begin{tabular}{lcccccccccccccc} 
\hline    
Star           & Ref.  &  FDU & [Fe/H]& [Na/Fe]& [Mg/Fe]& [ls$\dag$/Fe]& [hs$\dag$/Fe]&[hs$\dag$/ls$\dag$]& [Pb/Fe] & [Pb/hs$\dag$]& [La/Fe] & [Eu/Fe] & [La/Eu] & Type  \\
(1) & (2) & (3) & (4) & (5) & (6) & (7) & (8) & (9) & (10) & (11) & (12) & (13)& (14) & (15) \\
\hline
CS 22891--171  & M10   & yes   & -2.25 & -     & 0.70 &  -     &  2.22   &  -        &  1.85    &  -0.37  & 2.12  &  1.73    & 0.39     & sII/rII   \\
CS 22956$-$28  & S03   &  no   & -2.08 &   -   & -    &  1.38  &  0.42   &  -0.96    &  -       &  -      & -     & -        &  -       & sI/$-$        \\
"              & M10   &  "    & -2.33 &   -   & -    &   -    &  0.51   &  -        &  $<$1.33 & $<$0.82 &$<$0.50&$<$0.91   &   -      & "          \\
CS 22960$-$053 & A07   &  yes  & -3.14 & -     & 0.65 &   -    &  0.86   &  -        &  -       &   -     & -     &   -      &   -      & sI/$-$       \\
CS 22967$-$07  & L04   &  no   & -1.81 & 0.14  & 0.64 &  0.93  &  1.77   &  0.84     &  2.80    &  1.03   & 1.50  &  0.80    &   0.70   & sII       \\
CS 29495$-$42  & L04   &  no?  & -1.88 & -0.05 & 0.80 &  0.20  &  1.54   &  1.34     &  1.30    &  -0.24  & 1.30  &  0.80    &   0.50   & sI        \\
CS 29503$-$010$^{a}$ & A07   &  no   & -1.06 & 0.16  & 0.36 &  -     &  1.50   &  -        &  -       &   -     & -     &   -      &   -      & s(II)/$-$    \\
CS 29509$-$027 & S03   &  no   & -2.02 & -     & -    &  0.82  &  1.33   &  0.51     &  -       &   -     & -     &   -      &   -      & sI/$-$       \\
CS 30315$-$91  & L04   &  no?  & -1.68 & 0.21  & 0.77 &  0.26  &  1.23   &  0.97     &  1.90    &  0.67   & 0.90  &$\leq$0.4 &$\geq$0.5 & sI(/$-$)       \\
CS 30323$-$107 & L04   &  no   & -1.75 & -0.50 & 0.65 &  0.46  &  1.47   &  1.01     &  2.50    &  1.03   & 1.10  &$\leq$0.6 &$\geq$0.5 & sII(/$-$)    \\
CS 30338$-$089 & A07   &  yes  & -2.45 & 0.46  & 0.48 &   -    &  2.22   &  -        &  -       &   -     & -     &   -      &   -      & sII/$-$      \\
 "             & L04   &  "    & -1.75 & 0.20  & 0.35 &  0.61  &  1.71   &  1.10     &  3.70    &  1.99   & 1.60  &  1.80    &  -0.20   & sII(/rII)  \\
  G 18$-$24    & I10   &  no   & -1.62 & -0.30 & 0.23 &  0.58  &  1.17   &  0.59     &  -       &   -     &   -   &   -      &   -      & sI/$-$      \\ 
HE 0012$-$1441 & C06   &  no?  & -2.52 & -     & 0.91 &  -     &  1.15   &   -       &$\leq$1.92&$\leq$0.77& -     &  -       &   -      & sI/$-$      \\
HE 0024$-$2523 & L03   &  no   & -2.70 & 0.17  & 0.73 &  0.34  &  1.63   &  1.29     & 3.30     &  1.67   & 1.80  &$\leq$1.10&$\geq$0.7 & sII(/$-$)   \\
HE 0131$-$3953 & B05   &  no   & -2.71 & -     & 0.30 &  0.46  &  1.97   &  1.51     & -        &   -     &  1.94 &  1.62    &   0.32   & sII/rII     \\
HE 0206$-$1916 & A07   &  yes  & -2.09 & 0.34  & 0.52 &   -    &  1.97   &  -        &  -       &   -     &   -   &   -      &   -      & sII/$-$     \\
HE 0400$-$2030 & A07   &  no?  & -1.73 & 0.51  & 0.62 &   -    &  1.64   &  -        &  -       &   -     &   -   &   -      &   -      & sII/$-$      \\
HE 0441$-$0652 & A07   &  yes  & -2.47 & 0.32  & 0.35 &   -    &  1.11   &  -        &  -       &   -     &   -   &   -      &   -      & sI/$-$       \\
HE 0507$-$1653$^{a}$ & A07   &  yes  & -1.38 & 0.23  & 0.19 &   -    &  1.89   &  -        &  -       &   -     &   -   &   -      &   -      & sII/$-$      \\
HE 1001--0243  & M10   &  yes  & -2.88 & -     & 0.37 &   -    &  0.59   &  -        &$\leq$1.38&$\leq$0.79&0.55  & -0.04    &  0.59    & sI       \\
HE 1005$-$1439 & A07   &  yes  & -3.17 & 0.79  & 0.60 &   -    &  1.06   &  -        &  -       &   -     &   -   &   -      &   -      & sI/$-$       \\
HE 1157$-$0518 & A07   &  yes  & -2.34 & 0.34  & 0.50 &   -    &  2.14   &  -        &  -       &   -     &   -   &   -      &   -      & sII/$-$      \\
HE 1305+0132   & Sch08 &  yes  & -1.92 & -     & -    &   -    &  0.86   &  -        &  -       &   -     &   -   &   -      &   -      & sI/$-$      \\
HE 1319$-$1935 & A07   &  yes  & -1.74 & -     & 0.47 &   -    &  1.89   &  -        &  -       &   -     &   -   &   -      &   -      & sII/$-$      \\
HE 1410$-$0004 & C06   &  no?  & -3.02 & 0.48  & 0.58 &  0.18  &  1.06   &  0.88     &$\leq$3.17&$\leq$2.11&   -   &$\leq$2.40&  -      & sI/$-$       \\
HE 1419--1324  & M10   &  yes  & -3.05 & -     & 0.53 &   -    &  0.84   &  -        & 2.15     &  1.31   & 0.82  &  0.53    &  0.29    & sI       \\
HE 1429$-$0551 & A07   &  yes  & -2.47 & 0.65  & 0.52 &   -    &  1.57   &  -        &  -       &   -     &   -   &   -      &   -      & sII/$-$      \\
HE 1443+0113   & C06   &  yes  & -2.07 & 0.37  & 0.37 &   -    &  1.40   &  -        &  -       &   -     &   -   &   -      &   -      & sI/$-$       \\
HE 1447+0102   & A07   &  yes  & -2.47 & 0.67  & 1.43 &   -    &  2.70   &  -        &  -       &   -     &   -   &   -      &   -      & sII/$-$      \\
HE 1523$-$1155 & A07   &  yes  & -2.15 & -     & 0.62 &   -    &  1.72   &  -        &  -       &   -     &   -   &   -      &   -      & sII/$-$     \\
HE 1528$-$0409 & A07   &  yes  & -2.61 & 0.73  & 0.83 &   -    &  2.30   &  -        &  -       &   -     &   -   &   -      &   -      & sII/$-$      \\
HE 2221$-$0453 & A07   &  yes  & -2.22 & -     & 0.80 &   -    &  1.75   &  -        &  -       &   -     &   -   &   -      &   -      & sII/$-$      \\
HE 2227$-$4044 & B05   &  no?  & -2.32 & -     & 0.30 &  0.41  &  1.33   &  0.92     &  -       &   -     &  1.28 &   -      &   -      & sI/$-$       \\
HE 2228$-$0706 & A07   &  yes  & -2.41 & -     & 0.67 &   -    &  2.50   &  -        &  -       &   -     &   -   &   -      &   -      & sII/$-$      \\    
HE 2240$-$0412 & B05   &  no   & -2.20 & -     & 0.28 &  0.24  &  1.37   &  1.13     &  -       &   -     &   -   &   -      &   -      & sI/$-$       \\
HE 2330$-$0555 & A07   &  yes  & -2.78 & 0.58  & 0.67 &   -    &  1.22   &  -        &  -       &   -     &   -   &   -      &   -      & sI/$-$       \\ 
SDSS 0817+26   & A08   &  no   & -3.16 & -     & 0.43 &  0.14  &  0.77   &  0.63     &  -       &   -     &   -   &   -      &   -      & sI/$-$       \\
SDSS 0924+40   & A08   &  no   & -2.51 & 1.31  & 0.52 &  0.60  &  1.81   &  1.21     & 3.01     &  1.20   &   -   &   -      &   -      & sII/$-$      \\
SDSS 1707+58   & A08   &  no   & -2.52 & 2.71  & 1.13 &  2.25  &  3.40   &  1.15     &$\leq$3.72 &$\leq$0.32&   - &  -       &   -      & sII/$-$      \\
SDSS 2047+00   & A08   &  no   & -2.05 & 0.33  & 0.27 &  0.88  &  1.50   &  0.72     &   -      &   -     &   -   &   -      &   -      & sII/$-$      \\
\hline  
Range  &-&-&-1.1$\div$-3.2&-0.5$\div$2.7&0.2$\div$1.4&0.2$\div$2.3&0.4$\div$3.4&-1.0$\div$1.5&1.3$\div$3.7&-0.4$\div$2.0&0.6$\div$2.1&0.0$\div$1.7&-0.2$\div$0.7&-\\
\hline   
\multicolumn{15}{l}{$^{a}$ The two disc stars CS 29503--010 ([Fe/H] = $-$1.1) and HE 0507--1653 
([Fe/H] = $-$1.4), will be discussed in Paper III.}    \\
\end{tabular}} 
\end{center}           
\end{table*}

%% file: table4.tex
\begin{table*}
\caption{Number of stars belonging to different categories of 
CEMP-$s$ and CEMP-$s/r$.
Stars with high $s$-process enhancement are labeled 
`II' (CEMP-$s$II or CEMP-$s$II$/r$, with [hs/Fe] $\geq$ 1.5 -- 2);
stars with a mild $s$-process enhancement are labeled `I' 
(CEMP-$s$I or CEMP-$s$I$/r$I, with [hs/Fe] $<$ 1.5). 
Among CEMP-$s$ stars showing different $r$-process enhancements we distinguish
between CEMP-$s$II$/r$II and CEMP-$s$II$/r$I.
Stars with no europium measurement are labeled
CEMP-$s$I/$-$ and CEMP-$s$II/$-$.
We distinguish between stars having or having not suffered the FDU
(`no' or `yes', respectively).
CS 22183--015, for which discrepant atmospheric 
parameters have been measured by different authors 
\citep{cohen06,aoki07,JB02,lai07}, is labeled with `(?)'.
Note that none of the stars of the sample is classified as 
CEMP-$s$I/$r$I stars, likely because of our definition of CEMP-$s/r$ stars 
(see Section~\ref{CEMPs+r}).
We refer to Paper III for a detailed analysis of these stars.}
\label{class}
\begin{center}
\begin{tabular}{l|cccc}
\hline 
Class                                               & n. stars &  n. stars with limited number of data\\
(1) & (2) & (3) \\
\hline
CEMP-$s$II                                         & 3 no; 3 yes         & 3 no \\
CEMP-$s$I                                          & 3 no; 9 yes         & 2 no; 2 yes \\
\hline
CEMP-$s$II$/r$II with [r/Fe]$^{\rm ini}$ $\sim$ 2      & 5 no; 1 yes         & - \\
CEMP-$s$II$/r$II with [r/Fe]$^{\rm ini}$ $\sim$ 1.5    & 4 no; 5 yes; 1 (?)  & 1 no; 2 yes\\
CEMP-$s$II$/r$I with [r/Fe]$^{\rm ini}$ $\sim$ 1.0    & 3 no; 1 yes         & - \\
CEMP-$s$I$/r$I                                        & -                & - \\
\hline
CEMP-$s$II/$-$       		                             & 3 no; 3 yes         & 4 no; 11 yes \\ 
CEMP-$s$I/$-$ 				                             & 4 no; 7 yes         & 7 no; 7 yes\\
\hline                                                                                                
\end{tabular}
\end{center}
\end{table*}

%% file: table5_revised2.tex
\begin{center}
\begin{deluxetable}{lccccccc}
\tabletypesize{\scriptsize}
\tablewidth{0pt}
\tablecaption{Solar $s$-process contribution (N$_{\rm s}$) and residuals (N$_{\rm r}$) for isotopes from Cu to Bi.
In column~2 and~3 the solar abundances by \citet{AG89}, AG89, and \citet{lodders09},
L09, are listed, respectively. Values are normalised to the number of silicon atoms of
$N$(Si) = 10$^{6}$.
N$_{\rm s}$ by \citet{arlandini99} stellar model, A99, (column~4)
are compared with updated results (column~6) accounting for recent neutron capture cross sections, normalised 
to solar abundances by L09. 
The s contributions are normalised to $^{150}$Sm.
In columns~5 and~7 we report the percentages of the $s$-process solar contributions from 
A99 and from updated models, respectively.
The residual $r$-process percentages are listed in column~8 for isotopes from Ba to Bi.\label{ns}}
\tablecolumns{8}
\tablehead{
\colhead{Isotope}                  &
\colhead{Solar Abb.}                  &
\colhead{Solar Abb.}                 &   
\colhead{N$_{s}$}&
\colhead{s (\%)}               &         
\colhead{N$_{s}$}                  &
\colhead{s (\%)}               & 
\colhead{r (\%) }                 \\
\colhead{ }                  &
\colhead{AG89}                  &
\colhead{L09}                 &   
\colhead{A99}&
\colhead{A99}                 &
\colhead{Updated}               &
\colhead{Updated\tablenotemark{(a)}}                  &
\colhead{Updated}                 \\
\colhead{(1)}                  &
\colhead{(2)}                  &
\colhead{(3)}                  &
\colhead{(4)}                  &
\colhead{(5)}                  &
\colhead{(6)}                  &
\colhead{(7)}                  &
\colhead{(8)}                 \\
}       
\startdata
 $^{63}$Cu      &    3.61E+02    &  3.74E+02  &   2.95E+00      &  0.8                     &   2.73E+00       &   0.7    &         \\
 $^{65}$Cu      &    1.61E+02    &  1.67E+02  &   2.04E+00      &  1.3                     &   3.21E+00       &   1.9    &         \\
 Cu				&                &            &                 &  1.0                     &                  &   1.1$\pm$0.1   &         \\
                &                &            &                 &                          &                  &          &         \\
 $^{64}$Zn      &    6.13E+02    &  6.30E+02  &   9.21E$-$01    &  0.2                     &   8.39E$-$01     &   0.1    &         \\
 $^{66}$Zn      &    3.52E+02    &  3.62E+02  &   3.44E+00      &  1.0                     &   3.36E+00       &   0.9    &         \\
 $^{67}$Zn      &    5.17E+01    &  5.30E+01  &   7.78E$-$01    &  1.5                     &   7.58E$-$01     &   1.4    &         \\
 $^{68}$Zn      &    2.36E+02    &  2.43E+02  &   6.78E+00      &  2.9                     &   5.03E+00       &   2.1    &         \\
 $^{70}$Zn      &    7.80E+00    &  8.00E+00  &   2.36E$-$02    &  0.3                     &   1.20E$-$02     &   0.2    &         \\
 Zn				&                &            &      	        &  0.9                     &                  &   0.8$\pm$0.1    &         \\
                &                &            &                 &                          &                  &          &         \\
 $^{69}$Ga      &    2.27E+01    &  2.20E+01  &   8.73E$-$01    &  3.8                     &   7.71E$-$01     &   3.5    &         \\
 $^{71}$Ga      &    1.51E+01    &  1.46E+01  &   8.35E$-$01    &  5.5                     &   8.39E$-$01     &   5.7    &         \\
 Ga				&                &            &      	        &  4.5                     &                  &   4.4$\pm$0.2    &         \\
                &                &            &                 &                          &                  &          &         \\
 $^{70}$Ge      &    2.44E+01    &  2.43E+01  &   1.60E+00      &  6.6                     &   1.38E+00       &   5.7    &         \\
 $^{72}$Ge      &    3.26E+01    &  3.17E+01  &   2.47E+00      &  7.6                     &   2.38E+00       &   7.5    &         \\
 $^{73}$Ge      &    9.28E+00    &  8.80E+00  &   4.68E$-$01    &  5.0                     &   6.51E$-$01     &   7.4    &         \\
 $^{74}$Ge      &    4.34E+01    &  4.12E+01  &   2.62E+00      &  6.0                     &   3.68E+00       &   8.9    &         \\
 $^{76}$Ge      &    9.28E+00    &  8.50E+00  &   5.52E$-$03    &  0.1                     &   5.34E$-$03     &   0.1    &         \\
 Ge				&                &            &      	  	    &  6.0                     &                  &   7.1$\pm$0.7    &         \\
                &                &            &                 &                          &                  &          &         \\
 $^{75}$As      &    6.56E+00    &  6.10E+00  &   3.02E$-$01    &  4.6                     &   3.77E$-$01     &   6.2    &         \\
 As				&                &            &      	  	    &  4.6                     &                  &   6.2$\pm$0.6    &         \\
                &                &            &                 &                          &                  &          &         \\
 $^{76}$Se      &    5.60E+00    &  6.32E+00  &   8.62E$-$01    &  15.4                    &   8.48E$-$01     &   13.4   &         \\
 $^{77}$Se      &    4.70E+00    &  5.15E+00  &   3.15E$-$01    &  6.7                     &   3.35E$-$01     &   6.5    &         \\
 $^{78}$Se      &    1.47E+01    &  1.60E+01  &   1.57E+00      &  10.7                    &   2.36E+00       &   14.7   &         \\
 $^{80}$Se      &    3.09E+01    &  3.35E+01  &   2.73E+00      &  8.8                     &   2.87E+00       &   8.6    &         \\
 $^{82}$Se      &    5.70E+00    &  5.89E+00  &   3.39E$-$03    &  0.1                     &   4.03E$-$03     &   0.1    &         \\
 Se				&                &            &      	        &  8.9                     &                  &   9.5$\pm$0.7    &         \\
                &                &            &                 &                          &                  &          &         \\
 $^{79}$Br 	&    5.98E+00    &  5.43E+00  &   5.22E$-$01    &  8.7                     &   4.65E$-$01     &   8.6    &         \\
 $^{81}$Br 	&    5.82E+00    &  5.28E+00  &   5.41E$-$01    &  9.3                     &   5.69E$-$01     &   10.8   &         \\
 Br				&                &            &      	        &  9.0                     &                  &   9.7$\pm$1.5    &         \\
                &                &            &                 &                          &                  &          &         \\
 $^{80}$Kr      &    9.99E$-$01  &  1.30E+00  &   1.17E$-$01    &  11.7                    &   1.03E$-$01     &   7.9    &         \\
 $^{82}$Kr      &    5.15E+00    &  6.51E+00  &   1.91E+00      &  37.1                    &   1.55E+00       &   23.8   &         \\
 $^{83}$Kr      &    5.16E+00    &  6.45E+00  &   6.50E$-$01    &  12.6                    &   5.81E$-$01     &   9.0    &         \\
 $^{84}$Kr      &    2.57E+01    &  3.18E+01  &   3.54E+00      &  13.8                    &   3.71E+00       &   11.7   &         \\
 $^{86}$Kr      &    7.84E+00    &  9.61E+00  &   2.12E+00      &  27.0                    &   1.55E+00       &   16.2   &         \\
 Kr				&                &            &      	        &  19.0                    &                  &   13.4   &         \\
                &                &            &                 &                          &                  &          &         \\
 $^{85}$Rb      &    5.12E+00    &  5.12E+00  &   8.36E$-$01    &  16.3                    &   9.37E$-$01     &   18.3   &         \\
 $^{87}$Rb      &    2.11E+00    &  2.11E+00  &   7.46E$-$01    &  35.4                    &   6.18E$-$01     &   29.3   &         \\
 Rb				&                &            &      	        &  22.0                    &                  &   21.5$\pm$1.5   &         \\
                &                &            &                 &                          &                  &          &         \\
 $^{86}$Sr      &    2.32E+00    &  2.30E+00  &   1.09E+00      &  47.0                    &   1.36E+00       &   59.2   &         \\
 $^{87}$Sr      &    1.51E+00    &  1.60E+00  &   7.60E$-$01    &  50.3                    &   8.75E$-$01     &   54.7   &         \\
 $^{88}$Sr      &    1.94E+01    &  1.92E+01  &   1.79E+01      &  92.2                    &   -              &   +6.0\%\tablenotemark{(b)} &         \\
 Sr		&                &            &      	        &  85.0                    &                  &   97.3$\pm$6.8   &         \\
                &                &            &                 &                          &                  &          &         \\
 $^{89}$Y       &    4.64E+00    &  4.63E+00  &   4.27E+00      &  92.0                    &   -              &   +3.0\%\tablenotemark{(b)}  &         \\
 Y				&                &            &      	        &  92.0                    &                  &   +3.0\%$\pm$10.3\tablenotemark{(b)}    &         \\
                &                &            &                 &                          &                  &          &         \\
 $^{90}$Zr      &    5.87E+00    &  5.55E+00  &   4.24E+00      &  72.2                    &   4.70E+00       &   84.8   &         \\
 $^{91}$Zr      &    1.28E+00    &  1.21E+00  &   1.23E+00      &  96.1                    &   -              &   +5.5\%\tablenotemark{(b)}   &         \\
 $^{92}$Zr      &    1.96E+00    &  1.85E+00  &   1.83E+00      &  93.4                    &   -              &   +0.5\%\tablenotemark{(b)}  &         \\
 $^{94}$Zr      &    1.98E+00    &  1.87E+00  &   -             &+8.2\%\tablenotemark{(b)} &   -              &   +25.5\%\tablenotemark{(b)}  &         \\
 $^{96}$Zr      &    3.20E$-$01  &  3.02E-01  &   1.76E$-$01    &  55.0                    &   1.55E$-$01     &   51.3   &         \\
 Zr				&                &            &      	        &  83.0                    &                  &   96.0$\pm$9.6   &         \\
                &                &            &                 &                          &                  &          &         \\
 $^{93}$Nb      &    6.98E$-$01  &  7.80E-01  &   5.96E$-$01    &  85.4                    &   6.68E$-$01     &   85.6   &         \\
 Nb				&                &            &      	        &  85.0                    &                  &   85.6$\pm$8.6   &         \\
                &                &            &                 &                          &                  &          &         \\
 $^{94}$Mo      &    2.36E$-$01  &  2.33E-01  &   1.53E$-$03    &  0.6                     &   2.00E$-$03     &   0.9    &         \\
 $^{95}$Mo      &    4.06E$-$01  &  4.04E-01  &   2.25E$-$01    &  55.4                    &   2.81E$-$01     &   69.6   &         \\
 $^{96}$Mo      &    4.25E$-$01  &  4.25E-01  &   -             &+6.1\%\tablenotemark{(b)} &   -              &   +19.9\%\tablenotemark{(b)}  &         \\
 $^{97}$Mo      &    2.44E$-$01  &  2.45E-01  &   1.43E$-$01    &  58.6                    &   1.56E$-$01     &   63.7   &         \\
 $^{98}$Mo      &    6.15E$-$01  &  6.22E-01  &   4.66E$-$01    &  75.8                    &   5.11E$-$01     &   82.2   &         \\
 $^{100}$Mo     &    2.46E$-$01  &  2.50E-01  &   9.42E$-$03    &  3.8                     &   1.12E$-$02     &   4.5    &         \\
 Mo				&                &            &      	        &  50.0                    &                  &   57.7$\pm$5.8   &         \\
                &                &            &                 &                          &                  &          &         \\
 $^{99}$Ru      &    2.36E$-$01  &  2.27E-01  &   6.69E$-$02    &  28.3                    &   7.52E$-$02     &   33.1   &         \\
 $^{100}$Ru     &    2.34E$-$01  &  2.24E-01  &   2.23E$-$01    &  95.3                    &   -              &   +9.9\%\tablenotemark{(b)}  &         \\
 $^{101}$Ru     &    3.16E$-$01  &  3.04E-01  &   4.83E$-$02    &  15.3                    &   5.38E$-$02     &   17.7   &         \\
 $^{102}$Ru     &    5.88E$-$01  &  5.62E-01  &   2.53E$-$01    &  43.0                    &   2.81E$-$01     &   50.0   &         \\
 $^{104}$Ru     &    3.48E$-$01  &  3.32E-01  &   9.52E$-$03    &  2.7                     &   8.17E$-$03     &   2.5    &         \\
 Ru				&                &            &      	        &  32.0                    &                  &   37.3$\pm$2.2   &         \\
                &                &            &                 &                          &                  &          &         \\
 $^{103}$Rh     &    3.44E$-$01  &  3.70E-01  &   4.67E$-$02    &  13.6                    &   5.64E$-$02     &   15.2   &         \\
 Rh				&                &            &      	        &  14.0                    &                  &   15.2$\pm$1.5   &         \\
                &                &            &                 &                          &                  &          &         \\
 $^{104}$Pd     &    1.55E$-$01  &  1.51E-01  &   -             &+5.7\%\tablenotemark{(b)} &   -              &   +21.6\%\tablenotemark{(b)}  &         \\
 $^{105}$Pd     &    3.10E$-$01  &  3.03E-01  &   4.27E$-$02    &  13.8                    &   4.75E$-$02     &   15.7   &         \\
 $^{106}$Pd     &    3.80E$-$01  &  3.71E-01  &   1.95E$-$01    &  51.3                    &   2.17E$-$01     &   58.4   &         \\
 $^{108}$Pd     &    3.68E$-$01  &  3.59E-01  &   2.40E$-$01    &  65.2                    &   2.68E$-$01     &   74.6   &         \\
 $^{110}$Pd     &    1.63E$-$01  &  1.59E-01  &   5.93E$-$03    &  3.6                     &   4.73E$-$03     &   3.0    &         \\
 Pd				&                &            &      	        &  46.0                    &                  &   53.1$\pm$2.7   &         \\
                &                &            &                 &                          &                  &          &         \\
 $^{107}$Ag     &    2.52E$-$01  &  2.54E-01  &   3.77E$-$02    &  15.0                    &   4.21E$-$02     &   16.6   &         \\
 $^{109}$Ag     &    2.34E$-$01  &  2.36E-01  &   5.86E$-$02    &  25.0                    &   6.64E$-$02     &   28.1   &         \\
 Ag				&                &            &      	        &  20.0                    &                  &   22.1$\pm$1.1   &         \\
                &                &            &                 &                          &                  &          &         \\
 $^{108}$Cd     &    1.43E$-$02  &  1.40E-02  &   1.61E$-$05    &  0.1                     &   5.34E$-$05     &   0.4    &         \\
 $^{110}$Cd     &    2.01E$-$01  &  1.97E-01  &   1.95E$-$01    &  97.0                    &   -              &   +15.1\%\tablenotemark{(b)}  &         \\
 $^{111}$Cd     &    2.06E$-$01  &  2.01E-01  &   4.87E$-$02    &  23.6                    &   7.65E$-$02     &   38.0   &         \\
 $^{112}$Cd     &    3.88E$-$01  &  3.80E-01  &   2.05E$-$01    &  52.8                    &   2.84E$-$01     &   74.8   &         \\
 $^{113}$Cd     &    1.97E$-$01  &  1.92E-01  &   6.85E$-$02    &  34.8                    &   8.30E$-$02     &   43.2   &         \\
 $^{114}$Cd     &    4.63E$-$01  &  4.52E-01  &   2.95E$-$01    &  63.7                    &   4.06E$-$01     &   89.8   &         \\
 $^{116}$Cd     &    1.21E$-$01  &  1.18E-01  &   2.13E$-$02    &  17.6                    &   2.00E$-$02     &   17.0   &         \\
 Cd				&                &            &      	        &  52.0                    &                  &   69.6$\pm$4.9   &         \\
                &                &            &                 &                          &                  &          &         \\
 $^{113}$In     &    7.90E$-$03  &  8.00E-03  &   5.59E$-$08    &  0.0                     &   5.96E$-$08     &   0.0    &         \\
 $^{115}$In     &    1.76E$-$01  &  1.70E-01  &   6.43E$-$02    &  36.5                    &   7.53E$-$02     &   44.3   &         \\
 In				&                &            &      	        &  35.0                    &                  &   42.4$\pm$3.0   &         \\
                &                &            &                 &                          &                  &          &         \\
 $^{114}$Sn     &    2.52E$-$02  &  2.40E-02  &   4.75E$-$06    &  0.0                     &   5.14E$-$06     &   0.0    &         \\
 $^{115}$Sn     &    1.29E$-$02  &  1.20E-02  &   3.06E$-$04    &  2.4                     &   3.41E$-$04     &   2.8    &         \\
 $^{116}$Sn     &    5.55E$-$01  &  5.24E-01  &   4.76E$-$01    &  85.8                    &   5.04E$-$01     &   96.2   &         \\
 $^{117}$Sn     &    2.93E$-$01  &  2.77E-01  &   1.41E$-$01    &  48.1                    &   1.58E$-$01     &   56.9   &         \\
 $^{118}$Sn     &    9.25E$-$01  &  8.73E-01  &   6.67E$-$01    &  72.1                    &   7.04E$-$01     &   80.6   &         \\
 $^{119}$Sn     &    3.28E$-$01  &  3.09E-01  &   1.27E$-$01    &  38.7                    &   2.02E$-$01     &   65.3  &         \\
 $^{120}$Sn     &    1.25E+00    &  1.18E+00  &   9.77E$-$01    &  78.5                    &   9.93E$-$01     &   84.5   &         \\
 $^{122}$Sn     &    1.77E$-$01  &  1.67E-01  &   7.93E$-$02    &  44.8                    &   7.78E$-$02     &   46.6   &         \\
 Sn				&                &            &                 &  65.0                    &                  &   73.2$\pm$11.0   &         \\
                &                &            &                 &                          &                  &          &         \\
 $^{121}$Sb     &    1.77E$-$01  &  1.79E-01  &   6.78E$-$02    &  38.3                    &   7.40E$-$02     &   41.4   &         \\
 $^{123}$Sb     &    1.32E$-$01  &  1.34E-01  &   8.06E$-$03    &  6.1                     &   8.66E$-$03     &   6.5    &         \\
 Sb				&                &            &      	        &  25.0                    &                  &   26.4$\pm$4.0   &         \\
                &                &            &                 &                          &                  &          &         \\
 $^{122}$Te     &    1.24E$-$01  &  1.22E-01  &   1.09E$-$01    &  87.9                    &   1.19E$-$01     &   97.2   &         \\
 $^{123}$Te     &    4.28E$-$02  &  4.30E-02  &   3.83E$-$02    &  89.5                    &   4.18E$-$02     &   97.3   &         \\
 $^{124}$Te     &    2.29E$-$01  &  2.26E-01  &   2.08E$-$01    &  90.8                    &   -              &   +0.3\%\tablenotemark{(b)}   &         \\
 $^{125}$Te     &    3.42E$-$01  &  3.35E-01  &   6.80E$-$02    &  19.9                    &   7.44E$-$02     &   22.2   &         \\
 $^{126}$Te     &    9.09E$-$01  &  8.89E-01  &   3.68E$-$01    &  40.5                    &   3.99E$-$01     &   44.9   &         \\
 $^{128}$Te     &    1.53E+00    &  1.49E+00  &   2.47E$-$02    &  1.6                     &   5.69E$-$02     &   3.8    &         \\
 Te		&                &            &      	        &  17.0                    &                  &   19.6$\pm$1.4   &         \\
                &                &            &                 &                          &                  &          &         \\
 $^{127}$I      &    9.00E$-$01  &  1.10E+00  &   4.75E$-$02    &  5.3                     &   5.15E$-$02     &   4.7    &         \\
 I				&                &            &      	        &  5.3                     &                  &   4.7$\pm$1.0    &         \\
                &                &            &                 &                          &                  &          &         \\
 $^{128}$Xe     &    1.03E$-$01  &  1.22E-01  &   8.42E$-$02    &  81.7                    &   9.89E$-$02     &   81.1   &         \\
 $^{129}$Xe     &    1.28E+00    &  1.50E+00  &   4.03E$-$02    &  3.1                     &   4.25E$-$02     &   2.8    &         \\
 $^{130}$Xe     &    2.05E$-$01  &  2.39E-01  &   1.70E$-$01    &  82.9                    &   2.11E$-$01     &   88.1   &         \\
 $^{131}$Xe     &    1.02E+00    &  1.19E+00  &   6.65E$-$02    &  6.5                     &   7.92E$-$02     &   6.7    &         \\
 $^{132}$Xe     &    1.24E+00    &  1.44E+00  &   4.16E$-$01    &  33.5                    &   3.89E$-$01     &   27.1   &         \\
 $^{134}$Xe     &    4.59E$-$01  &  5.27E-01  &   2.22E$-$02    &  4.8                     &   1.91E$-$02     &   3.6    &         \\
 Xe		&                &            &      	        &  15.0                    &                  &   15.4   &         \\
                &                &            &                 &                          &                  &          &         \\
 $^{133}$Cs     &    3.72E$-$01  &  3.71E-01  &   5.39E$-$02    &  15.0                    &   5.80E$-$02     &   15.6   &         \\
 Cs				&                &            &      	        &  15.0                    &                  &   15.6$\pm$0.8   &         \\
                &                &            &                 &                          &                  &          &         \\
 $^{134}$Ba     &    1.09E$-$01  &  1.08E-01  &   1.07E$-$01    &  98.2                    &   -              &   +12.5\%\tablenotemark{(b)}   &         \\
 $^{135}$Ba     &    2.96E$-$01  &  2.95E-01  &   7.75E$-$02    &  26.2                    &   8.91E$-$02     &   30.2   &   69.8  \\
 $^{136}$Ba     &    3.53E$-$01  &  3.51E-01  &   -             &+0.3\%\tablenotemark{(b)} &   -              &   +13.7\%\tablenotemark{(b)}  &         \\
 $^{137}$Ba     &    5.04E$-$01  &  5.02E-01  &   3.30E$-$01    &  65.5                    &   3.38E$-$01     &   67.3   &   32.7  \\
 $^{138}$Ba     &    3.22E+00    &  3.21E+00  &   2.76E+00      &  85.7                    &   3.02E+00       &   94.2   &   5.8  \\
 Ba				&                &            &      	        &  81.0                    &                  &   88.7$\pm$5.3   &   11.3  \\
                &                &            &                 &                          &                  &          &         \\
 $^{139}$La     &    4.46E$-$01  &  4.57E-01  &   2.77E$-$01    &  62.1                    &   3.25E$-$01     &   71.0   &   28.9  \\
 La 			&                &            &      	        &  62.1                    &                  &   71.1$\pm$3.6   &   28.9  \\
                &                &            &                 &                          &                  &          &         \\
 $^{140}$Ce     &    1.01E+00    &  1.04E+00  &   8.36E$-$01    &  83.2                    &   9.33E$-$01     &   89.5   &   10.5  \\
 $^{142}$Ce     &    1.26E$-$01  &  1.31E-01  &   2.79E$-$02    &  22.1                    &   2.53E$-$02     &   19.3   &   80.7  \\
 Ce				&                &            &      	        &  77.0                    &                  &   81.3$\pm$4.1   &   18.7  \\
                &                &            &                 &                          &                  &          &         \\
 $^{141}$Pr     &    1.67E$-$01  &  1.72E-01  &   8.13E$-$02    &  48.7                    &   8.89E$-$02     &   51.7   &   48.3  \\
 Pr				&                &            &      	        &  49.0                    &                  &   51.7$\pm$3.6   &   48.3  \\
                &                &            &                 &                          &                  &          &         \\
 $^{142}$Nd     &    2.25E$-$01  &  2.31E-01  &   2.08E$-$01    &  92.4                    &   2.26E$-$01     &   97.6   &   2.4  \\
 $^{143}$Nd     &    1.00E$-$01  &  1.03E-01  &   3.16E$-$02    &  31.6                    &   3.37E$-$02     &   32.7   &   67.3  \\
 $^{144}$Nd     &    1.97E$-$01  &  2.03E-01  &   1.00E$-$01    &  50.8                    &   1.06E$-$01     &   52.2   &   47.8  \\
 $^{145}$Nd     &    6.87E$-$02  &  7.50E-02  &   1.89E$-$02    &  27.5                    &   1.97E$-$02     &   26.3   &   73.7  \\
 $^{146}$Nd     &    1.42E$-$01  &  1.47E-01  &   9.11E$-$02    &  64.2                    &   9.62E$-$02     &   65.5   &   34.5  \\
 $^{148}$Nd     &    4.77E$-$02  &  4.90E-02  &   9.05E$-$03    &  19.0                    &   9.41E$-$03     &   19.2   &   80.8  \\
 Nd				&                &            &      	        &  56.0                    &                  &   57.3$\pm$2.9   &   42.7  \\
                &                &            &                 &                          &                  &          &         \\
 $^{147}$Sm     &    3.99E$-$02  &  4.10E-02  &   8.25E$-$03    &  20.7                    &   1.08E$-$02     &   26.2   &  73.8  \\
 $^{148}$Sm     &    2.92E$-$02  &  3.00E-02  &   2.82E$-$02    &  96.6                    &   -              &   +2.2\%\tablenotemark{(b)}  &        \\
 $^{149}$Sm     &    3.56E$-$02  &  3.70E-02  &   4.45E$-$03    &  12.5                    &   4.68E$-$03     &   12.6   &  87.4  \\
 $^{150}$Sm     &    1.91E$-$02  &  2.00E-02  &   1.91E$-$02    &  100.0                   &   2.00E$-$02     &   100.0  &        \\
 $^{152}$Sm     &    6.89E$-$02  &  7.10E-02  &   1.58E$-$02    &  22.9                    &   1.64E$-$02     &   23.1   &  76.9  \\
 $^{154}$Sm     &    5.86E$-$02  &  6.00E-02  &   4.69E$-$04    &  0.8                     &   1.64E$-$03     &   2.7    &  97.3  \\
 Sm 			&                &            &      	        &  29.0                    &                  &   31.3$\pm$1.6   &  68.7  \\
                &                &            &                 &                          &                  &          &         \\
 $^{151}$Eu     &    4.65E$-$02  &  4.71E-02  &   3.04E$-$03    &  6.5                     &   2.81E$-$03     &   6.0    &  94.0  \\
 $^{153}$Eu     &    5.08E$-$02  &  5.14E-02  &   2.58E$-$03    &  5.1                     &   3.06E$-$03     &   5.9    &  94.1  \\
 Eu				&                &            &      	        &  5.8                     &                  &   6.0$\pm$0.3    &  94.0  \\
                &                &            &                 &                          &                  &          &         \\
 $^{152}$Gd     &    6.60E$-$04  &  7.00E-04  &   5.83E$-$04    &  88.3                    &   4.94E$-$04     &   70.5   &  29.5  \\
 $^{154}$Gd     &    7.19E$-$03  &  7.80E-03  &   6.85E$-$03    &  95.3                    &   6.86E$-$03     &   88.0   &  12.0  \\
 $^{155}$Gd     &    4.88E$-$02  &  5.33E-02  &   2.88E$-$02    &  59.0                    &   3.02E$-$03     &   5.7    &  94.3  \\
 $^{156}$Gd     &    6.76E$-$02  &  7.36E-02  &   1.15E$-$02    &  17.0                    &   1.24E$-$02     &   16.9   &  83.1  \\
 $^{157}$Gd     &    5.16E$-$02  &  5.63E-02  &   5.53E$-$03    &  10.7                    &   5.98E$-$03     &   10.6   &  89.4  \\
 $^{158}$Gd     &    8.20E$-$02  &  8.94E-02  &   2.25E$-$02    &  27.4                    &   2.42E$-$02     &   27.1   &  72.9  \\
 $^{160}$Gd     &    7.21E$-$02  &  7.87E-02  &   8.27E$-$04    &  1.1                     &   4.87E$-$04     &   0.6    &  99.4  \\
 Gd				&                &            &      	        &  15.0                    &                  &   13.5$\pm$0.7   &  86.5  \\
                &                &            &                 &                          &                  &          &         \\
 $^{159}$Tb     &    6.03E$-$02  &  6.34E-02  &   4.36E$-$03    &  7.2                     &   5.36E$-$03     &   8.4    &  91.6  \\
 Tb				&                &            &      	        &  7.2                     &                  &   8.4$\pm$0.6    &  91.6  \\
                &                &            &                 &                          &                  &          &         \\
 $^{160}$Dy     &    9.22E$-$03  &  9.40E-03  &   8.06E$-$03    &  87.4                    &   8.58E$-$03     &   91.3   &  8.7  \\
 $^{161}$Dy     &    7.45E$-$02  &  7.62E-02  &   4.12E$-$03    &  5.5                     &   3.95E$-$03     &   5.2    &  94.8  \\
 $^{162}$Dy     &    1.01E$-$01  &  1.03E-01  &   1.64E$-$02    &  16.2                    &   1.65E$-$02     &   16.0   &  84.0  \\
 $^{163}$Dy     &    9.82E$-$02  &  1.01E-01  &   3.52E$-$03    &  3.6                     &   4.36E$-$03     &   4.3    &  95.7  \\
 $^{164}$Dy     &    1.11E$-$01  &  1.14E-01  &   2.61E$-$02    &  23.5                    &   2.62E$-$02     &   23.0   &  77.0  \\
 Dy				&                &            &      	        &  15.0                    &                  &   14.8$\pm$0.7   &  85.2  \\
                &                &            &                 &                          &                  &          &         \\
 $^{165}$Ho     &    8.89E$-$02  &  9.10E-02  &   6.95E$-$03    &  7.8                     &   7.41E$-$03     &   8.1    &  91.9  \\
 Ho				&                &            &      	        &  7.8                     &                  &   8.1$\pm$0.6    &  91.9  \\
                &                &            &                 &                          &                  &          &         \\
 $^{164}$Er     &    4.04E$-$03  &  4.20E-03  &   3.34E$-$03    &  82.7                    &   3.13E$-$03     &   74.5   &  25.5  \\
 $^{166}$Er     &    8.43E$-$02  &  8.80E-02  &   1.25E$-$02    &  14.8                    &   1.40E$-$02     &   15.9   &  84.1  \\
 $^{167}$Er     &    5.76E$-$02  &  6.00E-02  &   4.92E$-$03    &  8.5                     &   5.49E$-$03     &   9.2    &  90.8  \\
 $^{168}$Er     &    6.72E$-$02  &  7.10E-02  &   1.90E$-$02    &  28.3                    &   2.03E$-$02     &   28.6   &  71.4  \\
 $^{170}$Er     &    3.74E$-$02  &  3.90E-02  &   2.69E$-$03    &  7.2                     &   4.81E$-$03     &   12.3   &  87.7  \\
 Er				&                &            &      	        &  17.0                    &                  &   18.2$\pm$0.9   &  81.8  \\
                &                &            &                 &                          &                  &          &         \\
 $^{169}$Tm     &    3.78E$-$02  &  4.06E-02  &   5.03E$-$03    &  13.3                    &   4.96E$-$03     &   12.2   &  87.8  \\
 Tm				&                &            &      	        &  13.3                    &                  &   12.2$\pm$0.9   &  87.8  \\
                &                &            &                 &                          &                  &          &         \\
 $^{170}$Yb     &    7.56E$-$03  &  7.60E-03  &   -             &+1.1\%\tablenotemark{(b)} &   6.89E$-$03     &   90.6   &  9.4  \\
 $^{171}$Yb     &    3.54E$-$02  &  3.61E-02  &   4.93E$-$03    &  13.9                    &   7.54E$-$03     &   20.9   &  79.1  \\
 $^{172}$Yb     &    5.43E$-$02  &  5.56E-02  &   1.65E$-$02    &  30.4                    &   2.44E$-$02     &   43.9   &  56.1  \\
 $^{173}$Yb     &    4.00E$-$02  &  4.13E-02  &   8.57E$-$03    &  21.4                    &   1.11E$-$02     &   26.9   &  73.1  \\
 $^{174}$Yb     &    7.88E$-$02  &  8.21E-02  &   3.91E$-$02    &  49.6                    &   4.97E$-$02     &   60.5   &  39.5  \\
 $^{176}$Yb     &    3.15E$-$02  &  3.33E-02  &   4.28E$-$03    &  13.6                    &   2.73E$-$03     &   8.2    &  91.8  \\
 Yb				&                &            &      	        &  33.0                    &                  &   39.9$\pm$2.0   &  60.1  \\
                &                &            &                 &                          &                  &          &         \\
 $^{175}$Lu     &    3.57E$-$02  &  3.70E-02  &   6.33E$-$03    &  17.7                    &   6.60E$-$03     &   17.8   &  82.2  \\
 $^{176}$Lu     &    1.04E$-$03  &  1.10E-03  &   -             &+25.0\%\tablenotemark{(b)}&  -               &   +1.2\%\tablenotemark{(b)}  &        \\
 Lu				&                &            &      	        &  20.0                    &                  &   20.2$\pm$1.0   &  79.8  \\
                &                &            &                 &                          &                  &          &         \\
 $^{176}$Hf     &    7.93E$-$03  &  8.10E-03  &   7.65E$-$03    &  96.5                    &   7.88E$-$03     &   97.3   &  2.7   \\
 $^{177}$Hf     &    2.87E$-$02  &  2.90E-02  &   5.29E$-$03    &  18.4                    &   4.98E$-$03     &   17.2   &  82.8  \\
 $^{178}$Hf     &    4.20E$-$02  &  4.25E-02  &   2.40E$-$02    &  57.1                    &   2.49E$-$02     &   58.5   &  41.5  \\
 $^{179}$Hf     &    2.10E$-$02  &  2.12E-02  &   7.74E$-$03    &  36.9                    &   8.70E$-$03     &   41.0   &  59.0  \\
 $^{180}$Hf     &    5.41E$-$02  &  5.47E-02  &   4.08E$-$02    &  75.4                    &   4.85E$-$02     &   88.8   &  11.2  \\
 Hf 			&                &            &      	        &  56.0                    &                  &   61.0$\pm$3.1   &  39.0  \\
                &                &            &                 &                          &                  &          &         \\
 $^{180}$Ta     &    2.48E$-$06  &  2.60E-06  &   1.21E$-$06    &  48.8                    &   1.96E$-$06     &   75.5   &  24.5   \\
 $^{181}$Ta     &    2.07E$-$02  &  2.10E-02  &   8.55E$-$03    &  41.3                    &   9.77E$-$03     &   46.5   &  53.5  \\
 Ta				&                &            &      	        &  41.0                    &                  &   46.5$\pm$4.7   &  53.5  \\
                &                &            &                 &                          &                  &          &         \\
 $^{180}$W      &    1.73E$-$04  &  2.00E-04  &   8.02E$-$06    &  4.6                     &   1.03E$-$05     &   5.1    &  94.9  \\
 $^{182}$W      &    3.50E$-$02  &  3.63E-02  &   1.60E$-$02    &  45.7                    &   2.20E$-$02     &   60.6   &  39.4  \\
 $^{183}$W      &    1.90E$-$02  &  1.96E-02  &   1.02E$-$02    &  53.7                    &   1.13E$-$02     &   57.9   &  42.1  \\
 $^{184}$W      &    4.08E$-$02  &  4.20E-02  &   2.88E$-$02    &  70.6                    &   3.27E$-$02     &   77.8   &  22.2  \\
 $^{186}$W      &    3.80E$-$02  &  3.90E-02  &   1.91E$-$02    &  50.3                    &   2.28E$-$02     &   58.5   &  41.5  \\
 W		&                &            &      	        &  56.0                    &                  &   64.8$\pm$6.5  &  35.2  \\
                &                &            &                 &                          &                  &          &         \\
 $^{185}$Re     &    1.93E$-$02  &  2.07E-02  &   4.78E$-$03    &  24.8                    &   5.68E$-$03     &   27.4   &  72.6  \\
 $^{187}$Re     &    3.51E$-$02  &  3.74E-02  &   6.65E$-$05    &  0.2                     &   3.74E$-$03     &   10.0   &  90.0  \\
 Re				&                &            &      	        &  8.9                     &                  &   16.2$\pm$1.6   &  83.8  \\
                &                &            &                 &                          &                  &          &         \\
 $^{186}$Os     &    1.07E$-$02  &  1.08E-02  &   1.04E$-$02    &  97.2                    &   -              &+11.6\%\tablenotemark{(b)} &        \\
 $^{187}$Os     &    8.07E$-$03  &  8.60E-03  &   6.58E$-$03    &  81.5                    &   3.43E$-$03     &   39.9   &  60.1  \\
 $^{188}$Os     &    8.98E$-$02  &  9.04E-02  &   1.72E$-$02    &  19.2                    &   2.68E$-$02     &   29.6   &  70.4  \\
 $^{189}$Os     &    1.09E$-$01  &  1.10E-01  &   4.70E$-$03    &  4.3                     &   4.93E$-$03     &   4.5    &  95.5  \\
 $^{190}$Os     &    1.78E$-$01  &  1.79E-01  &   2.14E$-$02    &  12.0                    &   2.61E$-$02     &   14.6   &  85.4  \\
 $^{192}$Os     &    2.77E$-$01  &  2.78E-01  &   2.86E$-$03    &  1.0                     &   9.73E$-$03     &   3.5    &  96.5  \\
 Os				&                &            &      	        &  9.4                     &                  &   12.3$\pm$1.0   &  87.7  \\
                &                &            &                 &                          &                  &          &         \\
 $^{191}$Ir     &    2.47E$-$01  &  2.50E-01  &   4.68E$-$03    &  1.9                     &   5.01E$-$03     &   2.0    &  98.0  \\
 $^{193}$Ir     &    4.14E$-$01  &  4.21E-01  &   4.40E$-$03    &  1.1                     &   5.54E$-$03     &   1.3    &  98.7  \\
 Ir				&                &            &      	        &  1.4                     &                  &   1.6$\pm$0.1    &  98.4  \\
                &                &            &                 &                          &                  &          &         \\
 $^{192}$Pt     &    1.05E$-$02  &  1.00E-02  &   1.03E$-$02    &  98.1                    &   8.71E$-$03     &   87.1   &  12.9  \\
 $^{194}$Pt     &    4.41E$-$01  &  4.20E-01  &   1.77E$-$02    &  4.0                     &   2.77E$-$02     &   6.6    &  93.4  \\
 $^{195}$Pt     &    4.53E$-$01  &  4.31E-01  &   7.53E$-$03    &  1.7                     &   1.11E$-$02     &   2.6    &  97.4  \\
 $^{196}$Pt     &    3.38E$-$01  &  3.22E-01  &   3.30E$-$02    &  9.8                     &   4.27E$-$02     &   13.2   &  86.8  \\
 $^{198}$Pt     &    9.63E$-$02  &  9.10E-02  &   2.33E$-$05    &  0.0                     &   3.06E$-$05     &   0.0    &  100.0 \\
 Pt				&                &            &      	        &  5.1                     &                  &   7.1$\pm$0.6    &  92.9  \\
                &                &            &                 &                          &                  &          &         \\
 $^{197}$Au     &    1.87E$-$01  &  1.95E-01  &   1.09E$-$02    &  5.8                     &   1.16E$-$02     &   5.9    &  94.1  \\
 Au				&                &            &      	        &  5.8                     &                  &   5.9$\pm$0.6    &  94.1  \\
                &                &            &                 &                          &                  &          &         \\
 $^{198}$Hg     &    3.39E$-$02  &  4.60E-02  &   -             &+2.4\%\tablenotemark{(b)} &   3.80E$-$02     &   82.6   &  17.4  \\
 $^{199}$Hg     &    5.74E$-$02  &  7.70E-02  &   1.52E$-$02    &  26.5                    &   1.67E$-$02     &   21.7   &  78.3  \\
 $^{200}$Hg     &    7.85E$-$02  &  1.06E-01  &   5.15E$-$02    &  65.6                    &   5.55E$-$02     &   52.4   &  47.6  \\
 $^{201}$Hg     &    4.48E$-$02  &  6.00E-02  &   2.22E$-$02    &  49.6                    &   2.38E$-$02     &   39.7   &  60.3  \\
 $^{202}$Hg     &    1.02E$-$01  &  1.37E-01  &   8.23E$-$02    &  81.1                    &   8.94E$-$02     &   65.3   &  34.7  \\
 $^{204}$Hg     &    2.33E$-$02  &  3.10E-02  &   2.07E$-$03    &  8.9                     &   2.49E$-$03     &   8.0   &  92.0  \\
 Hg				&                &            &      	        &  61.0                    &                  &   49.3$\pm$9.9  &  50.7  \\
                &                &            &                 &                          &                  &          &         \\
 $^{203}$Tl     &    5.43E$-$02  &  5.40E-02  &   4.06E$-$02    &  74.8                    &   4.49E$-$02     &   83.1   &  16.9  \\
 $^{205}$Tl     &    1.30E$-$01  &  1.29E-01  &   9.89E$-$02    &  76.3                    &   8.10E$-$02     &   62.8   &  37.2  \\
 Tl 			&                &            &      	        &  76.0                    &                  &   68.8$\pm$5.5   &  31.2  \\
                &                &            &                 &                          &                  &          &         \\
 $^{204}$Pb     &    6.11E$-$02  &  6.60E-02  &   5.76E$-$02    &  94.3                    &   6.38E$-$02     &   96.7   &         \\
 $^{206}$Pb     &    5.93E$-$01  &  6.14E-01  &   3.43E$-$01    &  57.8                    &   4.09E$-$01     &   66.6   &         \\
 $^{207}$Pb     &    6.44E$-$01  &  6.80E-01  &   4.10E$-$01    &  63.7                    &   3.93E$-$01     &   57.8   &         \\
 $^{208}$Pb     &    1.83E+00    &  1.95E+00  &   6.30E$-$01    &  34.5                    &   8.10E$-$01     &   41.6   &         \\
 Pb				&                &            &      	        &  46.0                    &                  &50.7$\pm$3.6 (87)\tablenotemark{(c)}   &  (13)\tablenotemark{(c)}   \\
                &                &            &                 &                          &                  &          &         \\
 $^{209}$Bi     &    1.44E$-$01  &  1.38E-01  &   7.07E$-$03    &  4.9                     &   8.71E$-$03     &   6.3      &         \\
 Bi				&                &            &      	        &  4.9                     &                  &6.3$\pm$0.6  (26)\tablenotemark{(c)}   &  (74)\tablenotemark{(c)}   \\
                &                &            &                 &                          &                  &          &         \\
\hline
\enddata
\tablenotetext{(a)}{The uncertainties provied in this column account of the solar abundance accuracy estimated by L09.}
\tablenotetext{(b)}{Overabundances with respect to solar (in percentage).}
\tablenotetext{(c)}{The values between brackets for Pb and Bi account for the contribution of the
strong component (see text).}
\end{deluxetable}
\end{center}                                                                                                        

%% file: table6.tex
\begin{table*}
\caption{Theoretical results of [La/Fe], [Eu/Fe] and [La/Eu] predicted in the
AGB envelope for models of $M^{\rm ini}_{\rm AGB}$ = 1.3 and 1.5 $M_{\odot}$, 
[Fe/H] = $-$2.6, and three choices of $^{13}$C-pocket (ST, ST/12 and ST/75).
Different $r$-process enrichments are considered ([r/Fe]$^{\rm ini}$ = 0.0, 
0.5, 1.0, 1.5, 2). Results similar to $M$ = 1.5 $M_{\odot}$ are obtained for
AGB models of initial mass $M$ = 2.0 $M_{\odot}$.}
\label{riniLaEu}
\begin{center}
\begin{tabular}{ll|c|cccc}
 \hline 
&  & \multicolumn{5}{c}{$M^{\rm ini}_{\rm AGB}$ = 1.3 $M_{\odot}$; [Fe/H] = $-$2.6} \\
\hline
  Case  & Predicted ratios & 
 [r/Fe]$^{\rm ini}$ = 0 & [r/Fe]$^{\rm ini}$ = 0.5 & 
 [r/Fe]$^{\rm ini}$ = 1 & [r/Fe]$^{\rm ini}$ = 1.5 & 
 [r/Fe]$^{\rm ini}$ = 2 \\ 
 (1) & (2) & (3) & (4) & (5) & (6)& (7)\\  
 \hline  
      & {[La/Fe]} &  0.73 & 0.73 & 0.86 & 1.12 & 1.51 \\  
ST    & {[Eu/Fe]} &  0.17 & 0.53 & 0.98 & 1.47 & 1.96 \\   
      & {[La/Eu]} &  0.56 & 0.20 & $-$0.12 &$-$0.35 & $-$0.42 \\  
\hline                                                              
       & {[La/Fe]} & 1.88 & 1.88 & 1.89 & 1.92 & 2.01 \\     
 ST/12 & {[Eu/Fe]} & 1.06 & 1.12 & 1.29 & 1.59 & 2.00 \\                                    
       & {[La/Eu]} & 0.82 & 0.76 & 0.60 & 0.33 & 0.01 \\   
\hline                                                                                       
       & {[La/Fe]} &  1.28 & 1.28 & 1.32 & 1.43 & 1.66 \\     
ST/75  & {[Eu/Fe]} &  0.29 & 0.59 & 1.00 & 1.47 & 1.96 \\    
       & {[La/Eu]} &  0.99 & 0.69 & 0.32 &$-$0.04 & $-$0.30\\    
\hline                                                                                                             
&  &\multicolumn{5}{c}{$M^{\rm ini}_{\rm AGB}$ = 1.5 $M_{\odot}$; [Fe/H] = $-$2.6} \\   
 \hline                                                                
       & {[La/Fe]} &  1.98 & 1.98 & 1.99 & 2.01 & 2.07 \\    
ST     & {[Eu/Fe]} &  1.19 & 1.23 & 1.35 & 1.59 & 1.95 \\     
       & {[La/Eu]} &  0.79 & 0.75 & 0.64 & 0.42 & 0.12 \\      
\hline                                                                                                        
       & {[La/Fe]} & 2.69 & 2.69 & 2.69 & 2.69 & 2.71 \\    
ST/12  & {[Eu/Fe]} & 1.72 & 1.73 & 1.77 & 1.87 & 2.10 \\
       & {[La/Eu]} & 0.97 & 0.96 & 0.92 & 0.82 & 0.61 \\   
\hline                                                                                                          
       & {[La/Fe]} &  1.50 & 1.50 & 1.52 & 1.58 & 1.73 \\ 
ST/75  & {[Eu/Fe]} &  0.43 & 0.63 & 0.97 & 1.40 & 1.88 \\ 
       & {[La/Eu]} &  1.07 & 0.87 & 0.55 & 0.18 & $-$0.15 \\     
\hline                                                                                                          
\end{tabular}
\end{center}
\end{table*}

%% file: table7.tex
\begin{table*}
\caption{Theoretical results of [El/Fe] predicted in the
AGB envelope for models of $M^{\rm ini}_{\rm AGB}$ = 1.3 and 1.5 
$M_{\odot}$, two different choices of the $^{13}$C-pocket 
(ST in Col.s~3 to~6 and ST/12 in Col.s~7 to~10),
 [Fe/H] = $-$2.6, and two initial $r$-process 
enrichments: [r/Fe]$^{\rm ini}$ = 0 and [r/Fe]$^{\rm ini}$ = 2.
The initial $r$-enhancement is applied for elements from Ba to Bi
(see text).
At the end of the Table, [ls/Fe], [hs/Fe], [hs/ls] and [Pb/hs] 
are also reported. Note that, with a pure $s$-process contribution
([r/Fe]$^{\rm ini}$ = 0), AGB models with $M^{\rm ini}_{\rm AGB}$ = 
1.3 and 1.5 $M_{\odot}$ predict [Eu/Fe]$_{\rm s}$ = 0.17 and 1.19
with case ST and 1.06 and 1.72 with case ST/12, respectively.}
\label{riniSrBi}
\begin{center}
\resizebox{17cm}{!}{\begin{tabular}{|llc|cc|cc|ccc|cc|}
\hline  
& & & \multicolumn{3}{c}{Case ST} & & & \multicolumn{3}{c}{Case ST/12} & \\
\hline
(1) & (2)&  &(3) & (4)&(5)&(6)& & (7)& (8)& (9)&(10) \\
& & & $M$ = 1.3 $M_{\odot}$   &  $M$ = 1.3 $M_{\odot}$  &   $M$ = 1.5 $M_{\odot}$ & $M$ = 1.5 $M_{\odot}$
& & $M$ = 1.3 $M_{\odot}$ &   $M$ = 1.3 $M_{\odot}$      &    $M$ = 1.5 $M_{\odot}$  &   $M$ = 1.5 $M_{\odot}$ 
 \\
${\rm [El/Fe]}$ &Z& &
[r/Fe]$^{\rm ini}$ = 0  &  [r/Fe]$^{\rm ini}$ = 2 &  
[r/Fe]$^{\rm ini}$ = 0  & [r/Fe]$^{\rm ini}$ = 2  & &
[r/Fe]$^{\rm ini}$ = 0  &  [r/Fe]$^{\rm ini}$ = 2 &
[r/Fe]$^{\rm ini}$ = 0  &  [r/Fe]$^{\rm ini}$ = 2 
\\
\hline
Sr &  38  & &   0.36    &   0.36  &  1.37   & 1.37   &  &    0.64   &   0.64   &   2.24   &   2.24  \\
Y  &  39  & &   0.35    &   0.35  &  1.40   & 1.40   &  &    0.80   &   0.80   &   2.41   &   2.41  \\
Zr &  40  & &   0.34    &   0.34  &  1.36   & 1.36   &  &    0.89   &   0.89   &   2.44   &   2.44  \\
\hline
Ba &  56  & &   0.74    &  1.22   &  1.96   & 2.00   &  &    1.85   &   1.91   &   2.72   &   2.73   \\
La &  57  & &   0.73    &  1.51   &  1.98   & 2.07   &  &    1.88   &   2.01   &   2.69   &   2.71   \\
Ce &  58  & &   0.82    &  1.37   &  2.16   & 2.20   &  &    2.05   &   2.11   &   2.76   &   2.77   \\
Pr &  59  & &   0.64    &  1.70   &  1.94   & 2.10   &  &    1.85   &   2.07   &   2.55   &   2.59   \\
Nd &  60  & &   0.71    &  1.65   &  2.04   & 2.15   &  &    1.94   &   2.10   &   2.61   &   2.64   \\
Sm &  62  & &   0.58    &  1.82   &  1.89   & 2.11   &  &    1.75   &   2.07   &   2.43   &   2.51   \\
Eu &  63  & &   0.17    &  1.96   &  1.19   & 1.95   &  &    1.06   &   2.00   &   1.72   &   2.10   \\
Gd &  64  & &   0.38    &  1.92   &  1.62   & 2.04   &  &    1.47   &   2.04   &   2.15   &   2.32   \\
Tb &  65  & &   0.25    &  1.95   &  1.38   & 1.98   &  &    1.23   &   2.01   &   1.90   &   2.18   \\
Dy &  66  & &   0.37    &  1.92   &  1.61   & 2.03   &  &    1.45   &   2.03   &   2.13   &   2.30   \\
Ho &  67  & &   0.24    &  1.95   &  1.37   & 1.98   &  &    1.21   &   2.01   &   1.88   &   2.17   \\
Er &  68  & &   0.45    &  1.90   &  1.75   & 2.08   &  &    1.56   &   2.05   &   2.25   &   2.38   \\
Tm &  69  & &   0.37    &  1.93   &  1.60   & 2.04   &  &    1.42   &   2.04   &   2.11   &   2.29   \\
Yb &  70  & &   0.70    &  1.79   &  2.09   & 2.23   &  &    1.86   &   2.11   &   2.57   &   2.62   \\
Lu &  71  & &   0.48    &  1.90   &  1.80   & 2.10   &  &    1.57   &   2.05   &   2.27   &   2.40   \\
Hf &  72  & &   0.86    &  1.64   &  2.27   & 2.33   &  &    2.05   &   2.17   &   2.74   &   2.76   \\
Ta &  73  & &   0.75    &  1.75   &  2.14   & 2.26   &  &    1.94   &   2.13   &   2.61   &   2.65   \\
W  &  74  & &   0.86    &  1.60   &  2.27   & 2.33   &  &    2.08   &   2.19   &   2.74   &   2.76   \\
Re &  75  & &   0.43    &  1.92   &  1.71   & 2.07   &  &    1.55   &   2.06   &   2.12   &   2.30   \\
Os &  76  & &   0.34    &  1.93   &  1.59   & 2.03   &  &    1.41   &   2.04   &   2.04   &   2.25   \\
Ir &  77  & &   0.06    &  1.98   &  0.77   & 1.92   &  &    0.63   &   1.99   &   1.18   &   1.96   \\
Pt &  78  & &   0.23    &  1.96   &  1.36   & 1.98   &  &    1.18   &   2.01   &   1.79   &   2.13   \\
Au &  79  & &   0.22    &  1.96   &  1.33   & 1.98   &  &    1.16   &   2.01   &   1.76   &   2.12   \\
Hg &  80  & &   0.92    &  1.74   &  2.35   & 2.42   &  &    2.17   &   2.28   &   2.77   &   2.79   \\
Tl &  81  & &   0.97    &  1.58   &  2.42   & 2.46   &  &    2.22   &   2.29   &   2.71   &   2.73   \\
Pb &  82  & &   3.19    &  3.19   &  4.06   & 4.06   &  &    2.99   &   2.99   &   3.25   &   3.26   \\
Bi &  83  & &   3.06    &  3.08   &  3.91   & 3.91   &  &    2.64   &   2.71   &   2.95   &   2.98   \\
\hline
{[ls/Fe]} &-& &   0.35    &  0.35   &  1.38   & 1.38   &  &    0.85   &   0.85   &   2.43   &   2.43   \\
{[hs/Fe]} &-& &   0.67    &  1.66   &  1.97   & 2.11   &  &    1.86   &   2.06   &   2.58   &   2.62   \\
{[hs/ls]} &-& &   0.32    &  1.31   &  0.59   & 0.73   &  &    1.01   &   1.21   &   0.15   &   0.19   \\
{[Pb/hs]} &-& &   2.52    &  1.53   &  2.09   & 1.95   &  &    1.13   &   0.93   &   0.67   &   0.64   \\
\hline                                                                                             
\end{tabular}}
\end{center}
\end{table*}

%% file: table8.tex
\begin{table*}                  
\caption{$\chi^2_{\small N}$ and distribution function $P$($\chi^2_{\small N}$)
(obtained from the related tables)
to test the goodness of the theoretical interpretation
of the giant HD 196944. 
$N$ is the number of elements considered. 
Case A: Y, Zr, La, Nd, Sm, Pb (N = 6).
Case B: Case A with Na and Mg (N = 8).
Case C: Na, Mg, Y, Zr, Ba, La, Ce, Nd, Sm, Eu, Dy, Er, Pb (N = 13).
Case D: Y, Zr, Ba, La, Ce, Nd, Sm, Eu, Dy, Er, Pb (N = 11).
Levels of confidence higher than 95\% are obtained with an AGB model
of initial mass $M$ = 1.5 $M_\odot$ and case ST/5.}
\label{chi}       
\begin{center}   
\resizebox{13cm}{!}{\begin{tabular}{lccccccccc}        
\hline
AGB Model		&case  & \multicolumn{2}{c}{Case A (N = 6)} &  \multicolumn{2}{c}{Case B (N = 8)} 
& \multicolumn{2}{c}{Case C (N = 13)} & \multicolumn{2}{c}{Case D (N = 11)}\\   
				&	 & $\chi^2_{\small N}$   &  P($\chi^2_{\small N}$)  &  $\chi^2_{\small N}$   &  P($\chi^2_{\small N}$)  
				&  $\chi^2_{\small N}$   &  P($\chi^2_{\small N}$)  & $\chi^2_{\small N}$   &  P($\chi^2_{\small N}$)    \\ 
(1)				& (2)    & (3) & (4) & (5) & (6) & (7) & (8) & (9) & (10) \\
\hline
$M$ = 1.3 $M_\odot$&	ST/12 &  4.7 &  58\% & 15.8 &  5\% & 18.2 & 15\% & 7.1 & 79\%\\
                  & ST/15 &  3.7 &  72\% & 14.8 &  6\% & 17.5 & 18\% & 6.3 & 85\%\\
                  & ST/18 &  4.9 &  56\% & 16.2 &  4\% & 19.6 & 11\% & 8.4 & 68\%\\
 \hline
$M$ = 1.5 $M_\odot$&	ST/3  &  3.0 &  81\% & 3.7  &  88\% & 6.7  & 92\% & 6.0 & 85\% \\
                  & ST/5  &  1.7 &  95\% & 3.5  &  90\% & 5.5  & 96\% & 3.7 & 98\% \\
                  & ST/6  &  2.0 &  92\% & 2.7  &  95\% & 6.1  & 94\% & 5.4 & 91\% \\
\hline                                           
$M$ = 2 $M_\odot$ &	ST/6   & 2.6 &  86\% & 7.7  &  46\% & 11.2 & 59\% & 6.0 & 87\%  \\
                &   ST/9   & 2.1 &  91\% & 7.8  &  45\% & 10.9 & 62\% & 5.1 & 93\%  \\
		        &   ST/12  & 3.1 &  80\% & 7.7  &  46\% & 11.6 & 56\% & 6.9 & 81\%  \\
\hline
\end{tabular}} 
\end{center}                               
\end{table*}

%% file: table9.tex
\begin{table*}
\caption{$^{12}$C/$^{13}$C, [C/Fe], [N/Fe] and [O/Fe] observed in CEMP-$s$ 
and CEMP-$s/r$ stars listed in Tables~\ref{table5_sindicator} (first group)
and~\ref{tablestellemancanti} (second group).
References and labels are the same as in Tables~\ref{table5_sindicator} and~\ref{tablestellemancanti}.
\citealt{goswami05}, G05, using low resolution spectra detected $^{12}$C/$^{13}$C in several CEMP-$s$.
For $^{12}$C/$^{13}$C measured by C06 \citep{cohen06}, the first value is detected
by using the C$_2$ band, the second by using the CH band. }
 \label{c12c13}
\centering
\resizebox{14cm}{!}{\begin{tabular}{lcccccccccccccccc} 
\hline
Stars     &  Ref.s &{\textit T}$_{\rm eff}$&log $g$& FDU&[Fe/H]& $^{12}$C/$^{13}$C & [C/Fe] & [N/Fe] & [O/Fe] &   Type \\
(1) & (2) & (3) & (4) & (5) & (6) & (7) & (8) & (9) & (10) & (11)   \\
\hline
BD +04$^{\circ}$2466&  P09	& 5100 & 1.8  & yes & -1.92  &15$^{+5}_{-3}$&   1.17  &  1.1	&  0.3	  & sI/$-$    \\  
CS 22183--015 &  C06        & 5620 & 3.4  & no? & -2.75  &  8--10       &   1.92  &  1.77   &   -         & sII/rII   \\        
CS 22880--074 &  A02,A07    & 5850 & 3.8  & no? & -1.93  &  $>$40       &   1.3   &  -0.1   &   -         & sI     \\  
CS 22881--036 &  PS01       & 6200 & 4.0  & no  & -2.06  &  :40	        &   1.96  &  1 	    &   -         & sII    \\
CS 22898--027 &  A02,A07    & 6250 & 3.7  & no  & -2.26  &  15$\pm$5    &   2.2   &  0.9    &   -         & sII/rII  \\
CS 22942--019 &  A02        & 5000 & 2.4  & yes & -2.64  &  8$\pm$2     &   2     &  0.8    &   -         & sI       \\
"             &  M10        & 5100 & 2.5  & "   & -2.43  & 12$\pm$1     &   2.14  & 1.15    &  0.97       & "       \\
CS 22948--27  &  BB05       & 4800 & 1.8  & yes & -2.47  &	-	&   2.43  &  1.75   &   -         & sII/rII \\        
"             &  A07        & 5000 & 1.9  & "   & -2.21  &  10$\pm$3    &   2.12  &  2.43   &   -         & "       \\      
CS 29497--030 &  I05        & 7000 & 4.1  & no  & -2.57  &  $>$10(S04)  &   2.3   &  2.12   &  1.48       & sII/rII  \\
CS 29497--34  &  BB05       & 4800 & 1.8  & yes & -2.9   &  -           &   2.63  &  2.38   &   -         & sII/rII   \\
"             &  A07        & 4900 & 1.5  & "   & -2.91  &  12$\pm$4    &   2.72  &  2.63   &   -         & "       \\
CS 30301--015 &  A02,A07    & 4750 & 0.8  & yes & -2.64  &  6$\pm$2     &   1.6   &  1.7    &   -         & sI   \\
CS 30322--023 &  M06        & 4100 & -0.3 & yes & -3.5   &  4$\pm$1     &   0.6   &  2.81   &  0.63       & sI     \\        
"             &  M10        & 4100 & -0.3 & yes & -3.39  &  4$\pm$1     &   0.8   &  2.91   &  0.63       & "    \\   
CS 31062--012 &  A02,A07,A08& 6250 & 4.5  & no  & -2.55  &  15$\pm$5    &   2.1   &  1.2    &   -         & sII/rII \\
CS 31062--050 &  A02        & 5600 & 3    & no? & -2.42  &  8$\pm$2     &   2     &  1.2    &   -         & sII/rII \\
HD 26         &  VE03       & 5170 & 2.2  & yes & -1.25  &$\sim$6(G05)  &   -     &   -     &   -         & sII  \\  
"             &  M10        & 4900 & 1.5  & yes & -1.02  &  9$\pm$2     &   0.68  &  0.94   &  0.36       & "       \\
HD 5223		  &  G06        & 4500 & 1.0  & yes & -2.06  &$\sim$6(G05)  &   1.57  &   -     &   -         & sII/$-$  \\
HD 187861     &  VE03       & 5320 & 2.4  & yes & -2.30  & -            &   -     &    -    &   -         & sII/rI  \\
"             &  M10        & 4600 & 1.7  & "   & -2.36  &  10$\pm$1    &   2.02  &  2.18   &  1.40       & "      \\
HD 196944     &  A02,A07    & 5250 & 1.8  & yes & -2.25  &  5$\pm$1     &   1.2   &  0.04   &   -         & sI     \\
"             &  M10        & 5250 & 1.7  & "   & -2.19  &  "           &   1.30  &  1.41   &  0.63       & "      \\
HD 206983     & M10        & 4200 & 0.6  & yes & -0.99 & 5$\pm$3;9(DP08)&   0.50  &  1.21   &  $<$0.23    & sI   \\
HD 209621	  &  GA10	    & 4500 & 2.0  & yes & -1.93  &$\sim$9(G05)  &   1.25  &   -     &   -         & sII/rI  \\
HD 224959     &  M10        & 4900 & 2.0  & yes & -2.06  & 4$\pm$2      &   1.77  &  1.88   &  1.10       & sII/rII  \\ 
HE 0143--0441 &  C06        & 6240 & 3.7  & no  & -2.31	 & $>$4		    &   1.98  &  1.73   &   -        & sII/rI  \\    
HE 0212--0557 &  C06        & 5075 & 2.15 & yes & -2.27  & 4.0$\pm$1.3  &   1.74  &  1.09   &   -         & sII/$-$   \\
HE 0336+0113  &  C06        & 5700 & 3.5  & no? & -2.68  & 2.5$\pm$1;7.5$^a$&   2.25  &  1.6    &   -         & sII     \\    
HE 0338--3945 &  J06        & 6160 & 4.13 & no  & -2.42	 & $\sim$10	    &   2.13  &  1.55   &  1.4        & sII/rII  \\
HE 1031--0020 &  C06        & 5080 & 2.2  & yes & -2.86  & 5$\pm$1.5    &   1.63  &  2.48   &    -        & sI/$-$   \\
HE 1305+0007  &  G06        & 4750 & 2.0  & yes & -2.03	 & 10           &   1.84  &   - 	&    -        & sII/rII    \\
"             &  Beers07    & 4560 & 1.0  & "   & -2.5	 & 9$\pm$2	    &   2.4   &  1.9	&  0.8        & "       \\
HE 1434--1442 &  C06	    & 5420 & 3.15 & yes & -2.39  & 5$\pm$1.5    &   1.95  &  1.4    &    -        & sI/$-$   \\
HE 1509--0806 &  C06        & 5185 & 2.5  & yes & -2.91  & 4$\pm$1.3    &   1.98  &  2.23   &    -        & sII(/$-$)  \\
HE 2148--1247 &  C03        & 6380 & 3.9  & no  & -2.3   &	     10	    &   1.91  &  1.65   &    -        & sII/rII  \\        
HE 2158--0348 &  C06        & 5215 & 2.5  & yes & -2.7   &6$\pm$1.8;3--5$^a$&   1.87  &  1.52   &    -        & sII    \\  
HE 2232--0603 &  C06        & 5750 & 3.5  & no? & -1.85  &   $>$6       &   1.22  &  0.47   &    -        & sI/$-$   \\          
LP 625--44    &  A02,A06    & 5500 & 2.5  & yes & -2.7	 &$\sim$20(A01) &   2.25  &  0.95   &  1.85:      & sII/rII  \\          
V Ari	      &  VE03       & 3580 & -0.2 & yes & -2.4	 & -            &    -	  &  -	    &	 -	  & sI/$-$ \\      													                              
"             &  Beers07    & 3500 & 0.5  & "   & -2.5   & 90$\pm$10    &   1.5   &  1.5    &  0.2        & "         \\                                                                                          
\hline                                                                                                   
\hline  
CS 22891--171 &  M10        & 5100 & 1.6  & yes & -2.25  &  6$\pm$2  &   1.56  &  1.67   &  $<$0.79     & sII/rII     \\                                                                                                                                 
CS 22956--28  &  M10        & 6700 & 3.5  & no  & -2.33  &  5$\pm$2  &   1.84  &  1.85   &  $<$2.47    &  sI/$-$       \\ 
"             &  S03        & 6900 & 3.9  & "   & -2.08  &  -        &   1.34  &   -     &  0.5:        & "           \\          
CS 22967--07  &  L04        & 6479 & 4.2  & no  & -1.81  &  $>$60    &   1.8   &  0.9    &  0.85        & sII(/$-$)    \\          
CS 29495--42  &  L04        & 5544 & 3.4  & no? & -1.88  &  7$\pm$2  &   1.3   &  1.3    &  0.64        & sI        \\          
CS 30315--91  &  L04        & 5536 & 3.4  & no? & -1.68  &  $>$60    &   1.3   &  0.4    &  0.51        & sI(/$-$)     \\          
CS 30323--107 &  L04        & 6126 & 4.4  & no  & -1.75  &  9$\pm$2  &   1.1   &  0.8    &  0.84        & sII(/$-$)      \\           
CS 30338--089 &  A07        & 5000 & 2.1  & yes &  -2.45 &  12$\pm$4 &   2.06  &  1.27   &	-           & sII(/rII) (L04)     \\
HE 0012--1441 &  C06        & 5730 & 3.5  & no? & -2.52  & $>$3      &   1.59  &  0.64   &   -          & sI/$-$      \\
HE 0024--2523 &  L03,C04    & 6625 & 4.3  & no  & -2.72  &6$\pm$1; $\sim$7(C04)&  2.6 & 2.1&  0.4       & sII(/$-$)   \\          
HE 0206--1916 &  A07        & 5200 & 2.7  & yes & -2.09  & 15$\pm$5	 &   2.1   &  1.61   &   -          & sII/$-$      \\
HE 0322--1504 &  Beers07    & 4460 & 0.8  & yes & -2	 & 6$\pm$2	 &   2.3   &  2.2	 &   - 	        & s/$-$          \\          
HE 0507--1430 &  Beers07    & 4560 & 1.2  & yes & -2.4	 & 9$\pm$2	 &   2.6   &  1.7	 &  1.1         & s/$-$          \\           
HE 0507--1653 &  A07        & 5000 & 2.4  & yes & -1.38&40$^{+20}_{-12}$; $\sim$7(G05)& 1.29&0.8&-      & sII/$-$       \\           
HE 1001--0243 &  M10        & 5000 & 2.0  & yes & -2.88  & 30$\pm$5  &   1.59  & 1.20    &  $<$1.92     & sI          \\   
HE 1045--1434 &  Beers07    & 4950 & 1.8  & yes & -2.5   &	20$\pm$2 &   3.2   &  2.8	 &  1.8         & s/$-$        \\           
HE 1157--0518 &  A07        & 4900 & 2    & yes & -2.34	 & 15$\pm$5	 &   2.15  &  1.56   &	-           & sII/$-$       \\
HE 1319--1935 &  A07        & 4600 & 1.1  & yes & -1.74  &	8$\pm$3	 &   1.45  &  0.46   &  -           & sII/$-$       \\        
HE 1410--0004 &  C06        & 5605 & 3.5  & no? & -3.02  & $>$3      &   1.99  &   -     &  1.18        & sI/$-$  \\          
HE 1419--1324 &  M10        & 4900 & 1.8  & yes & -3.05  & 12$\pm$2  &   1.76  &  1.47   & $<$1.19      & sI        \\  
HE 1429--0551 &  A07        & 4700 & 1.5  & yes & -2.47&30$^{+20}_{-10}$; $\sim$2(G05)&2.28&1.39&-      & sII/$-$  \\          
HE 1443+0113  &  C06        & 4945 & 1.95 & yes & -2.7   & 5$\pm$1.5 &   1.84  &  -      &   -          & sI/$-$       \\          
HE 1447+0102  &  A07        & 5100 & 1.7  & yes & -2.47  & 25$\pm$10 &   2.8   &  1.39   &   -          & sII/$-$      \\    
HE 1523--1155 &  A07        & 4800 & 1.6  & yes & -2.15  &  $\sim$2.5(G05)& 1.86  & 1.67 &   -          & sII/$-$    \\
HE 1528--0409 &  A07        & 5000 & 1.8  & yes & -2.61  & 12$\pm$5; $\sim$2.4(G05)&2.42 &  2.03   &-     & sII/$-$   \\                   
HE 2221--0453 &  A07        & 4400 & 0.4  & yes & -2.22  & 10$\pm$4; $\sim$13(G05)& 1.83 &  0.84   &-     & sII/$-$    \\          
HE 2228--0706 &  A07        & 5100 & 2.6  & yes & -2.41 &15$^{+5}_{-3}$& 2.32  &  1.13   &   -          & sII/$-$    \\             
\hline
\multicolumn{11}{l}{$^a$ The first value estimated using the C$_2$ band, the second using the CH band. }
\end{tabular}}
\end{table*}

%% file: table10.tex
\begin{table*}
\caption{Summary of theoretical interpretations for the stars listed
in Table 2. AGB initial mass, $^{13}$C-pocket, dilution factor, and 
initial $r$-enhancement are shown.
For stars, where Pb is not observed yet, we provide theoretical predictions.
The elements used for constraining the AGB models are listed in column~11 
(values in brackets are very uncertain).
If Eu is not detected we add the symbol (*) in column~9. 
In these cases, an average value of  [r/Fe]$^{\rm ini}$ = 0.5 is adopted
(see text).}
\label{summary1}
\begin{center}
\resizebox{14.5cm}{!}{\begin{tabular}{|l|llllllllllll|}
\hline
Stars            &  Ref.           & [Fe/H] & FDU  & Type & $M^{\rm AGB}_{\rm ini}$    & pocket   & dil  &  [r/Fe]$^{\rm ini}$     & [Pb/Fe]$_{\rm th}$  & NOTE                            \\ 
(1)              &  (2)            & (3)    & (4)  & (5)    & (6)  & (7) & (8)      &   (9) &     (10)  & (11)                    \\ %
\hline                                                                                                                        
BD +04$^\circ$2466& P09,I10,Z09   &  -1.92,-2.10 &  yes  & sI/$-$  & 1.3  & ST/9   & 0.9  &  0.5*    & -&  Na ([Fe/H]$_{\rm th}$ = -2)                                  \\ 
" & " & " & " & " & 1.5;2 & ST/3 &  1.8 &  0.5*    &  -  &   C, ([Fe/H]$_{\rm th}$ = -1.8)      \\
BS 16080--175    &  T05           &  -1.86 &  no & sII   & 1.5  & ST/3   & 1.2  &  0.7     & -& -             \\ 
" & " & "& "& "& 1.35 & ST/9 & 0.6 & "&  -& " \\  
" & " & "& "& "&  2   & ST/6 & 1.2 & "& -& " \\                         
BS 17436--058    &  T05           &  -1.90 &  yes & sI   & 1.5  & ST/5   & 1.6  &  0.7      & -& -                       \\ 
           "     &        "      &   "    &  "& "       & 1.4  & ST/9	& 1.2  &    -     & -&   -                                  \\ 
           "     &        "      &   "    &  "& "       & 1.3  & ST/12  & 0.7  &   -       & -& -                                    \\ 
           "     &        "      &   "    &  "& "       & 2    & ST/8   & 1.5  & - & -&   - \\  
CS 22183--015    &  A07,C06        &  -2.75  &  no?   & sII/rII & 1.3 & ST/12  & 0.0    &  1.5      & -&  Na, Sr, Y                                  \\ 
     "      &      JB02,T05,Lai07  &  -3.17  &  yes  & "        & 1.5 & ST/2 & 1.2 & -   &-& FDU \\   
     "      &     "                &  "      & "   & "        & 2   & ST/3 & 1.2 & -  & - &FDU  \\ 
CS 22880--074    &  A07,A02c,d     &  -1.93 &  no? & sI  & 1.3  & ST/6   & 0.9 &  0.0        &  -&Na, Y                                  \\ 
"   &  "   &  " &  "  & " & 1.2  & ST/6   & 0.4 &  -       &-&   "                                  \\ 
CS 22881--036    &  PS01           &  -2.06 &  no & sII   & 1.3  & ST/8 & 0.0    & 0.5      & 2.7&  Na                                  \\ 
CS 22887--048    &  T05           &  -1.70 &  no & sII/rI   & 1.4  & ST/2 & 0.0    &  1.0       &-& -                       \\ 
             "    &    "            &    "    &    " &" & 1.5;2  & ST$\times$1.2 & 0.3  &  "      & -& -                        \\ 
CS 22898--027    &  A07,A02c,d     &  -2.26 &  no  & sII/rII  & 1.3  & ST/12  & 0.0    &  2.0    & -& Na, ls                                  \\ 
CS 22942--019    &A02c,d,PS01,Sch08,M10&  -2.64,-2.43 &  yes & sI   & 2    & ST/50  & 0.7  &  0.5     & -& Na, FDU                  \\ 
CS 22948--27     &  BB05,A07 &  -2.47,-2.21 &  yes  & sII/rII  & 1.5  & ST/9   & 0.8 &   1.5      &  -&FDU  ([Na,ls/Fe]$_{\rm obs}$ low)           \\ 
              "   &      "           &    "   & " &   "   & (1.35) & (ST/15)  & (0.4)  &  -     & -& Na (FDU)                    \\ 
             "    &     "            &    "    &   " &"   & 2    & ST/18  & 0.7  &  -        &  - & FDU  ([Na,ls/Fe]$_{\rm obs}$ low)          \\ 
CS 22964--161A/B &  T08            &  -2.39 &  no  & sI  & 1.3  & ST/12  & 0.9  &  0.5     & -& Na                             \\ 
           "      &     "            & "       & "  & "  & 1.2  & ST/15  & 0.4  &     -     &   -&   -                                \\ 
CS 29497--030    &  I05            &  -2.57 &  no & sII/rII  & 1.35 & ST/9   & 0.0    &  2.0       & -&Na,Mg; (hs high)  \\ 
CS 29497--34     &  BB05,A07 &  -2.90 &  yes  & sII/rII  & 2  & ST/9   & 1.0    &  1.5      & -& FDU, Na    \\ 
CS 29513--032    &  R10            &  -2.08 &  no? & sI  & 1.2  & ST/9   & 0.3  &  0.3        &   -           -&                 \\ 
             "    &             "    &   "     &   "&  " & 1.3  & ST/9   & 1.4  &    -    & -& -                          \\ 
             "    &             "    &   "     &   "&  " & 1.5;2  & ST/3   & 2.4  &  -    & -& -                     \\ 
CS 29526--110    &A07,A02c,d,A08&-2.38,-2.06&  no  & sII/rII  & 1.3  & ST/6   & 0.0    &  1.5   & -& Na, Mg                                  \\ 
CS 29528--028    &  A07            &  -2.86 &  no  & sII/$-$  & 2    & ST/12  & 0.0    &  0.5*    &  3.8 & Na, high [ls,hs/Fe]$_{\rm obs}$                     \\ 
CS 30301--015    & A07,A02c,d      &  -2.64 &  yes  & sI  & 1.5  & ST/9   & 1.8  &  0.0       & -& Na, Mg                              \\ 
CS 30322--023    &  M06,A07,M10&-3.50,-3.25 & yes & sI  & 1.5;2  & ST/3   & 2.5  &  -1.0    & -& Na                                  \\ 
CS 31062--012    & A07,A02c,d,A08  &  -2.55 &  no & sII/rII   & 1.3  & ST/30  & 0.0    &  1.5     & -& Na, ([ls/Fe]$_{\rm obs}$ low)             \\ 
CS 31062--050&JB04,A07,A06,A02c,d &-2.42&  no? & sII/rII  & 1.3  & ST/12  & 0.2  &  1.6     & -& Na                  \\ 
HD 26            &  VE03,M10         &  -1.25,-1.02 &  yes &  disc (sII)   & 1.5;2  & ST/2   & 1.0    &  0.0       & -& FDU                                 \\ 
HD 5223          &  G06  &  -2.06 &  yes  & sII/$-$  & 2    & ST/15  & 1.2  &  0.5*    &-&  Na, FDU                             \\ 
HD 187861        
              &  (VE03)$^a$,M10 &  -2.36 &  yes  & sII/rI & 1.4   & ST/5   & 1.0  &  1.3     &  -&Mg \\  
HD 189711        &  VE03           &  -1.80 &  yes  & sI/$-$  & 1.5;2  & ST/24  & 0.9  &  0.5*     & -& FDU                                 \\ 
HD 196944        & A07,A02c,d,VE03,M10 &  -2.25 &  yes & sI   & 1.5  & ST/5   & 2.0 &  0.0         & -& Na                                  \\ 
HD 198269        &  VE03           &  -2.20 &  yes & sI/$-$   & 1.5;2  & ST/2   & 1.5  &  0.5*    & -& FDU                                 \\ 
HD 201626        &  VE03           &  -2.10 &  yes  & sII/$-$ & 1.5;2  & ST/3   & 1.3  &  0.5*    & -& FDU                                 \\ 
HD 206983        & M10, JP01       &  -0.99,-1.43 & yes   &  disc (sI) &  1.3 & $\sim$ ST  & 0.7  &   0.5   & -&- \\
"        & "       &  " & "   & " &  1.5;2  &$\sim$ ST & 1.6 &  "   & -&- \\
HD 209621        &  GA10           &  -1.93 &  yes  & sII/rI  & 2    & ST/15  & 0.9  &  1.0       & -& FDU; ([Na/Fe]$_{\rm obs}$ low)                         \\ 
HD 224959        &  VE03,M10           &  -2.20,-2.06 &  yes  & sII/rII   & 1.5;2  & ST/3   & 1.0    & 1.6    & -& FDU                                 \\ 
HE 0143--0441    &  C06            &  -2.31 &  no & sII/rI   & 1.3 & ST/9   & 0.0    &   1.0      & -& -                       \\ 
           "     &         "   &     "  & "  &" & 1.5  & ST/3   & 0.6  &  -        & -  &     "                             \\ 
           "     &         "   &     "  & " & "& 2    & ST/5   & 1    &  -        &    -&    "                             \\ 
HE 0202--2204    &  B05            &  -1.98 &  yes  & sI  & 1.3  & ST/9   & 0.7  &   0.0        &  2.0 & -            \\ 
            "  &       "        & "  & " &  "  & 1.5  & ST/3   & 1.6  &    -      &  - &                                 Mg  \\ 
            "  &       "        & "  & "&   "  & 2    & ST/6   & 1.7  &  -        & -&    -                                 \\ 
HE 0212-0557     &  C06            &  -2.27 &  yes & sII/$-$   & 2    & ST/8   & 0.8  &  0.5*    & 3.0 & FDU; ([Na,ls/Fe]$_{\rm obs}$ low)   \\ 
HE 0231-4016     &  B05            &  -2.08 &  no & sI/$-$   & 1.2 & ST/12  & 0.2  &   0.5*     & -& -                      \\ 
           "    &    "           &    "   & " &"  & 1.3  & ST/15  & 0.7  &     "     &  - &     -      \\ 
        "        &       "          &   "     &"    &"  & 1.5;2  & ST/5   & 1.6  &   -   & -&   -                                  \\  
HE 0336+0113     &  C06            &  -2.68 &  no? & sII & 1.4  & ST/55  & 0.0    &   0.5         & -& Mg                \\ 
             "    &       "        &  "     &" & "    & 1.5    & ST/45  & 0.2  &    -   &     -& - \\ 
             "   &        "       &   "    & " & "   & 2    & ST/45  & 0.3  &    -    &  -&  -                                  \\ 
HE 0338--3945    &  J06            &  -2.42 &  no  & sII/rII  & 1.3  & ST/11  & 0.0    &  2.0       & -& Na,Mg,ls                               \\ 
HE 0430--4404    &  B05            &  -2.07 &  no  & sI/$-$  & 1.2 & ST/9   & 0.0    &    0.5*    & -& -                       \\ 
           "    &    "           &    "   & " &"  & 1.3  & ST/15  & 0.7  &     "     &  - &   -         \\ 
             "    &    "             &   "     &  "   & - & 1.5;2  & ST/3   & 1.5  &   "    &    -&  -       \\ 
HE 1031--0020    &  C06            &  -2.86 &  yes  & sI/$-$  &  1.3  & ST/15   & 0.8 &    0.5*      & - &  FDU    \\
"                & "               & "     &  "   &  "     & 1.4  & ST/5   & 1.2  &    "     & -& FDU, (Mg)    \\ 
            "     &       "         &    "    &   "  &" & 2    & ST/5   & 1.6  &    "   &  -& FDU, (Mg)                         \\ 
HE 1105+0027     &  B05            &  -2.42 &  no  & sII/rII  & 1.3  & ST/9   & 0.0    &   1.8    & 3.0 & (Mg) \\ 
              "   &           "      & "       &   "  &" & 2    & ST/3   & 0.6  &    -   &  "&     -                         \\ 
HE 1135+0139     &  B05            &  -2.33 &  yes & sI   & 1.3  & ST/24   & 1.2  &  0.0         &  1.0 &         -           \\ 
              "  &       "          &   "     &  " &"   & 1.5;2  & ST/6  & 1.8  &  -        &    1.8 &    -                              \\ 
HE 1152--0355    &  G06  &  -1.27 &  yes  & disc (sI/$-$)  & 1.5;2  & ST/2   & 1.0;1.2    &  0.0*   & 2.0 & FDU                                 \\ 
HE 1305+0007     &  G06 &  -2.03 &  yes  & sII/rII   & 2    & ST/15  & 0.4 &  2.0    &  -&(FDU)           \\ 
HE 1430--1123    &  B05            &  -2.71 &  no  & sII/$-$ & 1.3  & ST/12  & 0.2  &    0.5*      & -& Mg                                  \\ 
           "     &   "              & "      &  "  &"  & 2    & ST/5   & 1    &   "   &  -&(Mg)                                  \\ 
HE 1434-1442     &  C06            &  -2.39 &  yes & sI/$-$   & 1.3  & ST/15  & 0.8  &  0.5*  & -& Na                                  \\ 
             "    &       "         &  "     &   "&"   & 1.4  & ST/12  & 1.2  &  -  & -&Na   \\  
HE 1509--0806    &  C06            &  -2.91 &  yes  & sII(/$-$)$^{a}$ & 1.4  & ST/18  & 0.7  &  (0.5*)$^{b}$  & -&FDU; (Mg)           \\ 
          "       &      "           &    "    & " &"    & 2    & ST/12  & 1.2  &  -   & -&-           \\ 
HE 2148--1247    &  C03        &  -2.30 &  no  & sII/rII  & 1.35 & ST/9   & 0.0    &  2.0     &  -&Mg,ls                                  \\ 
            "    &       "         &  "      & " &"   & (2)    & (ST/6)   & (0.7)  & -         &  -&(Mg)                                  \\ 
HE 2150--0825    &  B05            &  -1.98 &  no & sI/$-$   & 1.2  & ST/15  & 0.2  &   0.5*     & -& -                  \\ 
           "    &    "           &    "   & " &"  & 1.3  & ST/15  & 0.7  &     -     &  - &   -        \\ 
           "    &    "           &    "   & " &"  & 1.5;2  & ST/5   & 1.5 &   -   &-&    -               \\ 
HE 2158--0348    &  C06            &  -2.70 &  yes & sII  & 1.5  & ST/5   & 1.4  &  0.5      &  -&FDU                                 \\ 
  "     &     "            &  "      &   "  &" & 2    & ST/9   & 1.4  &  -        & -  &            -     \\ 
HE 2232--0603    &  C06            &  -1.85 &  no? & sI/$-$  & 1.2  & ST/12  & 0.4  &    0.5*      &-&  ls                     \\ 
             "   &  "               &  "      &  "  &"  & 1.3  & ST/12  & 1.0    &   -    &-& -                                     \\ 
HKII 17435--00532&  R08            &  -2.23 &  yes  & sI  & 1.5  & ST/12  & 1.8  &  0.3     &  1.4 &Na                                  \\ 
"                &  "              &   "    &  "    & "   &  "   & ST/5   & 2.1  & "        &  1.7 & Y  \\
LP 625--44        &  A02,A06        &  -2.70 &  yes & sII/rII   & 1.5  & ST/8   & 0.8  &  1.5  &  -&Na,Mg                               \\ 
V Ari             &  VE03           &  -2.40 &  yes & sI/$-$   & 1.5  & ST/30   & 0.9  &  0.5*  & -& FDU                                 \\ 
SDSS 0126+06     &  A08            &  -3.11 &  no  & sII/$-$  & 1.4  & ST/12  & 0.0    &  0.5*    & -& high [ls,hs/Fe], ([Na/Fe]$_{\rm obs}$ low)      \\ 
" & " & " & " &"& 1.5;2 & ST/6;ST/12 & 0.4 & - & -& (Na,Mg low) \\                         
SDSSJ 0912+0216  &  B10       &  -2.50 &  no & sII/rI    & 1.3  & ST/18  & 0.6  &  1.0    & -&  Na                                   \\ 
SDSSJ 1349--0229 &  B10       &  -3.00 &  no & sII/rII   & 1.35 & ST/15  & 0    &  1.5    &  -&  Na,Mg                                  \\ 
\hline  
\multicolumn{11}{l}{ $^{a}$ The solution provided here interprets the spectroscopic observations by \citet{masseron10}. 
The abundances by \citet{vaneck03} are more uncertain} \\
\multicolumn{11}{l}{due to the presence of molecular band contaminations in the spectra.}\\
\multicolumn{11}{l}{$^{b}$ For this star a low upper limit is measured for Eu.}        \\                                                                          
\end{tabular}}
\end{center}
\end{table*}

%% file: table11.tex
\begin{table*}                  
\caption{The same as Table~\ref{summary1}, but 
for the CEMP-$s$ and CEMP-$s/r$ stars listed in Table~\ref{tablestellemancanti}.
The symbol $\oplus$ indicates that all the solutions with AGB initial masses 
in the range 1.3 $\leq$ $M/M_\odot$ $\leq$ 2 are accepted.
$^{13}$C-pockets between brackets indicate very uncertain solutions due to
the limited number of spectroscopic observations.}
\label{summary2}       
\begin{center}   
\resizebox{17cm}{!}{\begin{tabular}{lccccccccc}                 
\hline    
Star               &Ref.&  [Fe/H]  & FDU  & Type & $M^{\rm AGB}_{\rm ini}$    & Pocket   & dil  &  [r/Fe]$^{\rm ini}$     & NOTE       \\
(1) & (2) & (3) &     (4) & (5) & (6) & (7) & (8) & (9) & (10) \\
\hline  
CS 22891--171       & M10&  -2.25  & yes &  sII/rII  & 2                  & ST/45       &  0.3       & 1.8    & (FDU)    \\                 
CS 22956$-$28      &M10,S03&-2.33,-2.08& no  & sI/$-$    & 1.3                & ST/80       &  0.0        & 0.5*  & [hs/ls] $\sim$ -0.6 \\                                                                                                                                                                                                                       
CS 22960$-$053      & A07&  -3.14  & yes & sI/$-$    & 1.5$^c$     & (ST/30)     &  (1.7)       & 0.5*   &   -    \\                                                                                                                                                                                                                   
CS 22967$-$07       & L04&  -1.81  & no  & sII(/$-$) & 1.3                & ST/9        &  0.0       &(0.0*)  & Na  \\        
CS 29495$-$42       & L04&  -1.88  & yes & sI     & 1.3                & ST/18       &  0.9       & 0.5   & Na, Sr \\                                                                                                                                                                                                                                 
CS 29503$-$010      & A07&  -1.06  & no  &  disc (s(II)/$-$) & 1.3$^c$     & ST/2 - ST/3 &  0.0       & 0.5*   &   Na \\                                                                                                                                                                                                           
CS 29509$-$027      & S03&  -2.02  & no  & sI/$-$    & 1.5$^c$     & ST - ST/18  &  0.0 - 1.5 & 0.5*   & - \\                                                                                                                                                                                                                   
CS 30315$-$91       & L04&  -1.68  & no? & sI/$-$    & 1.5$^c$     & ST/3        &  1.6       & (0.0*)    & - \\                                                                                                                                                                                                                          
CS 30323$-$107      & L04&  -1.75  & no  & sII(/$-$) & 1.3                & ST/3        &  0.3       & (0.0*)    & (Na)  \\                                                                                                                                                                                                                                 
CS 30338$-$089      & A07&  -2.45  & yes & sII/$-$   & 1.5 -- 2           & ST/2        &  $\sim$0.5 & 0.5*  &  FDU (Na)    \\                                                                                                                                                                                                                  
            "       & L04&  -1.75  & "   & sII(/rII) & "                  & ST/2 - ST/3 &  $\sim$0.5 & 1.8   &  FDU (Na)    \\                                                                                                                                        
G 18$-$24$^a$           & I10&  -1.62  & no  & (sI/$-$)$^a$    & 1.3$^c$      & ST/6       &  1.0       & 0.5*    & - \\                
HE 0012$-$1441      & C06&  -2.52  & no?  & sI/$-$    & 1.4 -- 2           & (ST/60)       &  0.0 -- 1.0   & 0.5*   &  Mg  \\                                                                                                                                                                                                                 
HE 0024$-$2523      & L03&  -2.70  & no  & sII(/$-$) & 1.3                & ST/9        &  0.0       & (0.0*) &  (Na),Mg   \\                                                                                                                                                                                                                    
HE 0131$-$3953      & B05&  -2.71  & no  & sII/rII  & 1.3           & ST/12       &  0.0       & 1.5    &  (Mg)  \\                                                                                                                                                                                                                      
HE 0206$-$1916      & A07&  -2.09  & yes & sII/$-$   & 1.5; 2             & (ST/5)        &  1.0       & 0.5*  &  FDU (Na)  \\                                                                                                                                                                                                                       
HE 0400$-$2030      & A07&  -1.73  & no?  & sII/$-$   & 1.5$^c$    & (ST/6)        &  1.9       & 0.5*  & -  \\                                                                                                                                                                                                                                 
HE 0441$-$0652      & A07&  -2.47  & yes & sI/$-$    & 1.5$^c$     & (ST/5)        &  (2.0)       & 0.5* &  - \\                                                                                                                                                                                                                         
HE 0507$-$1653      & A07&  -1.38  & yes & disc (sII/$-$)  & 1.5; 2             & (ST/2)        &  0.7       & 0.5*  &  FDU \\                                                                                                                                                                                                                         
HE 1001--0243       & M10&  -2.88  & yes & sI     &   1.3 - 2          & $\leq$ST/30 & 1.3-2.1    & 0.0   & -      \\   
HE 1005$-$1439      & A07&  -3.17  & yes & sI/$-$    & 1.5$^c$     & (ST/6)        &  (1.5)       & 0.5* &  -   \\                                                                                                                                                                                                                        
HE 1157$-$0518      & A07&  -2.34  & yes & sII/$-$   & 1.5; 2             & (ST/2 - ST/12)&$\sim$1.0   & 0.5*   &  FDU (Na)  \\                                                                                                                                                                                                             
HE 1305+0132        & Sch08& -1.92 & yes & sI/$-$    & 1.5$^c$     & (ST)        &  (1.2)       & 0.5*   &  -   \\  
HE 1319$-$1935      & A07&  -1.74  & yes & sII/$-$   & 1.5; 2             & (ST/2)       &$\sim$0.7  & 0.5*  &   FDU     \\
HE 1410$-$0004      & C06&  -3.02  & no?  & sI/$-$    & 1.2;1.4 & ST/24       &  0.2;1.5   & 0.5*  & Na  \\                                                                                                                                                                                                                           
HE 1419--1324       & M10&  -3.05  & yes & sI     & 1.5;2              & ST/2        & $\sim$2.0  & 0.5   & Mg      \\  
HE 1429$-$0551      & A07&  -2.47  & yes & sII/$-$   & 1.4 - 2            & (ST/5)      &  1.0 - 1.5 & 0.5*  &  FDU   \\                                                                                                                                                                                                                    
HE 1443+0113        & C06&  -2.07  & yes & sI/$-$    & 1.5$^c$     & ST/12       &  1.5       & 0.5*  &  FDU, (high Ba)\\                                                               
HE 1447+0102        & A07&  -2.47  & yes & sII/$-$   & 1.5; 2             & (ST/6)      &$\sim$0.5   & 0.5*  & (FDU, Na)   \\                                                                                                                                                                                                                     
HE 1523$-$1155      & A07&  -2.15  & yes & sII/$-$   & 1.5;  2            & (ST/2)       &  1.2       & 0.5*   & FDU  \\                                                                                                                                                                                                                          
HE 1528$-$0409      & A07&  -2.61  & yes & sII/$-$   & 1.5;  2            & (ST/2)      & $\sim$1.0   & 0.5*   &  FDU (Na, Mg)  \\                                                                                                                                    
HE 2221$-$0453      & A07&  -2.22  & yes & sII/$-$   & 1.5$^c$     & ST/2 - ST/24& $\sim$ 0.9  & 0.5*   &  -  \\                                                                                                                                                                                                                  
HE 2227$-$4044      & B05&  -2.32  & no? & sI/$-$    & 1.2 -- 2           & ST/12 - ST/6&  0.0  -- 1.5& 0.5*  & -  \\                                                                                                                                                                                                                     
HE 2228$-$0706      & A07&  -2.41  & yes & sII/$-$   & 1.5; 2             & ST/15       &  0.8       & 0.5*  &  FDU \\                                                                                                                                                                                                                
HE 2240$-$0412      & B05&  -2.20  & no  & sI/$-$    & 1.2 -- 2           & ST/12 - ST/6  &  0.0 -- 1.5      & 0.5*  &  -  \\                                                                                                                                                                                                                     
HE 2330$-$0555      & A07&  -2.78  & yes & sI/$-$    & 1.5$^c$     & (ST/5)        & (1.5)      & 0.5*  & -   \\                                                                                                                                                                                
(SDSS 0817+26)$^b$      & A08&  -3.16  & no  & (sI/$-$)  & -                  & -             & -          & -     &  -   \\ 
SDSS 0924+40	    & A08&  -2.51  & no  & sII/$-$   & 1.35 - 2           &ST/9 - ST/5 &  0.4 - 1.0 & 0.5*   & 	Na Mg\\                                                                                                                                                                                                           
SDSS 1707+58	    & A08&  -2.52  & no  & sII/$-$   & 2                  & ST/18       &  0.0       & 0.5* & high s	\\                                                                                                                                                                                                                
SDSS 2047+00	    & A08&  -2.05  & no  & sII/$-$   & 1.2; 1.3; 2        & ST/12 - ST/5 &  0.0 - 1.4 & 0.5*   & Na	\\                                                                                                                                                                                                           
\hline       
\multicolumn{10}{l}{ $^{a}$ We suggest caution in the intepretation of this star because no carbon and Eu have been detected. }   \\
\multicolumn{10}{l}{ $^{b}$ This star is excluded from CEMP-$s$ stars because no clear excess of carbon and $s$-process elements 
have been detected.} \\
\multicolumn{10}{l}{ $^{c}$ Solutions with AGB initial masses 
in the range 1.3 $\leq$ $M/M_\odot$ $\leq$ 2 are acceptable with a proper choice of the $^{13}$C-pocket and dilution.}          \\            
\end{tabular}}                     
\end{center}                               
\end{table*}